\documentclass[aps,notitlepage,twocolumn,superscriptaddress,nofootinbib,pra,10pt]{revtex4-1}
\usepackage[utf8]{inputenc}
\usepackage[pdftex]{graphicx,color}
\usepackage{amsmath}
\usepackage{amssymb}
\usepackage{epsfig}
\usepackage{amsthm}
\usepackage{tikz}
\usepackage{wrapfig}
\usepackage{dsfont}
\usepackage{latexsym,revsymb}
\usepackage{todonotes}
\usepackage{times}
\usepackage{enumerate,paralist}
\usepackage{mathtools}

\usepackage[hidelinks]{hyperref}
   \definecolor{jens}{rgb}{0,0,0}
  \newcommand{\new}[1]{{{\color{jens} #1}}}

\def\rh{\hat\varrho}

\def\p{\hat p}
\newcommand{\1}{\mathds{1}}
\newcommand{\id}{\1}

\DeclareMathOperator{\tr}{tr}
\newcommand{\ket}[1]{\left.\left|{#1}\right.\right\rangle}

\newcommand{\bra}[1]{\left.\left\langle{#1}\right.\right|}

\newcommand\vacket{{\ket{\emptyset}}}

\def\L{L}

\newcommand\f{\hat f}
\newcommand\fd{\hat f^\dagger}

\begin{document}
\title{Recovering quantum correlations in optical lattices from interaction quenches}
\author{Marek Gluza}\email{marekgluza@zedat.fu-berlin.de}\affiliation{\fu}
\author{Jens Eisert}\email{jense@zedat.fu-berlin.de}\affiliation{\fu}\affiliation{\hzb}

\newcommand{\fu}
  {{Dahlem Center for Complex Quantum Systems, 
    Freie Universit\"{a}t Berlin, 
    14195 Berlin, 
    Germany}}

\newcommand{\hzb}
{{Helmholtz-Zentrum Berlin f\"ur Materialien und Energie, 14109 Berlin, Germany}}

\begin{abstract}
Quantum simulations with ultra-cold atoms in optical lattices open up an exciting path towards understanding strongly interacting quantum systems. Atom gas microscopes are crucial for this as they offer single-site density resolution, unparalleled in other quantum many-body systems. However, currently a direct measurement of local coherent currents is out of reach. In this work, we show how to achieve that by measuring densities that are altered in response to quenches to non-interacting dynamics, e.g., after tilting the optical lattice. For this, we establish a data analysis method solving the closed set of equations relating tunnelling currents and atom number dynamics, allowing to reliably recover the full covariance matrix, including off-diagonal terms representing coherent currents. The signal processing builds upon semi-definite optimization, providing bona fide covariance matrices optimally matching the observed data. We demonstrate how the obtained information about non-commuting observables allows to lower bound entanglement at finite temperature which opens up the possibility to study quantum correlations in quantum simulations going beyond classical capabilities.
\end{abstract}

\maketitle

Quantum simulation experiments with {ultra-cold atoms} \cite{BlochSimulation} have lead to numerous insights 
into the physics of strongly correlated quantum systems, both in 
static \cite{Greiner-PRL-2001,StringOrder,HarperHofstadter,MonteCarloValidator} and in dynamical 
\cite{Trotzky_etal12,Kaufman,Emergence,Expansion,Hofferberth_etal07,Gring_etal12,SchmiedmayerGGE,1111.0776,Schweigler2017} regimes.
It is fair to say that there has been steady progress towards realizing the ambitious long-term goals set for quantum simulators \cite{Roadmap}.
Among them, the quest for understanding the precise mechanism underlying the physics of high-$T_c$ superconductors 
may take a particularly important role, driving forward significant experimental progress on quantum simulations with \emph{fermionic systems}
\cite{Mazurenko,ColdFermions,EsslingerReview,RoschTransport,Rom,PhysRevLett.120.243201,Koehl,PhysRevLett.122.110404}. In this line of research, achieving sufficiently cold temperatures is key and recently exciting progress has been reported, signified by an observation of very large anti-ferromagnets \cite{Mazurenko} with substantial evidence of string patterns \cite{Chiu251}. 
Thanks to advances towards alleviating this particular bottleneck~\cite{yang2020cooling}, it may in turn become a make or break issue to develop diagnostic tools regarding
genuine \emph{quantum correlations} in such systems.
Specifically, one can anticipate that not only methods for identifying the presence of entanglement will be needed, \new{which} can be done via entanglement witnessing, but it will be instrumental to have ways of unambiguously answering the overall physical question of \emph{how much entanglement} is there in a given quantum many-body system at finite temperature. 
Tools making this precise,  providing certification in this sense \cite{eisert2019quantum},
 should then offer to understand the role of quantum mechanical effects on the conductance of systems that have so far evaded modelling using numerical calculations.

In this work, we set out to present  diagnostic tools capable of tackling exactly these questions. 
They are based on information that is feasibly available via the so-called \emph{atom gas microscope}  \cite{Sherson-Nature-2010,Bakr-Science-2010,Weitenberg-nature-2011}.
Once this innovation had arrived, it allowed to observe string-order \cite{StringOrder,HarperHofstadter,MonteCarloValidator,Mazurenko}, time-dependent features of ordered \cite{Trotzky_etal12,Emergence,Greiner-PRL-2001,Expansion,1408.5148,1111.0776,Kaufman} and disordered models \cite{BlochMBL}. 
It should be stressed that these observations would have been much more limited without the atom gas microscope, say using just time-of-flight type measurements. 
The atom microscope has a strong limitation, however, as at any given time it provides information only about the \emph{local atom density}, which can be captured in terms of 
commuting operators in a quantum mechanical model. 
Because of that -- possibly surprisingly -- exploring \emph{quantum} correlations in optical lattices is by far not straightforward.

 In order to access expectation values of a set of non-commuting observables one must include some additional operations besides state preparation and direct measurements.
Tomographic schemes employing measurements along a single quantization axis in conjunction with Bloch sphere rotations 
constitute the simplest example.
In optical lattices, a sophisticated interference protocol \cite{Kaufman} has been demonstrated to reveal entanglement, but it is applicable to only small systems \new{(see also Ref.~\cite{bergschneider2019experimental})}.
Exploiting known time evolution in conjunction with feasible measurements
in recovery protocols in a general sense has been explored previously in 
Refs.~\cite{QuantumReadout,Efficient,PhysRevA.81.032126,PhysRevLett.120.050406,PhysRevLett.113.045303,ardila2018,qin2018charge,tarnowski2017characterizing,PhysRevA.89.061601,PhysRevA.98.033605Kollath,atala2014observation,PhysRevLett.117.170405,Shadows}. 

As we will show here, observing the density at various times after an interaction quench into a super-lattice enables new insights: It allows to recover expectation values of non-commuting observables and \emph{quantify} entanglement at finite temperature.
This can give clues as to \emph{why} a given system has a particular value of conductivity and what are the microscopic \emph{mechanisms} at play in the quantum system studied.
Put differently, understanding quantum correlations can allow for physical insights beyond the specific values of system parameters measured  by linear-response.
Linear-response measurements can be done both in quantum simulators and in materials. 
Concerning the latter, it has been possible to realize superconducting states at very high temperature.
If this will be done in optical lattices then by our method or its possible extensions it will be possible to investigate the role of coherent quantum mechanical effects in the system. 
This is typically not possible in materials and in fact access to sophisticated quantum observables can become one of the most important strengths of quantum simulations in optical lattices~\cite{Roadmap}.

\emph{Setting.}
The physical setting we have in mind is that
of \emph{fermionic atoms} in \emph{optical lattices} \cite{Rom,RoschTransport}. 
\new{That said, the technique carries over with litte modification
to any system in which excitation measurements 
and non-interacting dynamics are accessible.}
The discussion focuses on systems in one spatial dimension,
but it should be clear that similar ideas carry over to higher-dimensional lattices. 
It is also worth pointing out
that to an extent similar ideas have already proven highly useful and experimentally feasible in continuous quantum
field settings provided by cold bosonic atoms trapped on an atom chip \cite{QuantumReadout}.
Notation-wise, we denote fermionic annihilation operators associated with some degree of freedom at lattice site $x$
by $\f_x$. We put a focus on fermionic systems here but stress that the same machinery
works similarly for bosons as well.
The annihilation operators obey the canonical anti-commutation relations $\{\f_x,\fd_y\}=\f_x\fd_y+\fd_y\f_x =\delta_{x,y}$. 
The \emph{covariance matrix} $\Gamma$ of a state $\hat \varrho$ is defined as the collection of second moments  given by
\begin{align}
\Gamma_{x,y}= \langle\f^\dagger_x\f_y \rangle_{\hat \varrho}:=\tr[\f^\dagger_x\f_y {\hat \varrho}]\ .
\label{eq:cov}
\end{align}
This matrix is in general  a complex matrix  $\Gamma\in\mathbb C^{L\times L}$, $L$ being the
system size\footnote{If it was possible to directly measure currents then one would measure $\text{Re}[\Gamma_{x,y}]=\frac 12\langle\f^\dagger_x\f_y+ \f^\dagger_y \f_x \rangle_{\hat \varrho}$ and  $\text{Im}[\Gamma_{x,y}]=-\frac i2\langle \f^\dagger_x\f_y -  \f^\dagger_y \f_x \rangle_{\hat \varrho}$ (the latter vanishes oftentimes  given appropriate symmetries in the system).
}.
Additionally, we have $\Gamma = \Gamma^\dagger$ which means that it can be unitarily diagonalized by a Bogoliubov transformation of the type 
\begin{align}
\p_k = \sum_{x=1}^L U^*_{k,x}\f_x
\end{align} such that $\tilde\Gamma = U \Gamma U^\dagger$ with $\tilde \Gamma_{k,k'}= \langle\p^\dagger_k\p_{k'} \rangle_{\hat \varrho}$ is diagonal.
Noting that $\hat n_k = \p_k^\dagger \p_k$ are the number 
operators of the eigen-modes $\p_k$ we have that $\tilde \Gamma=\text{diag}(\lambda)$ has eigenvalues $0\le \lambda_k\le 1$ by the Pauli principle.
It is  useful to write $A\succeq B$ if $A-B$ is a matrix with a non-negative spectrum which yields
\begin{align}
  0\preceq \Gamma \preceq\id \ .
  \label{eq:cov_constr}
  \end{align}
This is a convex constraint that will be included in our reconstructions using \emph{semi-definite programming} methods~\cite{cvxpy}.
Due to statistical noise, a direct estimate $\Gamma^\text{(est)}$ of a covariance matrix $\Gamma$ may not fulfill this constraint, but the recovery procedure should find a \emph{physical} covariance matrix and hence taking into account Eq.~\eqref{eq:cov_constr} aids the reliability of the method.

A {non-interacting fermionic ({free}) evolution} conserving the particle number is generated by quadratic Hamiltonians
\begin{align}
  \hat H(h)=\sum_{x,y=1}^L h_{x,y} \f^\dagger_x \f_y
\end{align}
where $h=h^\dagger\in \mathbb C^{\L\times \L}$ is the coupling matrix. 
Most importantly, hopping on a line is captured by
\begin{align}
  \hat H_\text{NN}=\sum_{x=1}^{L-1}  \f^\dagger_x \f_{x+1}+\text{h.c.}
  \label{eq:NN}
\end{align}
where we use natural units in terms of the  tunnelling time throughout the note.
The Heisenberg evolution of mode operators reads
\begin{equation}
\label{eq:lin_opt}
\f_x(t)=e^{i t \hat H(h)}\f_x e^{-i t \hat H(h)}=\sum_{y=1}^L G^*_{x,y}(t) \f_y
\end{equation}
where $G^*(t)=e^{-i t h}$ is the \emph{propagator} matrix which can be computed efficiently in the system size $L$.
\new{Using Eq.\ \eqref{eq:lin_opt}} we see that the covariance matrix at time $t$ is 
\begin{align}
\Gamma(t)=G(t) \Gamma(0) G(t)^\dagger\ .
\label{eq:cov_t}
\end{align}
The geometry of the lattice is encoded in the propagator $G$ and by Eq.~\eqref{eq:cov_t} is imprinted in the correlations.
Our recovery method can be formulated independent of specifics of the lattice geometry.
However, for clarity only, we shall apply it to the setting of most immediate practical interest, namely for a chain with open boundary conditions \eqref{eq:NN}.

\emph{Tomographic read-out from interaction quenches.}
The core idea for reconstructing the covariance matrix $\Gamma$ is the following protocol.
The first step \new{ is to prepare} the state of interest:
\begin{align}
\text{(a) Prepare a fermionic state } \hat \varrho.
\end{align}
Indeed, we do not have to know anything about how the state is prepared precisely, 
specifically, whether during the preparation there are non-trivial interactions between the particles or not. 
The preparation, provides a density matrix and we would like to reconstruct the second moments $\Gamma$ of the possibly non-Gaussian state $\rh$.
\new{
In the second step, the task is:
\begin{align}
\text{(b) Quench to a free Hamiltonian } \hat H(h_\text{Quench})\ .
\end{align}
The ensuing coherent evolution should mix the information about the 
%<<<<<<< HEAD
currents into the occupation numbers dynamics. This quench should be rapid in terms
of the tunnelling times (which in practice means for ultra-cold atomic experiments that one
resorts to narrow Feshbach resonances), but does not have to be perfect.
%=======
%currents into the occupation numbers dynamics. 
%>>>>>>> 034ed628bd44d274dba16aaddc59e4f08cde3873
Finally, using the atom microscope, 
assess the occupation numbers $\hat N_x=\fd_x\f_x$
\begin{align}
  \text{(c) Measure } N_x(t) :=\langle \hat N_x\rangle_{\hat \varrho(t)}\ .
\end{align}
The complete tomographic protocol consists of performing the steps (a-c) for times $t=t_1,t_2,\ldots , t_K$, which can be chosen to be equidistant.

This prescription is at this point kept general on purpose, as it can be implemented in various setups and accordingly various quench Hamiltonians $\hat H(h_\text{Quench})$ are possible.
In this main text, we show that a suitable such choice is  
\begin{align}
\hat H(h_\text{Quench})=\hat H_\text{NN}+\sum_x\mu_x \hat N_x
\end{align}
where $\mu_x= x$ represents a gradient of the chemical potential.
Additional simulations presented in the appendix demonstrate 
that quenches into a doubled-up lattice or involving artificial magnetic fields can be also advantageous, 
in settings where this is feasible. This works because in (c) we are acquiring information about currents due to the equation
\begin{align}
  N_x(t) = \Gamma_{x,x}(t)= \sum_{y,y'=1}^L G_{x,y}(t)G_{x,y'}^*(t) \Gamma_{y,y'}(0)\ .
\end{align}
We find that generically the right-hand side will  depend on the off-diagonal matrix elements in the initial covariance matrix.
The reconstruction procedure makes use of the reversed direction of this equality and can be intuited as harvesting information about the currents from the response of the particle number dynamics following the quench.}

\begin{figure}[t]
\includegraphics[width=.9\columnwidth]{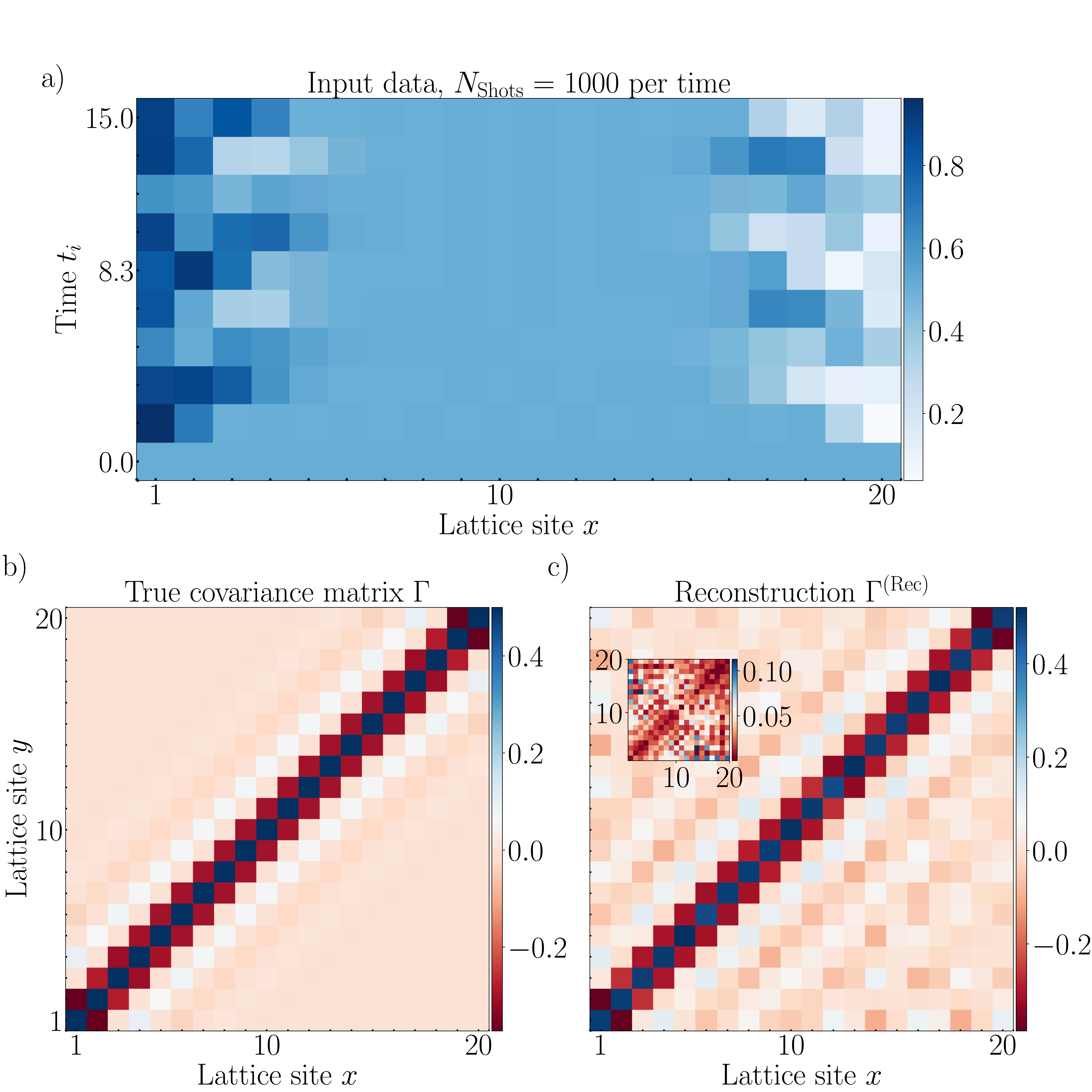}
\caption{\new{\textbf{Tomographic reconstruction.} \emph{a)} Input data for the reconstruction based on out-of-equilibrium data of local particle numbers $N_x(t_i)$ measured at $K=10$ equidistant times after the quench to nearest-neighbour hopping in the tilted lattice.
We model statistical fluctuations  by sampling occupation numbers $\mathcal N_\text{Shots} = 10^3$ times for each $t_i$ and estimating $N_x(t_i)$ via the empirical mean.
\emph{b)}~The input data have been obtained by evolving a thermal covariance of $\hat H_\text{NN}$ with inverse temperature $\beta=3$ such that there are relatively large currents to be recovered. 
\emph{c)} Results of the reconstruction $\Gamma^{(\rm Rec)}$. 
  The extent of deviations shown in the inset is $\max|\Gamma_{x,y} - \Gamma^{(\rm Rec)}_{x,y}|\approx 0.1$ and is explained by the fact that among the data some $N_x(t_i)$ can fluctuate statistical by two or three
  standard deviations.
}}
\label{fig:results}
\end{figure}

The reconstruction  is based on an algorithm that in a nut-shell takes a guess for the covariance matrix $\Gamma'$, evolves forward to the times $t_i$ where the particle number data has been measured and checks whether the extrapolated distribution of the particles $N_x(t_i;\Gamma')=\Gamma'_{x,x}(t_i)$ reproduces the data $N_x(t_i;\Gamma')\approx N_x(t_i)$.
In the next step, an improved guess $\Gamma''$ is obtained,
 such that the new observables $N_x(t_i;\Gamma'')$ are {closer} to the data
\begin{align}
  |N_x(t_i;\Gamma'') - N_x(t_i)| \le |N_x(t_i;\Gamma') - N_x(t_i)|\ .
\end{align}
By iterating this, the algorithm solves the following optimization task.
We collect all measured data into a vector $b$, and define a linear map $\mathcal A$ which from an input covariance matrix $\Gamma'$ produces the respective occupation numbers $N_x(t_i;\Gamma')$ in the same ordering as in $b$.
The reconstruction $\Gamma^\text{(Rec)}$ is \new{the} optimal solution to the 
optimization problem
\begin{align}
 \min_{0\preceq \Gamma' \preceq \id} \| \mathcal A(\Gamma') -b\|_2\ .
 \label{eq:opt}
\end{align}
The cost function is the 2-norm so we need to perform a least-square recovery problem with a positivity constraint \cite{cvxpy}.
Convexity of the problem guarantees an efficient convergence to a globally optimal solution with a polynomial runtime in the system size $L$ and desired accuracy $\epsilon>0$ \cite{cvxpy} and in practice takes a few seconds for $L\approx 40$.

\new{To exemplify the functioning of the method we consider thermal states $\rh_\beta = e^{-\beta \hat H_\text{NN}}/Z_\beta$, where $Z_\beta$ is the partition function, $\beta>0$ is the inverse temperature and  $\Gamma^{(\beta)}$ is the corresponding covariance matrix.
The results of the numerical reconstruction \cite{github_tomography} are presented in Fig.~\ref{fig:results}.}
The particle number measurement need not be perfect and  Fig.~\ref{fig:results} discusses reconstructions that include statistical noise \new{from necessarily finite numbers of measurements.}

In step (a), additional assumptions can be included such as translation invariance in the initial state or a finite correlation length.
\new{The quench is motivated by existing control functionalities, see, e.g., Ref.\ \cite{Preiss1229} for an experimental study showcasing a superb degree of coherence in the system when tilting the optical lattice.
We remark that other quenches that lead to non-trivial dynamics  (the covariance matrix is not a steady state of the quench Hamiltonian) can be considered. 
The reconstruction code \cite{github_tomography} does not depend on the quench Hamiltonian being nearest-neighbour, or whether there is a trap present so other variants are possible, see the appendix.}
If the couplings $h$ are real, then the tomography reconstructs only the real part of the currents.
This is enough for thermal states of quadratic Hamiltonians with no magnetic fields, see the 
appendix for reconstructions in their presence.
Finally, note that similar ideas have successfully been applied in an atom chip experiment for a continuum system  \cite{QuantumReadout} where some unknown stray interactions have been present \cite{Recurrence} but have been negligible in short time windows.

Quantum simulation studies of low-temperature systems in presence of Hubbard interactions with strength $U$ are of particular interest.
\new{Our method can be used to measure the second moments of interacting states by switching-off \new{$U\mapsto 0$} as fast as possible, e.g., by choosing narrow Feshbach resonances, see the appendix for comparisons of ramps of varying duration compared to the tunnelling time.}
Crucially though, it follows directly from the Lieb-Robinson bound \cite{LR} that even if the quench has a finite duration then only the {local} correlations will be affected but, e.g., the presence of \emph{long-range order} can be reliably inferred.
If the quench has a negligible duration compared to the relevant time-scales in the system then even local correlations will be faithfully reconstructed implying  the possibility of measuring also the \emph{kinetic} energy in addition to the Hubbard interaction term that can be measured with the atom microscope.
\new{
Our method does not assume translation invariance or thermality of the unknown state which are the corner stones but also limitations of existing methods \cite{PhysRevLett.125.113601,Nichols383,Brown379,Takasueaba9255,PhysRevA.99.033609,PhysRevX.7.031025,PhysRevResearch.2.023210,PhysRevX.10.011042} and hence can pave the way towards reading out the results of variational quantum simulations \cite{kokail2019self} in optical lattices for systems with a complicated connectivity graph.}

\emph{Quantitatively estimating fermionic mode entanglement.}
Let us now show how to analyze the second moments $\Gamma$ of a possibly interacting or non-equilibrium state to lower bound the so-called \emph{entanglement cost} \cite{PureBipartiteBennett,Horodecki}.
This statement is particularly appealing as it goes beyond merely showing that there is entanglement present, but provides an answer to the question of ``how much'' entanglement is there in the system~\cite{quant-ph/0607167,Audenaert06,Guehne,1302.4897}.
The entanglement cost $E_C$  quantifies mixed-state entanglement \cite{PureBipartiteBennett,Horodecki} as it is the asymptotic rate at which maximally entangled pairs must be used for the creation of a given state $\hat \varrho$ using {local operations with classical communication} (LOCC).
$E_C$ is broadly studied in quantum information theory but is not easy to access in practice and therefore, especially in context of experiments, lower bounds by means of practically measurable quantities are needed. 

We consider $\rh$ to describe a bipartite system  $A\cup B$, e.g., some number of sites in an optical lattice, and will explain how to lower bound $E_C(\hat \varrho)$.
Firstly, if the asymptotically optimal creation of the state $\hat \varrho$ via LOCC necessitates maximally entangled pairs at rate $E_C(\hat \varrho)$  then it could be that even more entangled pairs are needed for a system consisting of fermionic particles.
Indeed, any physical operation in this case is subject to the fermionic parity and total number super-selection rules (SSR) \cite{Earman,WignerWightmanWick} which further restrict LOCC.
However, in  Ref.~\cite{quant-ph/0404079}, it has been shown that the asymptotic rates do not change, i.e., $E_C^{\rm SSR}(\rh) = E_C(\rh)$ (see the appendix).
Secondly, we use that entanglement cost is lower bounded by distillable entanglement $E_D$ 
\cite{PureBipartiteBennett,Horodecki}
\begin{equation}
	E_C(\rh)\geq E_D ( \rh)\ .
\end{equation}
\new{In fact, in our discussion, we can equally well also refer to the distillable entanglement.}
Thirdly, \new{the} distillable entanglement is lower bounded by virtue of the \emph{hashing bound} \cite{HashingBound}
\begin{equation}
	E_D ( \rh ) \geq S ( \rh_A) - S(\rh),
\end{equation}
where for any state $\hat \sigma$ the von Neumann entropy is $S(\hat \sigma) = -\tr[\hat\sigma \log_2(\hat \sigma) ]$ and the subscript in $ \rh_A$ indicates the the reduction to subsystem  $A$.
Finally, the right hand side can be lower bounded by the same expression but now in terms of Gaussian entropies.
Specifically, let us denote by $S^{(\Gamma)} = S(\rh_\Gamma)$ to be the von Neumann entropy of a fermionic Gaussian state $\rh_\Gamma$ with the same second moments $\Gamma$ as $\rh$.
As shown in Refs.~\cite{Pastawski2016,GaussianChannel} we have
\begin{equation}
	S ( \rh_A) - S(\rh)\geq S^{(\Gamma_A)} - S^{(\Gamma)}:= E_G(\Gamma)\ .
\end{equation}
The Gaussian entropy $S^{(\Gamma)}$ can be easily computed from the recovered covariance matrix $\Gamma^\text{(Rec)}$, see the appendix for details.
Summarizing, for any bipartite state $\rh$ whose second moments $\Gamma$ one can measure using our method we have found a lower bound to the entanglement cost $E_G(\Gamma) \le E_C(\rh)$.

\begin{figure}[t]
\includegraphics[width=0.9\columnwidth]{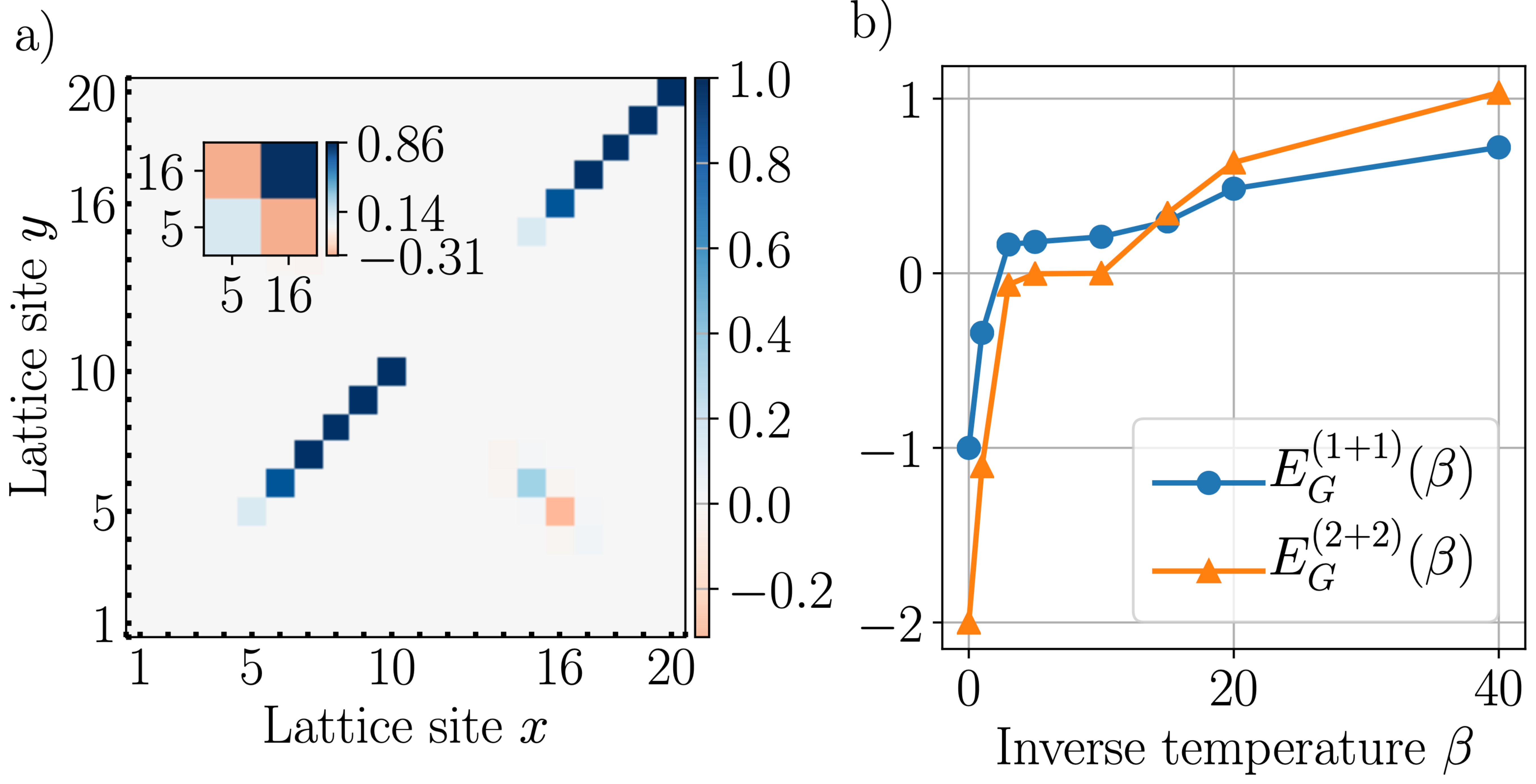}
\caption{\textbf{Entanglement cost at finite temperature.} \emph{a)} \new{The} covariance matrix $\Gamma^{(D)} = U_D \Gamma^{(\beta)}U_D^\dagger$ for \new{the inverse temperature} $\beta = 3$ after a \new{suitable} local transformation $U_D = U_A\oplus U_B$ \new{preserving the entanglement}. 
\new{The inset shows the sub-matrix of the covariance matrix reflecting one mode in subsystem $A$ and one in $B$.}
%Two modes have distinct correlations, the inset shows the covariance matrix after selecting one mode in $A$ and one mode in $B$.
% which is used to compute $E_G^{(1+1)}(\beta)$.
\emph{b)} \new{This figure shows how the lower bound for the entanglement depends on 
whether it is applied on one or two
modes in systems $A$ and $B$ each. Selecting merely one mode each provides a positive entanglement cost for relatively large temperatures ($E_G^{(1+1)}$), while  two modes are better suited to detect entanglement at low temperatures ($E_G^{(2+2)}(\beta)$).}}
%(\beta\approx 3)>0$ while selecting two modes indicates larger values at low temperatures where large rates $E_C$ may % be required.}
%Selecting one mode in each subsystem gives a non-trivial entanglement cost lower for relatively large temperatures $%E_G^{(1+1)}(\beta\approx 3)>0$ while selecting two modes indicates larger values at low temperatures where large rates %$E_C$ may be required.}
\label{fig:E_G}
\end{figure}

\emph{Entanglement cost at finite temperatures.}
\new{In what follows we discuss the application of the witness to again assess thermal states of $\hat H_\text{NN}$.}
As detailed in the appendix without increasing $E_C$ we can perform a \emph{local} unitary Bogoliubov transformation in subsystems $A$ and $B$ individually.
In Fig.~\ref{fig:E_G}a) we show the covariance matrix for $\beta=3$ after such a local transformation showing that essentially two modes are non-trivially correlated.
In Fig.~\ref{fig:E_G}b) we depict the entanglement cost lower bound $E_G(\beta)\new{:=} E_G(\Gamma^{(\beta)})$ 
as a function of inverse temperature.
We select either one or two modes in each subsystem and find a non-trivial lower bound $E_G(\beta)>0$ for sufficiently low temperatures.
Choosing one mode gives a non-trivial lower bound for higher temperatures than for two modes because the total entropy in the latter case tends to be larger at high temperature.
In contrast, at extremely low temperatures, the one 
mode lower bound saturates at its maximum value $E_G^{(1+1)}(\beta)\le 1$ while the two mode witness indicates that the entanglement cost of preparing $\rh_\beta$ is asymptotically larger than that of one maximally entangled pair.
Going beyond the Gaussian case, note that the second moments of low-temperature states will vary continuously with the strength of the many-body interaction \cite{kraus2010generalized}.
Thus we can be  confident that a non-trivial $E_C$ lower bound will be obtained for sufficiently weak interactions and low temperatures. Such states, e.g., in two spacial dimensions become difficult to treat numerically in practice, however, our reconstruction and entanglement quantification methods remain applicable for \new{quantum simulations.}

{\it Outlook.} \new{In this work, we
have shown how to recover the full covariance matrix of quantum states in
optical lattices by unifying atom microscope measurements with suitable quenches and
performing efficient reconstructions using semi-definite programming.
The method introduced here is reliable and versatile as it
does not depend on the geometry of the quench: Other ways of inducing visible particle number dynamics specific to a given setup can also be considered and some other ideas are discussed in the appendix. 
The prospect of advancing our data analysis by making use of recent theoretical ideas such as shadow estimation \cite{Shadows} or random operations that can be efficiently classically back-tracked \cite{Efficient,Elben} should also be noted.}
%In some ways, the idea of classically undoing a time evolution is in some aspects 
%reminiscent of notions of shadow estimation \cite{Shadows} or other recovery schemes making use of
%random operations that can be efficiently classically back-tracked \cite{Efficient,Elben}, here in a
%many-body regime involving non-linear quantities in the state.}
Building on the accessibility of full covariance matrices, including coherent currents, we have
\new{exploited an
 entropic witness} that allows to lower bound the entanglement cost and the \new{distillable entanglement}. 
This is a \new{quantitative} measure of entanglement \new{implying} a \new{substantially} 
stronger statement than merely showing its presence.
We have shown that our witness can give non-trivial values at finite temperatures.
This will remain true also for weak interactions as second moments should vary continuously in the strength of interactions.
We hence, have established a method to recover and quantify quantum correlations in optical lattice quantum simulations which is applicable even in regimes where numerical calculations cease being practical.

\emph{Acknowledgements.}
We thank M.\ Greiner, C.\ Gross, P.\ Preiss and R.\ Sweke for stimulating discussions.
This work has been supported by the ERC (TAQ), 
 the DFG (CRC 183, FOR 2724, EI 519/7-1), \new{the BMBF (DAQC),}
 and the Templeton  Foundation.   This  work  has  also  received  funding  from  the  European  Union's  Horizon2020  research  and innovation  programme  under  grant  agreement  No.\ 
 817482 (PASQuanS).

%\bibliography{References,QuantumSimulationReferences}

\newpage
\appendix
\new{
\section*{Appendix}
This appendix provides substantial detail of the arguments of the main text, and corroborates
the conclusions drawn there by further analysis. It is organized as follows.
In Section~\ref{sec:addrec}, we provide additional statistical and systematic analysis if the ``tilting'' protocol presented in the main text. We also show the results of recoveries based on two other examples of initial conditions, this time non-translation invariant.
In Section~\ref{sec:doubling}, we discuss an additional protocol based on doubling of the optical lattice.
In Section~\ref{sec:otherprotocols}, we provide more simulations testing a line of ideas for other protocols.
In Section~\ref{sec:checks}, we introduce some ideas for benchmarking in the experiment the doubling protocol, with the aim to experimentally infer how well it works.
In Section~\ref{sec:interactions}, we show additional simulations for the analysis of finite durations of switching off the interactions in a Hubbard chain.
Finally, in Section~\ref{sec:entanglement}, we go into great detail presenting further material on the entropic entanglement witness presented in the main text and its numerical evaluation.
\section{Additional reconstructions} 
\label{sec:addrec}
\subsection{Analysis of performance for finite sample sizes and single-shot statistics}

In this section, we provide a description of the procedure we have used to simulate the single-shot outcomes,
beyond the mere use of expectation values. This is motivated by experiments employing the atom microscope, e.g., 
the expectation value
$\langle \hat N_x(t)\rangle\approx 0.5$  has to be estimated by averaging a series of individual outcomes which are either $0$ (particle is absent) and $1$ (particle is present). We next describe how to simulate such single-shot measurements.

At a fixed time $t>0$, the occupation number operators commute $[\hat N_x(t), \hat N_y(t)]=0$, a feature that
implies that the measurement of one does not influence the measurement outcome of the other.
For this reason, the single-shot outcomes of the measurement of each $\hat N_x(t)$ can be ``parallelized'' and
obtained from the reduced density matrix on the mode located at $x$.
This is a $2\times 2$ matrix which can be taken to be
\begin{align}
\hat\varrho_x(t) = \text{diag}(p,q):= \begin{pmatrix} \langle \hat N_x(t)\rangle &0\\0&1-\langle \hat N_x(t)\rangle\end{pmatrix} \ .
\end{align}
Thus, to simulate the outcomes of atom number measurements at a position $x$ and time $t_i$ in the optical lattice in $\mathcal N_\text{sample}(t_i)$ experimental runs we can simply sample $\mathcal N_\text{sample}(t_i)$ times from the Bernoulli distribution with 
\begin{align}
p=\langle \hat N_x(t_i)\rangle.
\end{align}
Performing this step for all lattice sites, we obtain a simulation of the random outcomes of the measurements in the entire system.
Finally, this should be repeated for each measurement time $t_i$ to simulate the what would be the outcome of an experiment.
In order to assess the performance of the tomographic recovery procedure, we iterate over all times $t_i$ and positions $x$ and compute the empirical estimate of the mean particle number 
\begin{align}
\langle \hat N_x(t_i)\rangle \approx \frac 1 {\mathcal N_\text{sample}(t_i)} \sum_{j=1}^{\mathcal N_\text{sample}(t_i)} N_x^{(j)}(t_i)
\end{align}
 where $N_x^{(j)}(t_i)\in \{0,1\}$ is the $j$-th outcome of either having or not having the particle at position $x$ and time $t_i$.
The estimated atom numbers $N_x(t_i)$ are then taken as the input to the reconstruction.

We then repeat this approach a sufficiently large number of times and each time we run the tomography.
The finite number of single-shot outcomes leads to statistical errors in the estimation of the genuine 
expectation values $\langle \hat N_x(t_i)\rangle$ which then propagates into the fidelity of the reconstructed matrix.
What we find is that generically the noise leads to the introduction of stray off-diagonal correlations as shown in the main text, while the overall correlation pattern for entries of the covariance matrix of substantial magnitude are not overturned by noise.
For example, for $\mathcal N_\text{Total} = 10.000$ total state preparations distributed over $\mathcal N_\text{Times} =10$ measurement times we find that the relevant currents in the thermal covariance matrix of the nearest-neighbour Hamiltonian can be clearly discerned with good signal to noise ratio.
These parameters have been used in the reconstruction result shown in the main text based on a random sample of occupation numbers.
In Fig.~\ref{fig:results_statistics},
we show the cumulative results after running the tomography 500 times to assess the typical and worst-case results of the procedure.
We show that repeating the procedure many times it is possible that a rare event can happen in that a matrix element of the reconstructed covariance matrix will largely deviate from the true value.
However, such rare events are just a result of random sampling repeated many times (unlikely but possible events will happen eventually if one tries sufficiently often)  and most results concentrate around the median or mean taken over the maximal deviations.

\begin{figure}[h]
\includegraphics[width=0.9\columnwidth]{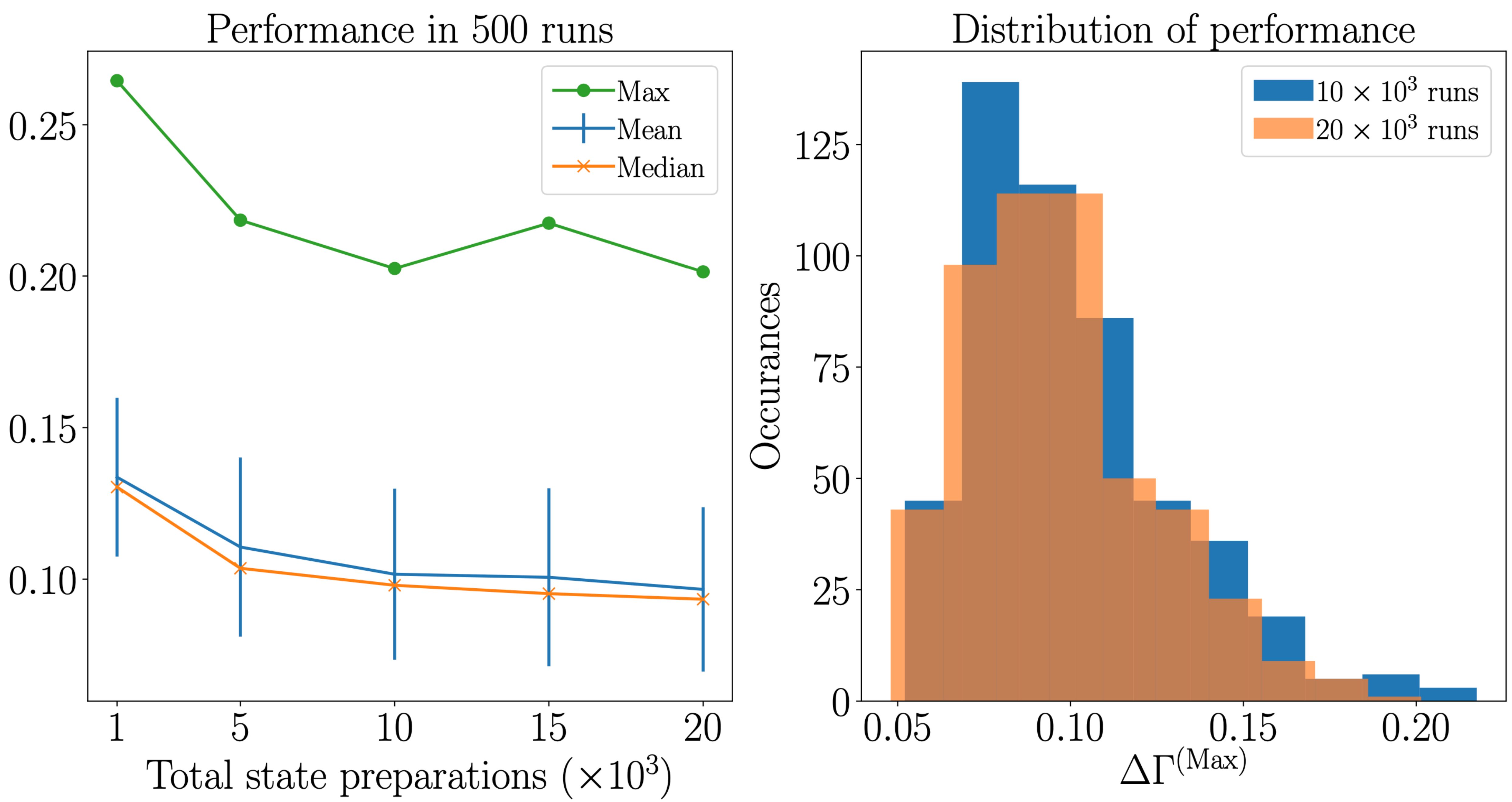}
\caption{\new{\textbf{Dependence of the reconstruction fidelity on the sampled atom occupation numbers.}
For each run $j=1,\ldots,  500$ of the tomography we consider the maximal deviation $\Delta \Gamma^{(j)} = \max|\Gamma_{x,y} - \Gamma^{(\rm {Rec}, j)}_{x,y}|$ of the reconstructed covariance matrix $ \Gamma^{(\rm {Rec}, j)}$  from the true covariance matrix $\Gamma$.
In panel \emph{a)}, we show the mean ($\mathbb E_j[\Delta \Gamma^{(j)}]$) and median ($\text{Median}[\Delta \Gamma^{(j)}]$) of the maximal deviation in dependence of the total number of experimental runs.
The presence of large maximal deviation ($\max[\Delta \Gamma^{(j)}]$) can be explained by the many repetitions of the reconstruction procedure and as is shown in the histograms in panel \emph{b)} most results concentrate around the mean and median while the large deviations happen only rarely.
}}
\label{fig:results_statistics}
\end{figure}

\subsection{Dependence on the number of measurement times}

Another aspect to consider, one we dedicate this section to, 
is how to best choose the discrete and equally spaced
measurement times. There are two parameters to consider when working with equidistant times. These
are on the one hand the total evolution time $T$ and on the other hand the 
number of measurement times $K$.
The quench evolutions that we consider in this work feature an effective causal cone confined 
by the \emph{Lieb-Robinson bound} \cite{HastingsKoma06}, 
which in turn is implied by the locality of the Hamiltonian
\begin{align}
  |G_{x,y}(t)| \le C_\text{LR} e^{-|x-y|+v_\text{LR}t}.
\end{align}
Here, we have denoted by $v_\text{LR}>0$ the Lieb-Robinson velocity. 
For this reason one can develop an intuition that the
total evolution time should be sufficiently long such that the particle number dynamics in the 
relation discussed in the main text
\begin{align}
  N_x(t) &=  \sum_{y,y'=1}^L G_{x,y}(t)G_{x,y'}^*(t) \Gamma_{y,y'}(0)\\&
  \approx \sum_{|y-x| \le 2v_\text{LR} t }^L\sum_{|y'-x| \le 2v_\text{LR} t }^L G_{x,y}(t)G_{x,y'}^*(t) \Gamma_{y,y'}(0)\ 
\end{align}
has a chance to be large.
Here, the approximation made has 
neglected the exponentially suppressed terms in the radius of an 
enlarged Lieb-Robinson cone $d_\text{LR cone}(t) = 2 v_\text{LR}t$.
For this reason, it seems that the evolution time $T>0$ should be large enough such that the influence of the relevant currents is not exponentially suppressed.
Even for ideal data, in such a case  a reconstruction should be possible in principle,
but issues of numerical stability can come into play and either way, in the 
presence of statistical noise one wants to have as visible dynamics as possible.
Fig.~\ref{fig:results_N} shows the result for $T=15$ and $L=10$.
We have performed a tomographic recovery 
based on measurements taken at varying interval spacing with the intention to see how the time spacing 
of the measurements influences the quality of the 
reconstruction.
As we see, quite quickly the reconstruction becomes accurate and the relevant currents 
narrow down 
around their true values.

\begin{figure}[h]
\includegraphics[width=0.8\columnwidth]{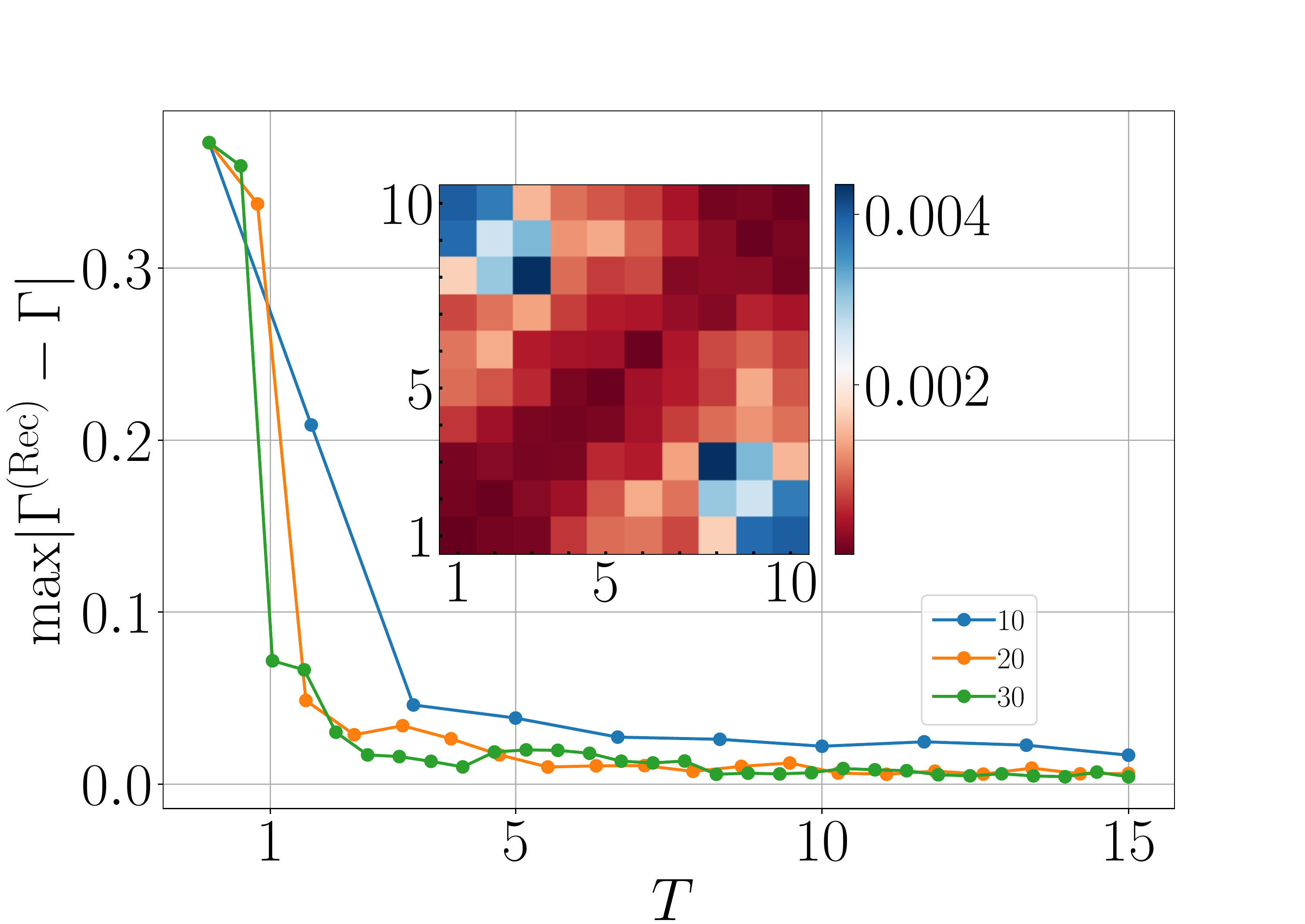}
\caption{\new{\textbf{Dependence of the reconstruction fidelity on the number of measurement times.}
For $L=10$ and $T=15$ we consider the time spacing $\Delta T:=  T/K =2/3,1,3/2$ ($K=10,20,30$, respectively).
For each $\Delta$ we perform a tomographic recovery without the presence of statistical noise for varying number of consecutive points with the respective bullet corresponding to the end of the interval.
We see a rapid decay of the reconstruction residue, reflecting the fact 
that the bulk features of the correlations can be inferred with relatively little input.
The slight non-monotonicity in the $K=30$ curve can be explained that for tomographically 
incomplete inputs there is a number of covariance matrices consistent with the input data and the slightly increased deviation when providing more input can be an unlucky nudge in the wrong direction in this set of covariance matrices which are equivalent on the level of $N_x(t)$. Such behaviour becomes even less pronounced for sufficiently large numbers of inputs.
The inset shows the full reconstruction residue matrix with elements 
$|\Gamma^{(\rm Rec)}_{x,y}-\Gamma_{x,y}|$ for $K=30$ input times and indicates 
that the source of the deviations is the difficulty in constraining the far-away 
correlations based on the input dynamics. When appropriate, suppressing such artefacts 
can be done by making use of prior knowledge and penalizing additionally 
correlations between far-way points.
}}
\label{fig:results_N}
\end{figure}

\subsection{Reconstruction of a disordered initial state}

In the example discussed in this section,
we consider a thermal state of the Anderson insulator model
\begin{align}
  \hat H_\text{Anderson} = \hat H_\text{NN} + \sum_{x=1}^L \nu_x \hat N_x
\end{align}
where $\nu_x\in [-0.3,0.3]$ has been chosen independently and identically at random for each $x$.
Using such a Hamiltonian, we construct a thermal ensemble 
\begin{align}
\rh_\text{Anderson}(\beta) = e^{-\beta \hat H_\text{Anderson}}/Z_\beta, 
\end{align}
where $Z_\beta>0$ again denotes the partition function.
We take the inverse temperature to be $\beta=3$ and denote by $\Gamma^{(\rm Anderson)}$ 
the covariance matrix of $\rh_\text{Anderson}(\beta)$.

Fig.~\ref{fig:Anderson} shows a reconstruction for one realization of such a random initial condition.
This is an instructive initial condition because we see that the reconstruction correctly
detects  the regions where the particles are delocalized within the small regions allowed by the typical Anderson localization length.
In those restricted regions, the particles roam freely 
as evidenced by the off-diagonal correlations in the initial state but also in the reconstructed state, despite the random noise coming from taking into account a finite number of state preparations.

This demonstrates that our tomographic reconstruction method can detect the coherences in random ensembles where there is essentially no prior information to the character of the correlations.
To the best of our knowledge, there is no other method delivering these important observables as very many methods make simplifying assumptions such as translation invariance or thermal equilibrium of a known model.
It is beyond the scope of this work to illustrate the versatility of the method in all of its ramifications,
but it is worth pointing out that one can equally well perform such reconstruction for the second moments of interacting states and thus study many-body localization.
In this context, the covariance matrix has been shown to reveal insights about this genuinely non-Gaussian phenomenon~\cite{FHM}.

\begin{figure}[h]
\includegraphics[width=0.9\columnwidth]{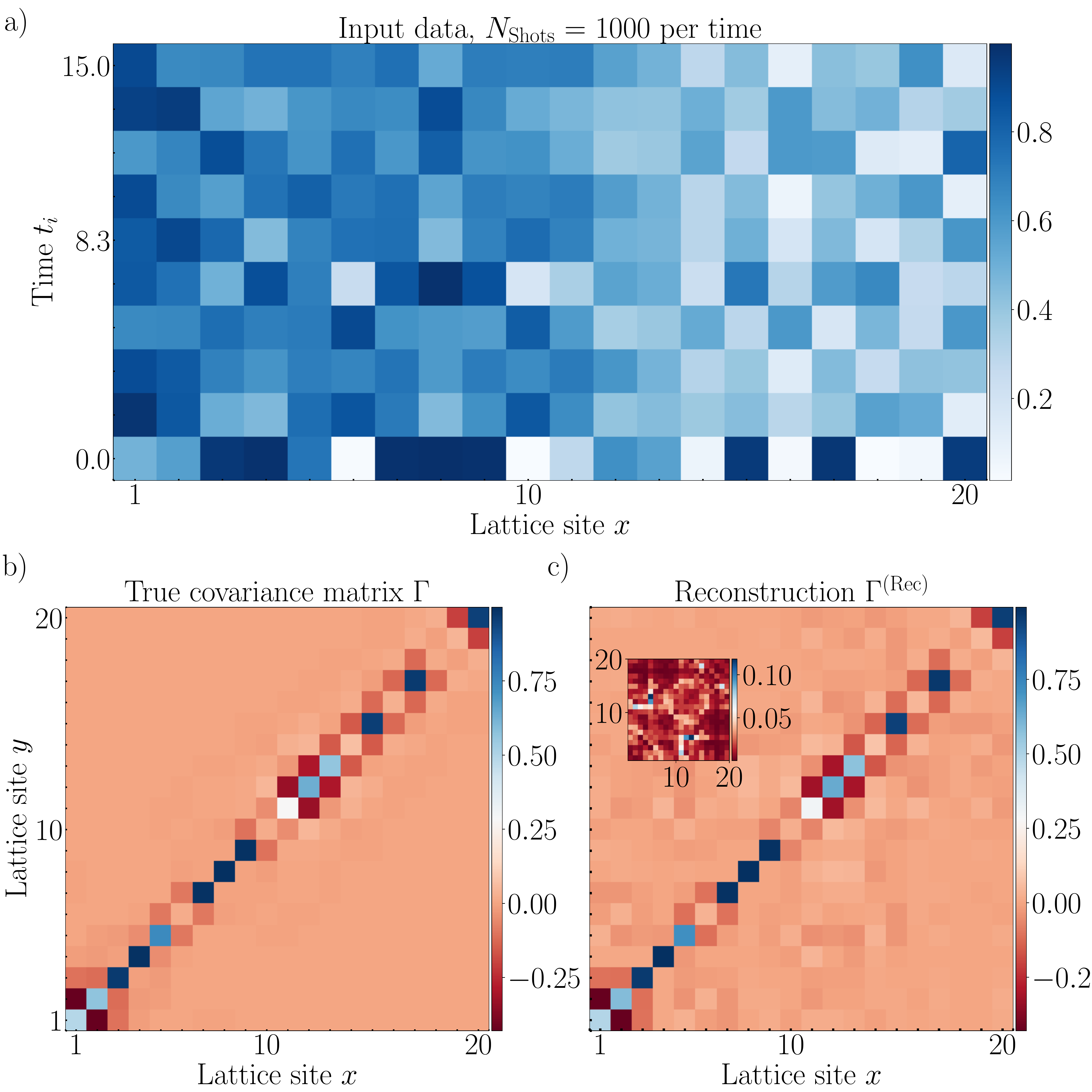}
\caption{\new{\textbf{Reconstruction of a disordered initial condition.} After sampling the on-site disorder $\nu_x$ we have computed the thermal covariance matrix of the Anderson model with $\beta =3$. 
The reconstructed covariance matrix (panel \emph{c)} closely matches the true covariance matrix (panel \emph{b)} with the coherences being accurately reproduced with a clear signal to noise ratio.
}}
\label{fig:Anderson}
\end{figure}

\subsection{The role of compressibility when reconstructing based 
on gradients of the chemical potential}
In this section, we comment on a curious feature of the particle number 
dynamics that can be seen in the figure shown in the main text. 
This is the feature that the bulk of the system is not influenced by the quench adding a 
gradient of the chemical potential, but rather only the edges are affected by this.
One could think that the lack of visible atom 
number dynamics would hint at an absence of currents.
Indeed, if we consider any internal quench in a non-interacting and isolated system, 
then states entirely lacking currents would feature any visible dynamics.
This can be seen by translating the above statement into some Greens function $t\mapsto G_\text{Quench}(t)$,
 the unitary single-particle Greens function.
Homogenous states lacking currents feature 
covariance matrices which are multiples of the identity $\Gamma^{(\rm No Currents)}=\alpha \id$
and for any $0<\lambda<1$ we have that there will be no dynamics 
\begin{eqnarray}
\Gamma^{(\rm No Currents)}(t)&=& G_\text{Quench}(t) \Gamma^{(\rm No Currents)}(0)G_\text{Quench}(t)^\dagger\nonumber\\
&=&\Gamma^{(\rm No Currents)}(0)G_\text{Quench}(t) G_\text{Quench}(t)^\dagger\nonumber\\
&=:&\Gamma^{(\rm No Currents)}(0).
\end{eqnarray}
We are then lead to the following conclusions: Covariance matrices invariant under \emph{all} non-interacting quenches in an isolated system are those reflecting product states (or infinite temperature states).
In particular, these states are incompressible, and gradients of the chemical potential do not modify their density distribution.

In the main text, we have seen 
that the bulk of the system remains homogenous, but as it turns out, it reveals the information that the system is homogenous everywhere and has currents in the bulk matching the currents present at the edges.
This can be seen in Fig.~\ref{fig:inhomogenoustest} where we consider a quench into the nearest-neighbour Hamiltonian together with a chemical potential gradient (exactly as in the main text), but now for a fiducial state that can be thought of as a composition of a chain in the superfluid state, then a very hot chunk in the infinite temperature state and then again a superfluid. As shown in the figure, the system responds with markedly different atom number dynamics than a homogeneous state.
We conclude from this that the atom number dynamics influenced by the gradients of the chemical potential 
is sensitive to the presence or absence of initial correlations.
The depletions and increases of the density of the gas over time we can interpret as ``breathing'' 
of the gas related to spatially varying compressibility. By the equation of state compressibility is related to 
the temperature. 

\begin{figure}[h]
\includegraphics[width=0.9\columnwidth]{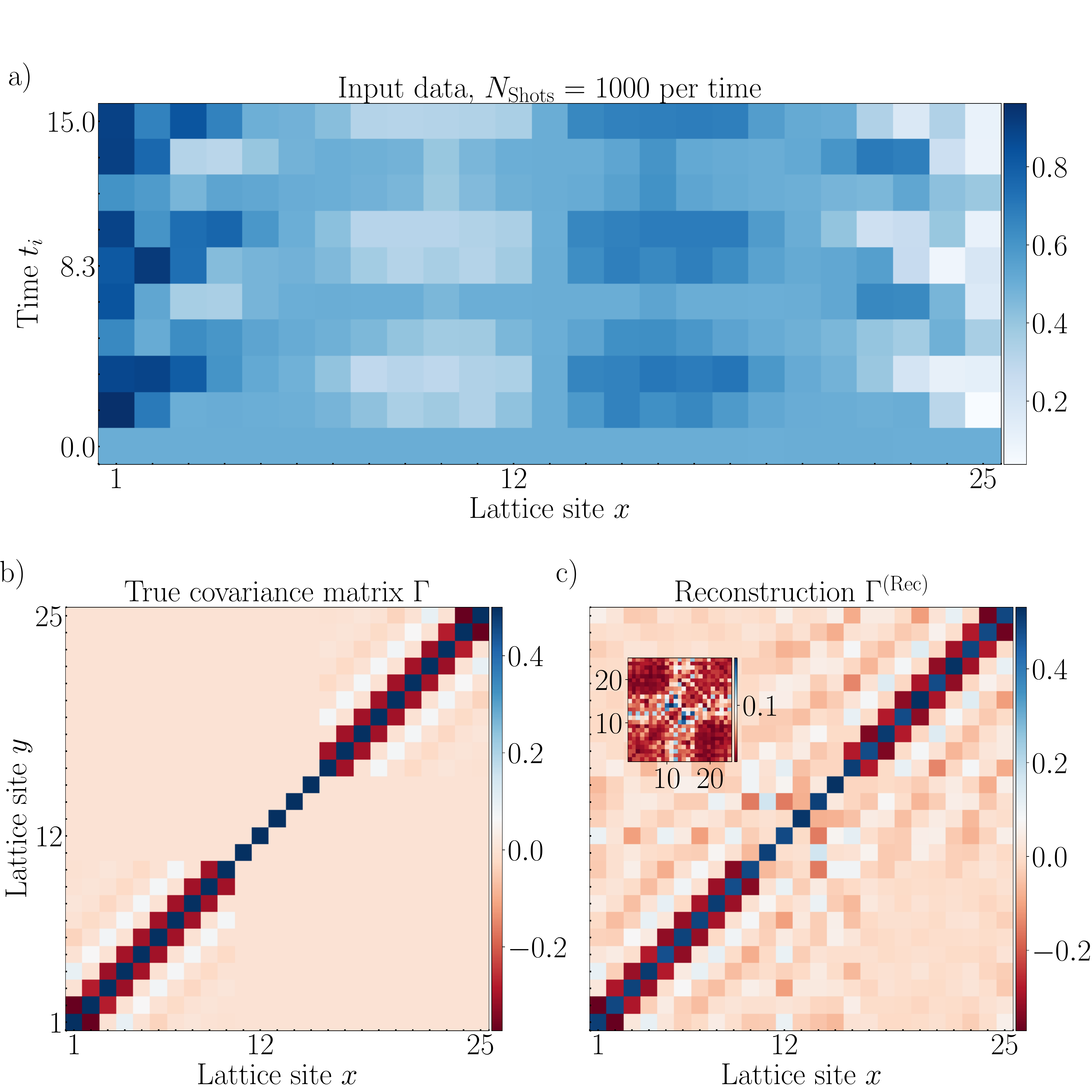}
\caption{\new{\textbf{Reconstruction of an inhomogeneous state.}  We exemplify the reconstruction of an initial condition which includes a block with no correlations. \emph{a)} The presence of this block visibly modifies the atom number dynamics. \emph{b)} The initial covariance matrix is faithfully reconstructed \emph{(c)} even despite the noise (see inset for the difference).}
}
\label{fig:inhomogenoustest}
\end{figure}

\section{Reconstructions using a quench into a sub-lattice}
\label{sec:doubling}

The full recovery of the covariance matrix $\Gamma$ is actually possible by means of suitable 
choices of the quench involved in the protocol. In this section, we provide some ideas for a more sophisticated protocol which involves a step of quenching into a sub-lattice.
To state the protocol, we follow along the lines presented in the main text.
The first step is again preparing the state of interest:
\begin{align}
\text{(a) Prepare a fermionic state } \hat \varrho.
\end{align}
We then split the quench into two steps.
As the second step, the task is:
\begin{align}
\text{(b') Double-up the lattice locally } \f_{x}\mapsto \f_{2x-1}\ .
\end{align}
In Fig.~\ref{fig:results_main_sublattice}b), we illustrate this by assuming that the system has initially been in a thermal state of Eq.~\eqref{eq:NN} with a translationally invariant covariance matrix and a finite correlation length.
We assume the doubling is fast and the \emph{in-between} sites are still unoccupied while the correlations between the original sites have remained unchanged which gives rise to a distinct checker-board correlation pattern.
The fast doubling is in many experimental situations actually a highly plausible assumption and perfectly
feasible.

Next, we shall use coherent evolution under Eq.~\eqref{eq:NN} to mix information about the coherent current into the particle number occupation operators:
\begin{align}
\text{(b'') Quench to a free Hamiltonian } \hat H_\text{NN}.
\end{align}
In  Fig.~\ref{fig:results_main_sublattice}a), we depict the resulting atom number dynamics.
In the last step, again the local atom numbers should be  measured
\begin{align}
  \text{(c) Measure } N_x(t) :=\langle \hat N_x\rangle_{\hat \varrho(t)}\ .
\end{align}
As shown in  Fig.~\ref{fig:results_main_sublattice}c), 
an accurate reconstruction can be achieved from this protocol even in the presence of finitely many state preparations.

\begin{figure}[h]
\includegraphics[width=0.9\columnwidth]{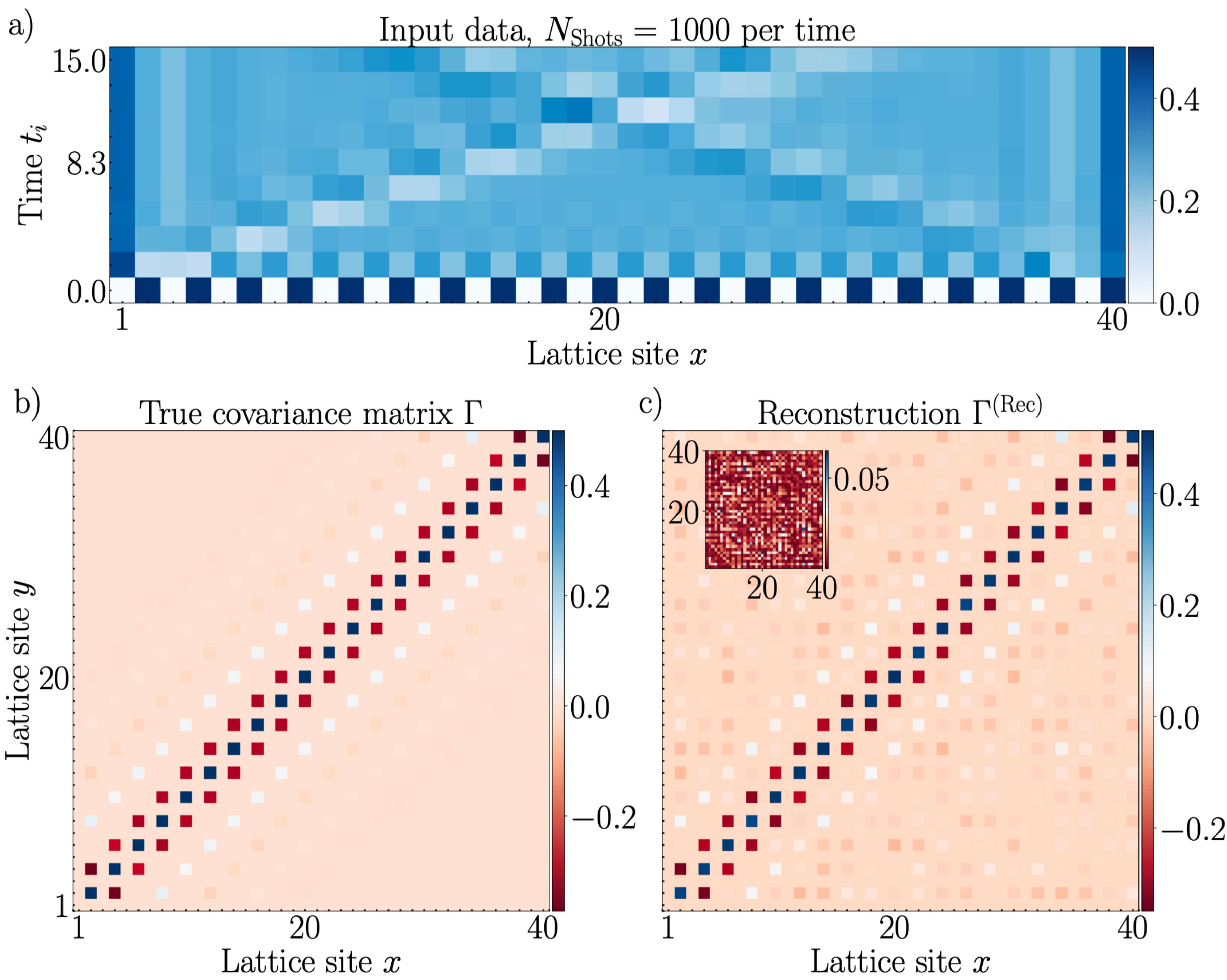}
\caption{\new{\textbf{Tomographic reconstruction.} \emph{a)} Input data for the reconstruction based on out-of-equilibrium data of local particle numbers $N_x(t_i)$ estimated from $\mathcal N_\text{Shots}=10^3$ shots per measurement time $t_i$ with in total $K=10$ equidistant times after the quench to nearest-neighbour hopping in the superlattice.
\emph{b)} The input data have been obtained by evolving a thermal covariance with inverse temperature $\beta=3$. We 
have chosen a temperature so that there are relatively large currents to be recovered. 
The covariance matrix $\Gamma$ is shown after step (b') after the sub-lattice has been created. 
Note that besides the new checker-board pattern, the correlations between sites are assumed to be exactly preserved.
\emph{c)} Results of the reconstruction $\Gamma^{(\rm Rec)}$ and the extent of the
deviations shown in the inset is $\max|\Gamma_{x,y} - \Gamma^{(\rm Rec)}_{x,y}|\approx 0.05$.
}}
\label{fig:results_main_sublattice}
\end{figure}

\subsection{Ramping up instead of quenching into doubled lattice}

The protocol involving the step (b') where the lattice is doubled is more complex, 
in that it involves an additional step. This step is what we assess in this section with respect to its
feasibility. From the perspective of current experimental implementations, performing this step seems rather straightforward: The periodicity of the lattice is controlled by the trapping lasers which create the optical lattice and can be tuned rather fast.
In what follows, we model  the systematic imperfection stemming from such a doubling ramp as follows.
For this purpose, we introduce a new fermionic mode in-between each consecutive pair of modes in the initial lattice.
We initialize that mode in a vacuum state. This, in particular, 
implies that there is no correlations to any other site and so for simulating a finite 
duration of the doubling the initial covariance matrix will be 
\begin{equation}
\Gamma^{\rm (b')} = \Gamma^{(\text{Ini})}\otimes \begin{pmatrix} 0&0\\0&1\end{pmatrix}.
\end{equation}
Here, we  involve a method using the Kronecker product to concisely write the effect of the doubling in (b') and $\Gamma^{\rm (b')}$ if the doubling was perfect this would be the perfect state preparation.
We then model the gradual appearance of the in-between sites over time $t$ by performing an interpolation of the Hamiltonians
\begin{align}
  \hat H_\text{Doubling}(t) = (1-t/T_\text{Doubling}) \hat H_\text{NNN}+t/T_\text{Doubling}) \hat H_\text{NN}.
\end{align}
Here, $T_\text{Doubling}>0$ parametrizes the duration of the doubling and $\hat H_\text{NNN}$ is 
the next-next-nearest neighbour hopping Hamiltonian which arises from the fact that the next-nearest neighbour sub-lattices were initially the system of interest and the auxiliary unoccupied sites. We then perform an evolution to time of $\Gamma^{\rm (b')}$ under the time-dependent Hamiltonian $\hat H_\text{Doubling}(t)$ using a standard Trotterization scheme and obtain the covariance matrix $\Gamma^{\rm (b')}(T_\text{Doubling})$ which encodes the imperfections stemming from the finite duration of the doubling.
This systematic deformation $\Gamma^{\rm (b')} \mapsto \Gamma^{\rm (b')}(T_\text{Doubling})$ would not be detected by the tomographic reconstruction as it would assume an instantaneous quench (though if the ramp is known it can be included in the parametrization of the reconstruction).

In Fig.~\ref{fig:app_doubling_ramp}, we display the results of an analysis of the fidelity of the state preparation involved in step (b') by performing an additional numerical analysis.
Qualitatively, one observes that the checker-board pattern melts rapidly whenever $T_\text{Doubling}\approx 1$ is of the order of the tunnelling time.
The dynamics occurs again locally, so most of the dynamics is concentrated around the diagonal.
We do not go into the details of this evolution but rather show global figures of merit: We consider the matrix of the absolute values of  differences between the entries of the ramped and ideal covariance matrices 
$|\Gamma^{\rm (b')} - \Gamma^{\rm (b')}(T_\text{Doubling})|$ 
and convert them into a scalar by taking a maximum, mean or a median over all entries.
We find that if the doubling occurs at a small fraction of the tunnelling time $T_\text{Doubling}\ll 1$, 
then the ramp will not make a difference and will not substantially impact the state that will be reconstructed.

This modelling does not account for other possible imperfections involved in the doubling of the lattice.
In particular, it seems that ensuring that the atoms remain in the lowest band of the lattice at each time of the state preparation (b') is the most immediate concern.
Not accounting for this (or possibly not mitigating for it by optimal control) would possibly result in 
an instance of a ``heating'' 
in the experiment, at least unless precise band mapping would uncover the band excitations.
Having said that, given that the atoms at each lattice site do not have a large extension within 
the local well the fraction of band excitations should be small.
Qualitatively, it should be related to the spatial ``squeezing'' of the local atom wave-function which 
is needed to restrain the particles into wells of a reduced size.
Further analysis is beyond the scope of this appendix, 
but in principle one could once more extend our modelling to couple the 
system of primary interest to auxiliary unoccupied ``sites'' of the higher excitation bands.
The coupling constants for this would however depend on the optimality 
of the time-dependent ramp involved which would depend on whether 
there is a global harmonic trap present or it has been flattened in the 
region of interest by additional light fields using a digital-mirror device.

\begin{figure}[h]
\includegraphics[width=0.65\columnwidth]{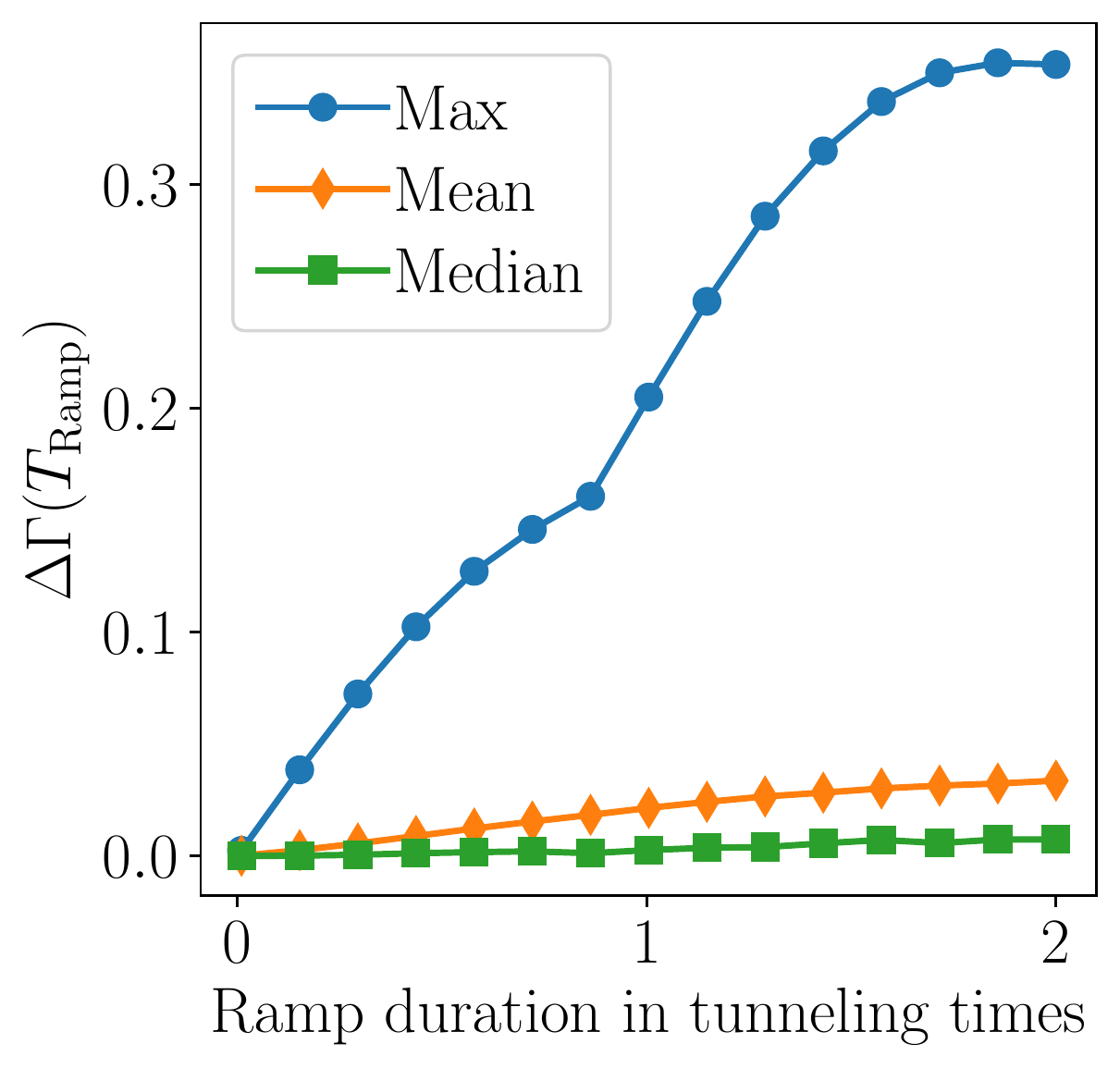}
\caption{\new{\textbf{Effect of the final ramp for the doubled lattice on initial correlations.}  
If one assumes a finite time $T_\text{Ramp}$ for the duration of the transition between the initial and the doubled-up lattice then the initial correlations will be deformed.  The extent of deviations is $\Delta \Gamma = \max|\Gamma_{x,y} - \Gamma^{(\rm Rec)}_{x,y}|$ and is seen to be negligible for small $T_\text{Ramp}\ll 1$ (much less than a tunnelling time) and $T_\text{Ramp}$ not negligible compared to the tunnelling time.}
}
\label{fig:app_doubling_ramp}
\end{figure}

\section{Choosing an appropriate quench Hamiltonian}
\label{sec:otherprotocols}

In this section, we come back to some relevant 
observations regarding the choice of the quench Hamiltonian  $h_\text{Quench}$.
This material is intended to provide further insights into
 how one can see that a quench does not provide enough atom number dynamics to facilitate a reconstruction for researchers interested in implementing the reconstruction based on the response of the gas to other protocols than the tilting or doubling of the lattice as presented above.
Our first example is motivated by notions of time of flight measurements: What happens, after all, 
if we just simply let the gas expand?
Fig.~\ref{fig:app_expansion_ballistic} shows that such an approach is not sufficient because in the optical lattice the gas expands ballistically outwards and the particles moving at the front do not mix with the ones propagating behind them. 
We see that according with this intuition the reconstruction fails to detect currents for the thermal initial condition shown in the main text.
That said, if this approach does not quite work do to a lack of mixing, then how about 
 mixing with an auxiliary system?
After all, we are interested in coherent tunnelling events which could in some formalisms be related to non-zero derivatives of an appropriately defined ``phase'' and hence interference could possibly be used to uncover the currents.
Fig.~\ref{fig:app_interferenceCDW} shows that choosing an incoherent charge-density wave as the auxiliary system does not actually lead to a correct reconstruction.
A possible explanation is that the tunnelling currents should be seen as ``coherences'' and the mixing to the incoherent system is insensitive to these correlations.
Finally, as we show in Fig.~\ref{fig:app_wallexpansion}, an expansion constrained  by a hard-wall implemented by a sudden increase of the chemical potential allows for reconstructions of the coherent tunnelling currents.
We have chosen a quench Hamiltonian where the gas can expand through three 
sites to its left and right via nearest-neighbour hopping but then suddenly 
the hopping is obstructed by a sudden jump in the chemical potential
\begin{align}
  \mu_x =\begin{cases}
100, & x< -3\\
0, & -3\le x \le L+3\\
100, & x> L+3. 
  \end{cases}
\end{align}
Summarizing, caution is necessary when choosing the precise form of the quench Hamiltonian.
In particular, quenches inducing ``coherent'' mixing within the system turn out to be 
necessary to uncover the ``coherences'', i.e., the tunnelling currents. This intuition seems to be well grounded in the examples provided above.
Having said that, we would like to point out that a full characterization of the tomographic resourcefulness of a given quench protocol occurs to be an interesting open problem which seems to be challenged by the wealth 
of possible quench ideas that one can consider as exemplified above.

\begin{figure}[h]
\includegraphics[width=0.9\columnwidth]{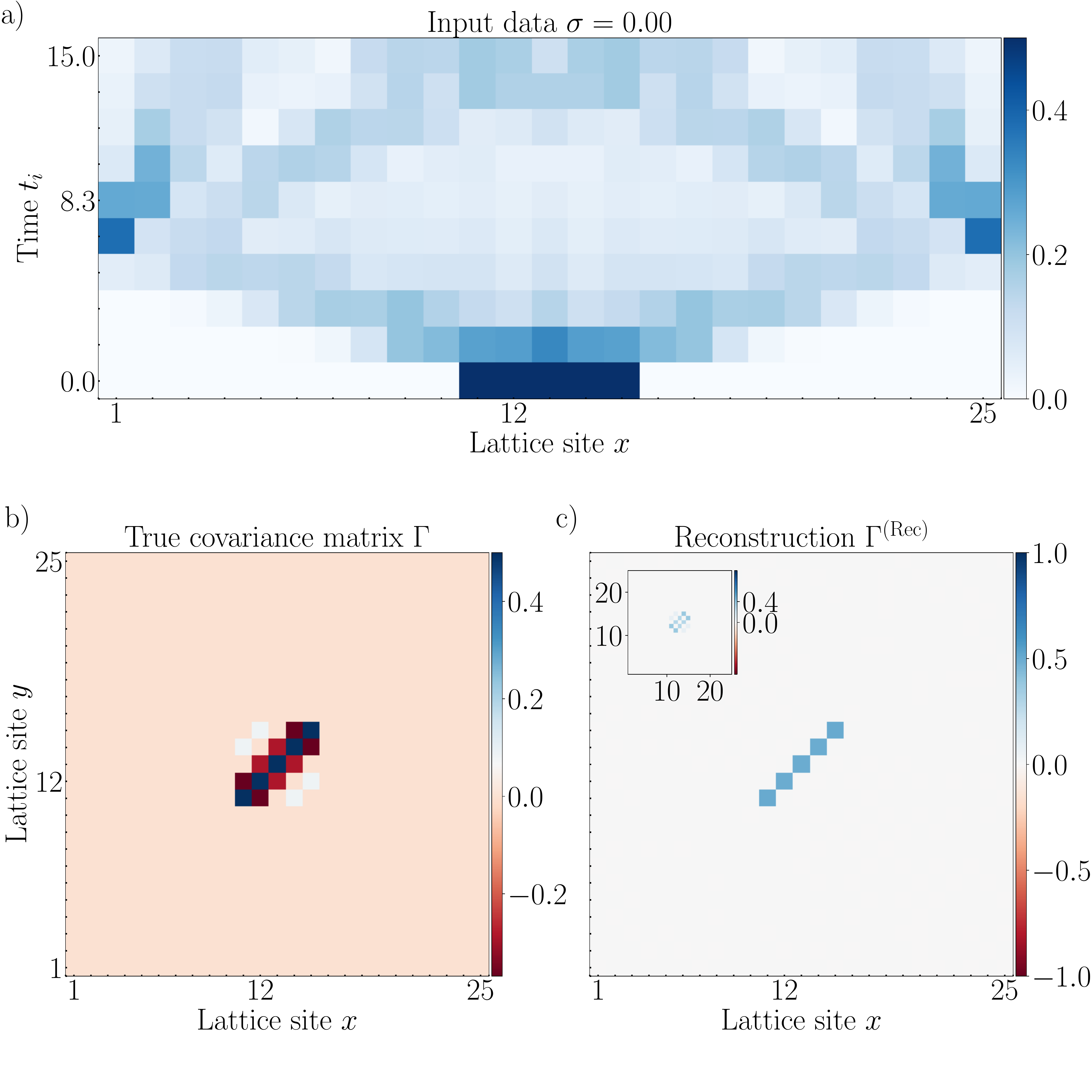}
\caption{\new{\textbf{Ballistic expansion.}  
Allowing the gas to suddenly expand is not an appropriate choice for $h_\text{Quench}$ because the atoms just move outwards quickly and do not mix together in the process (panel \emph{a)}).
As in the main text, we use the thermal initial condition of the nearest 
neighbour hopping Hamiltonian (panel \emph{b)} as the initial condition and the reconstruction does not uncover the currents in the state. This indicates that in the ballistic expansion the atom occupation numbers are 
not influenced by the presence or absence of tunnelling currents.}
}
\label{fig:app_expansion_ballistic}
\end{figure}

\begin{figure}[h]
\includegraphics[width=0.9\columnwidth]{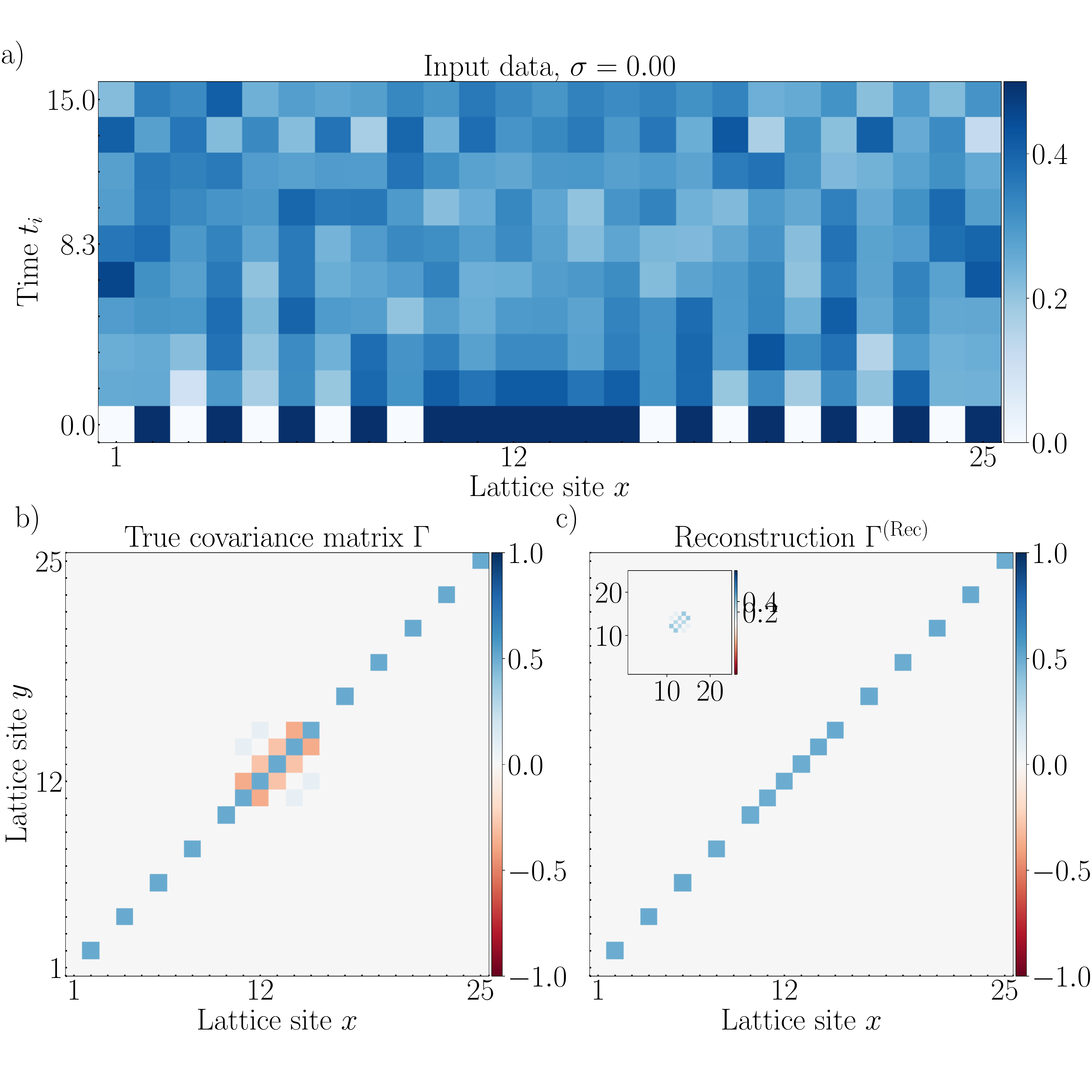}
\caption{\new{\textbf{Mixing with a charge-density wave.}  
We consider the possibility of measuring the currents by interfering a system with atoms that were held at fixed positions at its sides. As before, we make use of a thermal initial condition of the nearest neighbour hopping Hamiltonian (panel \emph{b)} as the initial condition. Once more, for this candidate choice of tomographic quench the reconstruction 
does not uncover the currents in the state. This indicates that in the mixing with charge-density waves on the sides is incoherent in the sense that the atom number dynamics is insensitive to the ``coherence'', i.e., the tunnelling currents.}
}
\label{fig:app_interferenceCDW}
\end{figure}

\begin{figure}[h]
\includegraphics[width=0.9\columnwidth]{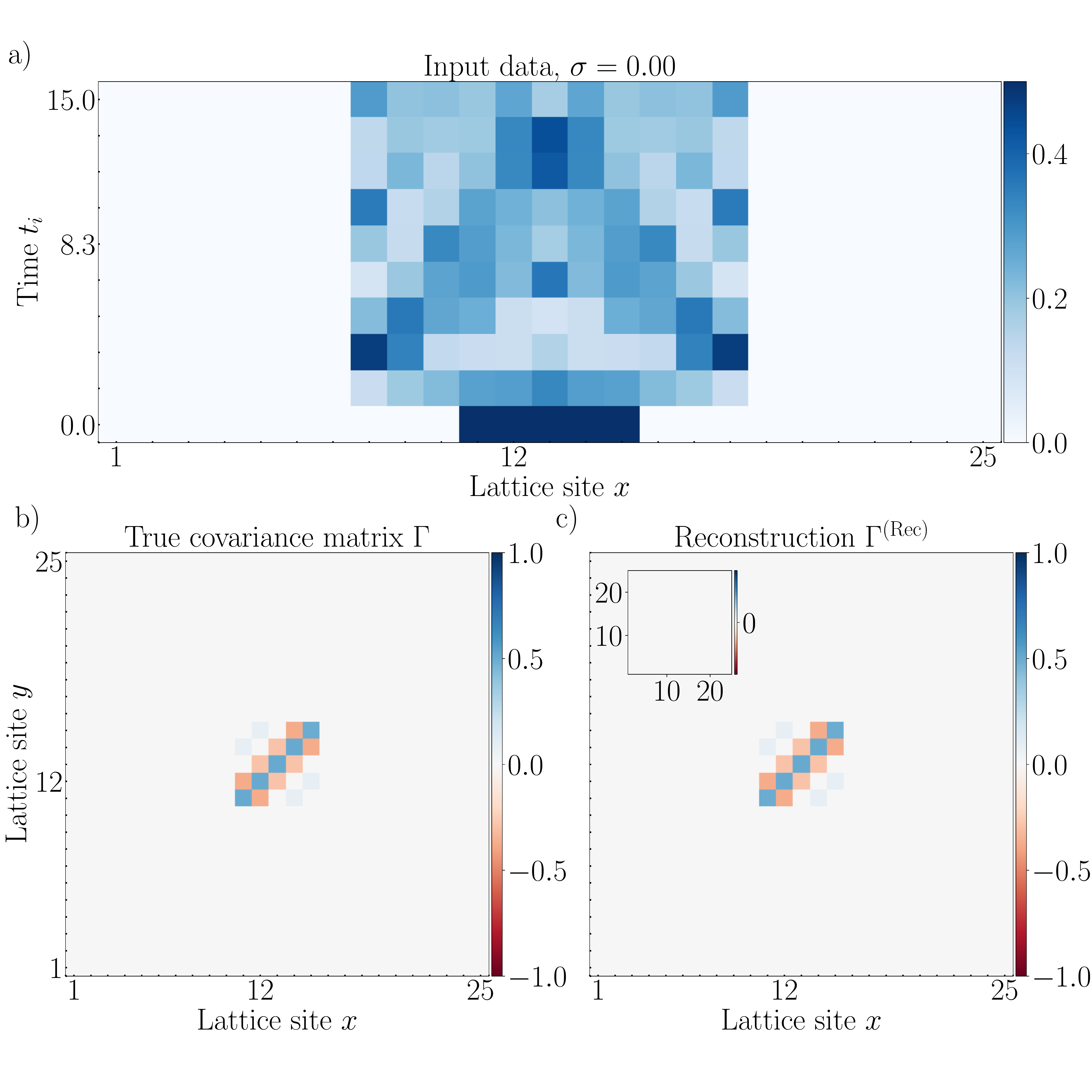}
\caption{\new{\textbf{Expansion into a hard-wall trap.}  
We consider once again a ballistic expansion, but now constrained by hard walls on both sides of the system.
As in the previous examples we use the thermal initial condition of the nearest neighbour 
hopping Hamiltonian (panel \emph{b)} as the initial condition.
In this case, we find that the reconstruction has successfully uncovered the currents in the state implying that in the mixing induced by the reflections from the walls is sensitive to the currents.
This protocol can be viewed as an abstraction of the quench shown in the main text involving the tilting of the lattice: In both cases the atoms are moving on top of each other with reflections occurring during the dynamics.
}
}
\label{fig:app_wallexpansion}
\end{figure}

}
\subsection{Symmetries of the Hamiltonian}

Various symmetries of the Hamiltonian may lead to some aspects of the state to remain hidden in the tomographic recovery procedure, or more precisely some correlation functions may be unconstrained by the observed particle number dynamics. Such symmetries are discussed here. Some simple examples are the following.

\emph{(i)} Hopping Hamiltonians mix correlations within the tunnelling correlation sector only. Therefore pairing correlations such as $\langle \f_x\f_y\rangle +\text{h.c.}$ can be arbitrary and their presence or absence does not modify the input to the tomographic
recovery procedure.

\emph{(ii)} If  both the initial state and the quench Hamiltonian are translation 
invariant then even if there are non-trivial currents in the covariance matrix, the particle number dynamics remains unchanged.
This can be seen by observing that  both $h$ and $\Gamma(0)$ can be simultaneously 
diagonalized by a Fourier transform and so their commutator vanishes at all times.
This implies that $N_x(t) = \Gamma_{x,x} (t)=\Gamma_{x,x} (0) = N_x(0)$ and the currents are unconstrained.
We resolved this issue by doubling up the lattice which implies that every other site is unoccupied and necessarily the state is not translation invariant.

The case when the couplings $h$ are real is related to the absence of magnetic fields.
We will now show   that if  one quenches to a Hamiltonian with such couplings, 
then only the real part of the currents can be reconstructed.
That is to say, let the state have some second moments $\Gamma = \Gamma^{(\rm Re)}+i \Gamma^{(\rm Im)}$ with $ \Gamma^{(\rm Re)}, \Gamma^{(\rm Im)}$ being the real and imaginary parts, respectively.
Then the tomography performed using the measurement of particle numbers will not constrain the imaginary part covariance matrix.

Let us now demonstrate that the particle number dynamics does not depend on the initial imaginary part of the currents if the couplings $h$ are real. To see this, we note that $\Gamma^{(\rm Im)}=-{\Gamma^{(\rm Im)}}^T$ because $\Gamma=\Gamma^\dagger$.
Secondly, we will use that $h=h^T$ implies $G(t)^*=G(-t)$ and $G(t)^T=G(t)$.
The particle numbers at time $t$ are given by 
\begin{align}
  N_x(t) = \Gamma_{xx}(t) = G(t)\Gamma^{\rm (Re)}G(t)^\dagger+i G(t)\Gamma^{\rm (Im)}G(t)^\dagger .
\end{align}
The first term related to the real part of the currents will in general influence the particle number dynamics.
For the second term by transposing twice we see  that
\begin{align}
\Gamma^{\rm (Im)}(t) :&=  G(t)\Gamma^{\rm (Im)}G(t)^\dagger \\
&=  \left(G(t)^*(\Gamma^{\rm (Im)})^T G(t)^T\right)^T \nonumber \\
&= - \left(G(-t)\Gamma^{\rm (Im)} G(-t)^\dagger\right)^T \nonumber \\
&= - \Gamma^{\rm (Im)}(-t)^T \nonumber 
\end{align}
and
\begin{align}
\Gamma^{\rm (Im)}(t)^* &=  G(t)^*(\Gamma^{\rm (Im)})^* (G(t)^\dagger)^* \\
&=  G(-t)\Gamma^{\rm (Im)}G(-t)^\dagger  \nonumber \\
&=  \Gamma^{\rm (Im)}(-t) \  \nonumber  \nonumber .
\end{align}
The only way for the imaginary part of currents $\Gamma^{\rm (Im)}$ to contribute to 
particle number dynamics is via the diagonal matrix elements of  its real part after the time evolution 
\begin{align}
  \text{Re}[\Gamma^{\rm (Im)}(t)] &=\frac12(\Gamma^{\rm (Im)}(t)+\Gamma^{\rm (Im)}(t)^*)\\
  &= \frac12(\Gamma^{\rm (Im)}(t)-\Gamma^{\rm (Im)}(t)^T) \ .\nonumber 
  \end{align}
The last relation proves that the real part of $\Gamma^{\rm (Im)}(t)$ is an antisymmetric matrix.
This implies that its diagonal matrix elements are vanishing and hence the imaginary part of the currents cannot be detected by the tomography after a quench to an evolution with real couplings so 
\begin{align}
\langle \hat N_x(t_i)\rangle_{\rh_{\Gamma}}=\langle \hat N_x(t_i)\rangle_{\rh_{{\rm Re}[\Gamma]}}\ .
\label{eq:im0}
\end{align}
The next section discusses this property further in relation to practical examples.

\section{Strategies for experimentally benchmarking  the systematics of the 
sub-lattice quench on charge density waves}
\label{sec:checks}
\begin{figure}[t]
\includegraphics[width=0.9\columnwidth]{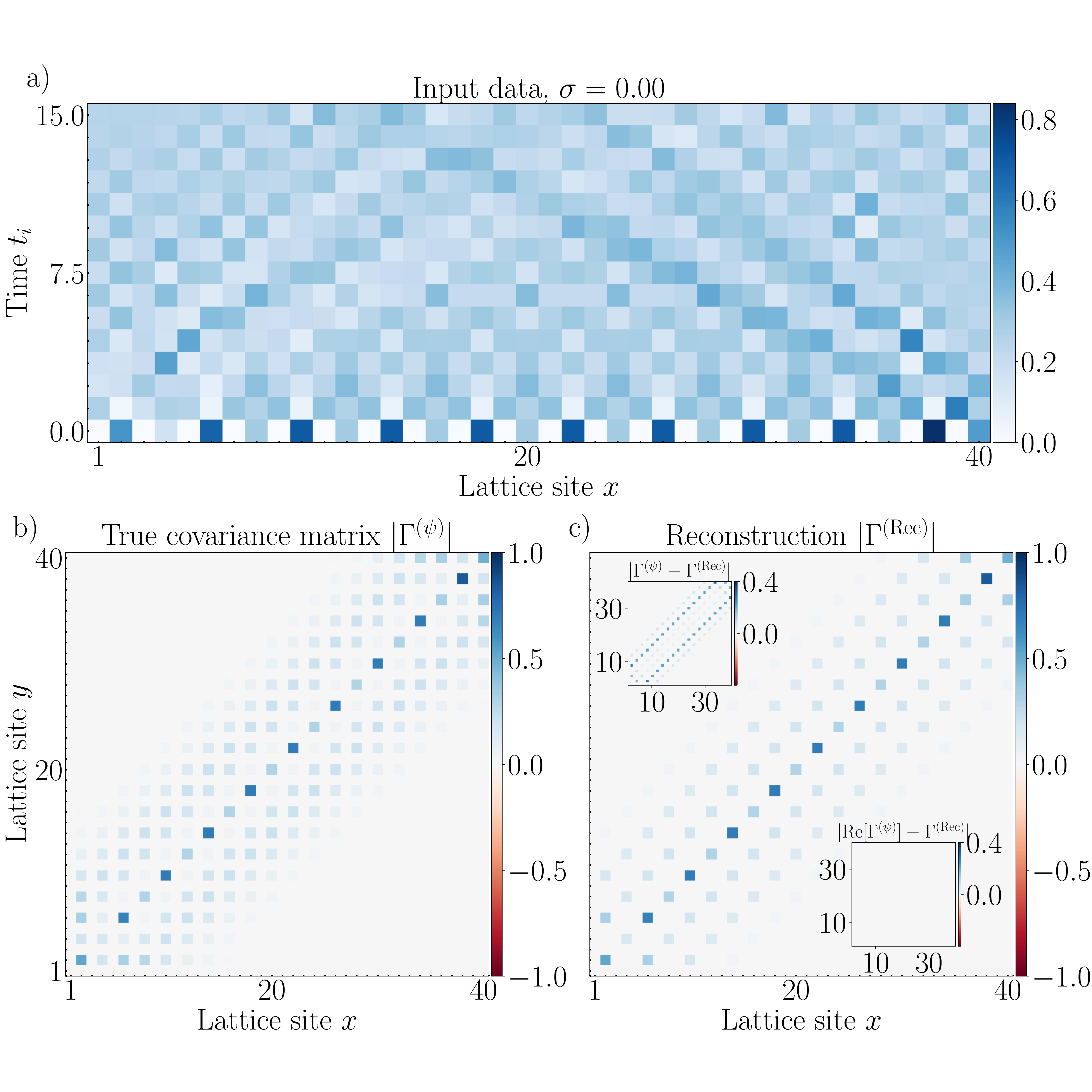}
\caption{{\bf Reconstruction for quenched CDW.}
\emph{ a)} The particle number dynamics used in the quench.
\emph{ b)} The same covariance matrix as above is to be reconstructed.
\emph{ c)} The reconstruction has not detected the presence of a complex-valued current.
The upper inset shows that certain tunnelling correlation are missing, these precisely the imaginary part of the currents as shown in the bottom inset \new{ showing that the real part of the correlations has been correctly recovered.}
}
\label{fig:tomography_app}
\end{figure}

In Ref.~\cite{Trotzky_etal12}, it has been 
shown how to prepare experimentally a charge density wave (CDW) which is just a Fock state vector with alternating particle occupation numbers $\ket{\phi} = \ket{0,1,0,1\ldots}$ prepared on $L$ sites and quench to free evolution.
The covariance matrix of this state is then $\Gamma^{(\phi)} = \text{diag}({0,1,0,1\ldots})$. \new{In what follows, 
we will discuss that this state} can be used to build trust in the reconstruction method.

First of all, we can make sure that the finite duration of the super-lattice creation when preparing the CDW does not induce correlations between sites, i.e., just using measurements using the atom microscope we can make sure that the covariance matrix is $\Gamma^{(\phi)} = \text{diag}({0,1,0,1\ldots})$.
This can be argued by a \emph{fidelity witnessing} argument \cite{FW,arute2020hartree}.  
If we have $L$ sites and $N_x$ does 
not deviate from $0$ or $1$ by more than $\epsilon$, then the fidelity of the unknown state in the laboratory $\hat \varrho_p$ with respect 
to the CDW Fock state vector $\ket \phi$ is lower bounded by
\begin{align}
  F(\hat \varrho_p) = \bra \phi \hat \varrho_p \ket \phi \ge 1- \varepsilon = 1-\epsilon L\ ,
\end{align}
where $L$ as before is the number of lattice sites.
Having said that, to benchmark our method it is not necessary to evaluate the fidelity to the full density matrix of $\ket\phi$ but it will suffice to bound the deviation in the experiment from the covariance matrix from $\Gamma^{(\phi)}$ using the Cauchy-Schwarz inequality
\begin{align}
  |\Gamma_{x,y}| \le \sqrt{N_x N_y},
\end{align}
which readily implies that $|\Gamma_{x,y}|\le \sqrt\epsilon$ whenever $N_x\le \epsilon$.
If at a given site $x$ the particle number is measured to be $N_x = 1-\epsilon$ then after a Bogoliubov transformation swapping particles and holes we again obtain $|\Gamma_{x,y}|\le \sqrt\epsilon$.
We conclude that if the CDW preparation is extremely good and the atom microscope measurement, too, then just from this data one can conclude that there are \emph{no} currents in the system with the strength of a quantum certification test \cite{FidelityEstimation,eisert2019quantum}.
Such a certified Fock state with a precise bound on fidelity can be used as a starting point for benchmarking of the dynamics with or without interactions.

Note that the step of creating a state with no currents can be verified by the static information about particle numbers alone using the atom microscope.
Once we know that there is no currents in the initial state, then we can create them by known dynamics.
If this is done and the dynamics can be simulated numerically then one obtains an experimental state with well-grounded prior knowledge what currents to expect to be present in the system and one can verify the functioning of the tomographic reconstruction.
The next section shows an example where we present results for reconstructions of time-evolved CDW states.

\subsection*{Check \#1: Reconstructing a known current assuming $\hat H_\text{NN}$}
 Assuming the Hamiltonian is $\hat H_\text{NN}$ we can consider some fixed time $t_0$ and attempt to reconstruct the state
 vector 
 \begin{equation}
 \ket\psi = e^{-i t_0\hat H_\text{NN}} \ket{\phi}
 \end{equation}
 which has the covariance matrix $\Gamma^{(\psi)} = G(t_0)\Gamma^{(\phi)}G(t_0)^\dagger$.
If we assume the free nearest-neighbour 
hopping evolution to be exact, then we can be sure that the covariance matrix in the laboratory is $\epsilon$-close to $\Gamma^{(\psi)}$ if we verified that the experimental preparation of the CDW has been $\epsilon$-close to $\Gamma^{(\phi)}$.
But then knowing that, we know precisely which off-diagonal currents should be reconstructed.

The {check} reads: One prepares a CDW and performs 
quenches to times $t=t_0, t_0+\Delta t, \ldots, t_0+ T$ equi-distributed in steps $\Delta t$. From the measured data, 
one recovers the covariance matrix assuming that 
the data have been taken at times $t=0, \Delta t, \ldots, T$.
The tomography should return a covariance matrix close to $\Gamma^{(\psi)}$.
Note that when running the tomography here we are bypassing the step \emph{(b)} from the main text.
We can learn from the atom microscope measurements in step \emph{(d)} whether the initial state has 
indeed been prepared with high fidelity. 
And it is not necessary to double up the lattice anymore as the initial state is not translation invariant by construction.
Having said that, the super-lattice creation is going to be important when studying homogeneous thermal states that will be of interest in future quantum simulation experiments and hence it is important to map out the systematic influence of this step on the tomography which we describe next.

\subsection*{Check \#2: Assessing the systematic influence of doubling up the lattice}
 The above check tests the reconstruction of a state with known currents but with the input 
 influenced by statistical errors.
 In the check the steps \emph{(b-d)} have been on purpose bypassed as much as possible.
 The lattice doubling has been employed only in the sense of the state preparation in step \emph{(a)}.

It is possible to check how step \emph{(b)} influences the reconstruction in the experiment.
Again the task is to prepare the CDW with covariance matrix $\Gamma^{(\phi)}$ and evolve it under the nearest-neighbour hopping Hamiltonian of $L$ sites to time $t_0$ obtaining the covariance matrix $\Gamma^{(\psi)}$.
Then the super-lattice from step \emph{(b)} should be created resulting in the 
checker-board covariance matrix  $\Gamma^{(\psi,b)}$.
In  step \emph{(c)} the system should be evolved under the nearest-neighbour hopping Hamiltonian (now on the doubled lattice) to times $t=0, \Delta t, \ldots, T$.
As always in the last step \emph{(d)} local particle numbers at each of the times are to be estimated to obtain the input to the reconstruction.

The results for the tomographic recovery in this scenario with $t_0=1$ are shown in Fig.~\ref{fig:tomography_app}.
The input is depicted in Fig.~\ref{fig:tomography_app}a) after being subjected to 
noise modelling statistical errors.
Note that the system considered here is 
 far from being in thermal equilibrium and the particle numbers are changing over time-scales that are longer
 than those considered in the  scenario discussed in the main text where the system was closer to thermal equilibrium.
The checker-board covariance matrix  $\Gamma^{(\psi,b)}$ is shown in Fig.~\ref{fig:tomography_app}b).
In Fig.~\ref{fig:tomography_app}c) we show the results of the reconstruction.
Importantly, the reconstruction does not recover the true covariance matrix as shown in the upper inset of Fig.~\ref{fig:tomography_app}c).
As explained above, this is because this reflects a non-equilibrium situation with a non-trivial imaginary part of currents $\text{Im}[\Gamma^{(\psi,b)}]\neq 0$.
On the other hand, as shown in the lower inset of Fig.~\ref{fig:tomography_app}c) the 
output of the reconstruction closely matches the real part of currents $\text{Re}[\Gamma^{(\psi,b)}]$.
In fact, these are reliably recovered as they influence non-trivially the dynamics of the 
particle number and the deviations stem from the random noise realization that 
has been added to the input and is shown in 
Fig.~\ref{fig:tomography_app}a).

\subsection*{Check \#3: Benchmarking the reconstructions in the presence of artificial 
magnetic fields}
In this last section, 
we show how to check the method experimentally if one wishes to reconstruct also the imaginary part of the currents.
For this, it is necessary to have complex tunnelling amplitudes present during the tomographic quench in step \emph{(c)}.
There does not seem to be a canonical choice which model to choose as this depends on the way the optical lattice is modulated in order to obtain the artificial magnetic field.
We hence show a minimal example where the initial system is a two-site optical lattice with a single particle.
Again we perform an evolution  with $t_0=1$ which leads to a single tunnelling current $\Gamma_{1,2}\neq0$ which turns out to be purely complex in this case.
We then  double up the lattice obtaining 4 sites and as above the particle numbers are to be measured after evolutions in the superlattice at equidistant times.

\begin{figure}[h]
\includegraphics[width=1\columnwidth]{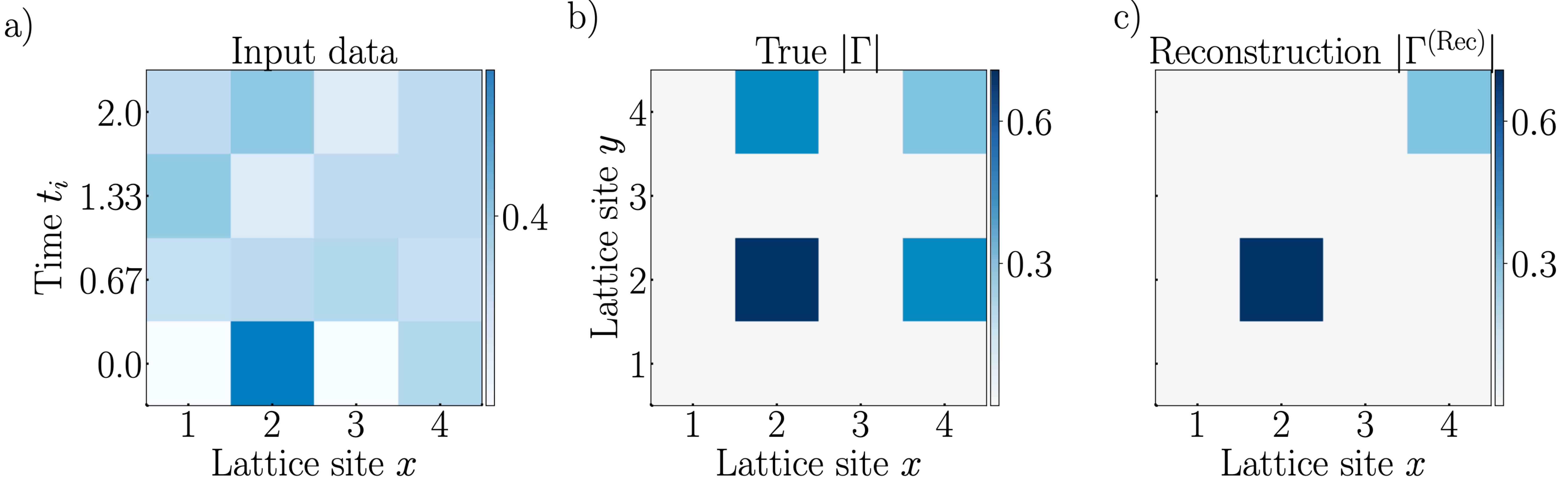}
\caption{{\bf Tomographic recovery of the two site system to a superlattice on four sites}.
\emph{a)} The particle number dynamics used as input as shown above does not depend on the complex part of the covariance matrix because we use the nearest-neighbour hopping Hamiltonian $\hat H_\text{NN}$ with real couplings.
\emph{b)} The absolute value of the covariance matrix. The off-diagonal current is purely imaginary.
\emph{c)} As discussed the imaginary part of the covariance matrix is unconstrained and  hence the reconstruction is missing the off-diagonal current. The  evolution of the reconstructed covariance matrix yields particle numbers 
exactly matching the input as proven 
analytically.
}
\label{fig:2x2noB}
\end{figure}

\begin{figure}[h]
\includegraphics[width=1\columnwidth]{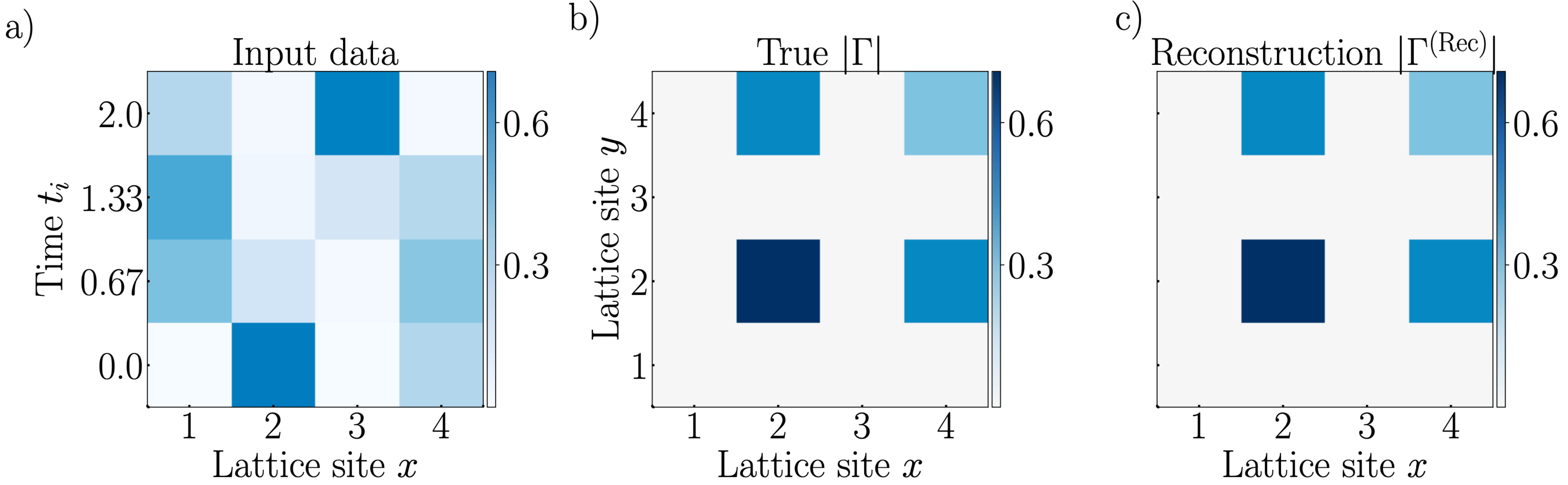}
\caption{{\bf Recovery in the presence of artificial magnetic fields.}
\emph{a)} The particle number dynamics is influenced by the complex part of the covariance matrix if we use a Hamiltonian $\hat H = \hat H_\text{NN} +i\sum_{x=1}^3 \fd_x\f_{x+1}+\text{h.c}$ with complex couplings.
\emph{b)} The same covariance matrix as above is to be reconstructed.
\emph{c)} The reconstruction now has detected the presence of the complex-valued current.
}
\label{fig:2x2B}
\end{figure}

Fig.~\ref{fig:2x2noB} shows that again simple hopping evolution does not uncover anything about the complex current.
In contrast Fig.~\ref{fig:2x2B} shows that adding a simple complex hopping amplitude to the quench Hamiltonian in  step \emph{(c)} leads to a full reconstruction of the current.
In both examples we did not add noise on the particle numbers so that the difference in the dynamics can be more easily compared between the evolution with and without the magnetic fields.

Summarizing this check, we point out that if it was possible to reconstruct using some Hamiltonian which has real couplings $\hat H_{\rm (c), real}$ and additionally using a Hamiltonian that \emph{adds} imaginary couplings without modifying the previous ones $\hat H_{\rm (c), gen} =  \hat H_{\rm (c), real} +\hat  H_{\rm (c), imag}$ then one can do two series of data taking and the tomography using  $\hat H_{\rm (c), gen} $ should give a similar real part of the covariance matrix to that obtained using $\hat  H_{\rm (c), real} $.


\new{
\section{Systematic influence of ramps when quenching many-body interactions}
\label{sec:interactions}
In practice, the switching off of the interactions by means of Feshbach resonances will not be 
instantaneous. Narrow Feshbach result in relatively fast quenches, and there are no
fundamental limits to a fast switching. That said, any switching that is not instantaneous will
have some effect. In this section, we discuss the impact of finite ramps, reflecting the process
of switching off interactions. Specifically, to provide insight into the resulting effects, 
we consider the case of fermions with spin and will compute the covariance matrix of thermal states of the Hubbard chain using exact diagonalization.
We denote the annihilation operator at $x$ of a spin up fermion by $\f_{x,\uparrow}$ and for spin down  by $\f_{x,\downarrow}$.
In this notation the Hamiltonian of a Hubbard chain on $L$ sites with open boundary conditions reads
\begin{align}
  \hat H_\text{Hubbard} = -\sum_{\sigma=\uparrow,\downarrow}\sum_{x=1}^{L-1} \fd_{x,\sigma}\f_{x+1,\sigma}+\text{h.c.}+U\sum_{x=1}^{L}\hat N_{x,\uparrow}\hat N_{x,\downarrow}
\end{align}
where $\hat N_{x,\sigma} := \fd_{x,\sigma}\f_{x,\sigma}$ for $\sigma=\uparrow,\downarrow$.
To ease the notation, we define 
\begin{align}
	\hat a_{2x-1} := \f_{x,\uparrow},\,\,\hat a_{2x}:=\f_{x,\downarrow} 
\end{align}
which allows us to concisely write down the Jordan-Wigner transformation
\begin{align}
  \hat a_x = Z^{\otimes x-1}\otimes S^- \otimes \id_2^{\otimes 2L-x}
  \label{eq:JW}
\end{align}
with
\begin{align} 
Z = \begin{pmatrix} 1&0\\0&-1\end{pmatrix}
\end{align} referring to the Pauli-$z$ matrix and 
\begin{align} 
S^- = (S^+)^\dagger := \begin{pmatrix} 0&1\\0&0\end{pmatrix}.
\end{align} 
Using the ordering we have chosen, we find
\begin{align}
  \hat H_\text{Hubbard} = -\sum_{x=1}^{2L-2} \hat a^\dagger_{x}\hat a_{x+2}+\text{h.c.}+U\sum_{x=1}^{L}\hat N_{2x-1}\hat N_{2x}
\end{align}
which under the Jordan-Wigner transformation turns into
\begin{align}
  \hat H_\text{Hubbard} = \sum_{x=1}^{2L-2} S^+_{x}Z_{x+1} S^-_{x+2}+\text{h.c.}+U\sum_{x=1}^{L}Z_{2x-1}Z_{2x}
\end{align}
where we have used the concise notation 
\begin{align}
\Xi_x := \id_2^{\otimes x-1}\otimes \Xi \otimes  \id_2^{\otimes 2L- x}
\end{align}
 with $\Xi := Z, S^+$ or $S^-$.
Using exact diagonalization, we compute the thermal density matrix
\begin{align}
  \rh_\beta(U) = e^{-\beta \hat H_\text{Hubbard}}/Z_\beta
\end{align}
with $Z_\beta:=\tr[e^{-\beta \hat H_\text{Hubbard}}]$.
Using the Jordan-Wigner transformation we then can compute the respective covariance matrix
with entries for $x<y$
\begin{align}
  \Gamma^{(\rm Hubbard)}_{x,y} &= \tr[ \hat a_x^\dagger \hat a_y \rh_\beta(U)]\\
  &= - \tr[ S^+_x Z_{x+1}\ldots Z_{y-1} S^-_y \rh_\beta(U)].\nonumber
\end{align}
Again using exact diagonalization, we have computed the thermal state for $\beta = 3$ and additionally the covariance matrix.
The results for the superfluid phase with $U=1$ are shown in Fig.~\ref{app_SF} together with a comparison the covariance matrix for just the nearest-neighbour hopping with $U=0$.
We find that the difference is very small which means that the Gaussian model is representative for the second moments also of the interacting thermal state if we neglect a small deviation.
For this reason the effect of ramping down the interaction in finite time should be less severe because the state will not be strongly affected by it.

For the case of the Mott phase with $U=10$ we find, see Fig.~\ref{appMott}, a very substantial difference of the second moments when compared to those obtained from a Gaussian model.
To perform a meaningful comparison we consider a heuristic mean-field ansatz for a non-interacting Hamiltonian to yield a Gaussian state to compare to the second moments arising from the interacting Hamiltonian.
We chose
\begin{align}
  \hat H_{\text{MF},\downarrow}(N^{(\uparrow)}) = -\sum_{x=1}^L \fd_{x+1,\downarrow}\f_{x,\downarrow} + U / 8 \sum_{x=1}^L N^{(\uparrow)}_x \hat N_{x,\downarrow}
\end{align}
where the spin-up density $ UN^{(\uparrow)}/8$ plays the role of a vector of the effective chemical potential for the spin-down band.
Qualitatively, the correlations are restricted to nearest neighbour sites while in the non-interacting case they span a larger region.
Here a finite duration of ramping down the interaction can sizeably affect the correlations.

The simplest protocol for tomographically reconstructing the second moments $\Gamma^{(U)}$ of an interacting state using our method is to ramp down the interaction in time $T_\text{Ramp}$ and at the end of the ramp tilt the optical lattice as explained above.
The ramp will have affected the covariance matrix leading to some new second moments $\Gamma^{(U\rightarrow 0)}$ and the tomography will recover these rather than  $\Gamma^{(U)}$.
The result will hence be accurate if  $\Gamma^{(U\rightarrow 0)}\approx \Gamma^{(U)}$.
In Fig.~\ref{apprampint}, we show the effect on the second moments of ramping down the interaction starting from $U=10$ to $U=0$.
We depict covariance matrices at the end of ramps of varying length $T_\text{Ramp}=0.5,1$ and $2$, so smaller than the tunnelling time, equal, and larger than it.
We find that the ramp time does influence 
the correlations to certain degree but importantly the overall pattern is not overhauled.
The changes in the values of the currents increase monotonically with the increase of the duration of the ramp.
Due to the limitations of system sizes available in exact diagonalization, we have provided results for only small systems while in the thermodynamical limit one should expect defects appearing due to the Kibble-Zurek mechanism.
It seems that it can be avoided, however, by separating a large system into smaller portions such that energy gaps never become smaller than a certain threshold and perform the tomography for each individually.
The smaller systems should be larger than the correlation length and such subdivision may be advisable also when dealing with a finite number of total state preparations.

Ramps of this order of duration can be thought of as being implemented by a narrow 
Feshbach resonance allowing for a relatively fast switching time.
With the same approach one could study the case of ramps larger durations but it seems that ramps an order of magnitude longer than the tunnelling time are ill-advised for the sake of correlation read-out using the response to non-interacting dynamics. The reason for this is that for short ramps, as mentioned in the main text, Lieb-Robinson bounds allow to narrow down the influence of the ramp, restricting the action of the its dynamics to the close-by sites and crucially this statement is state independent so can be also employed in cases that are not classically tractable in practice.
In contrast, for long ramps, the Lieb-Robinson bounds would not offer a non-trivial estimate of the effect of the ramp. 

\begin{figure}[h]
\includegraphics[width=1\columnwidth]{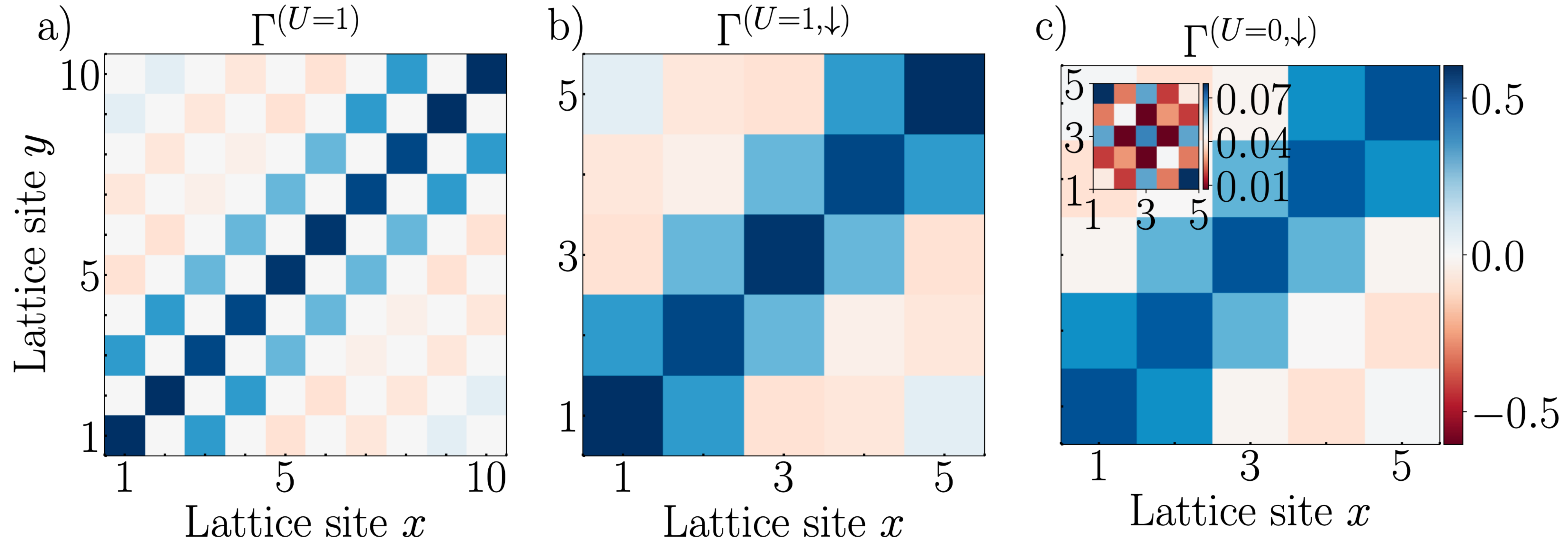}
\caption{\new{\textbf{Second moments in the superfluid phase.}  
panel \emph{a)} shows the covariance matrix of the Hubbard model for $U=1$, panel \emph{b)} the restriction spin-$\downarrow$ modes.
panel \emph{c)} depicts the covariance matrix of the nearest-neighbour hopping with the same temperature.
The inset shows the difference to the covariance matrix of interacting state in absolute values.
}}
\label{app_SF}
\end{figure}
\begin{figure}[h]
\includegraphics[width=1\columnwidth]{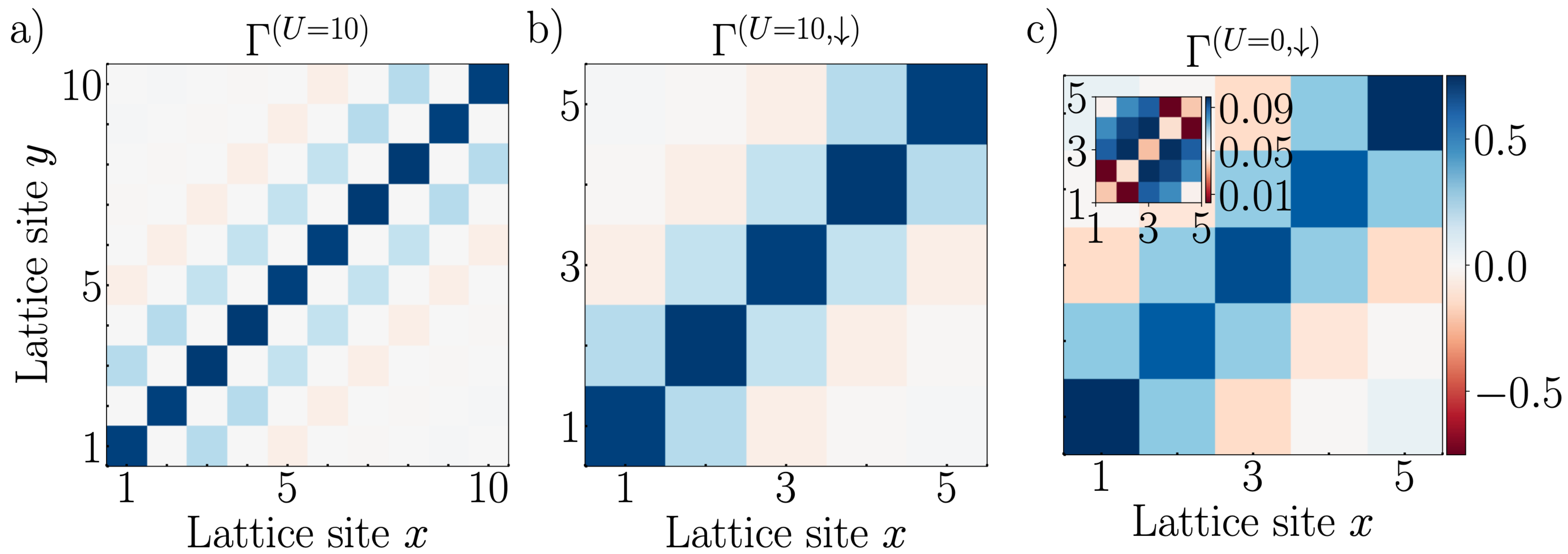}
\caption{\new{\textbf{Second moments in the Mott phase.}  
panel \emph{a)} shows the covariance matrix of the Hubbard model for $U=10$, panel \emph{b)} the restriction spin-$\downarrow$ modes. panel \emph{c)} illustrates the covariance matrix of the nearest-neighbour hopping with the same temperature.
The inset shows the difference to the covariance matrix of interacting state in absolute values.
}}
\label{appMott}
\end{figure}

\begin{figure}[h]
\includegraphics[width=1\columnwidth]{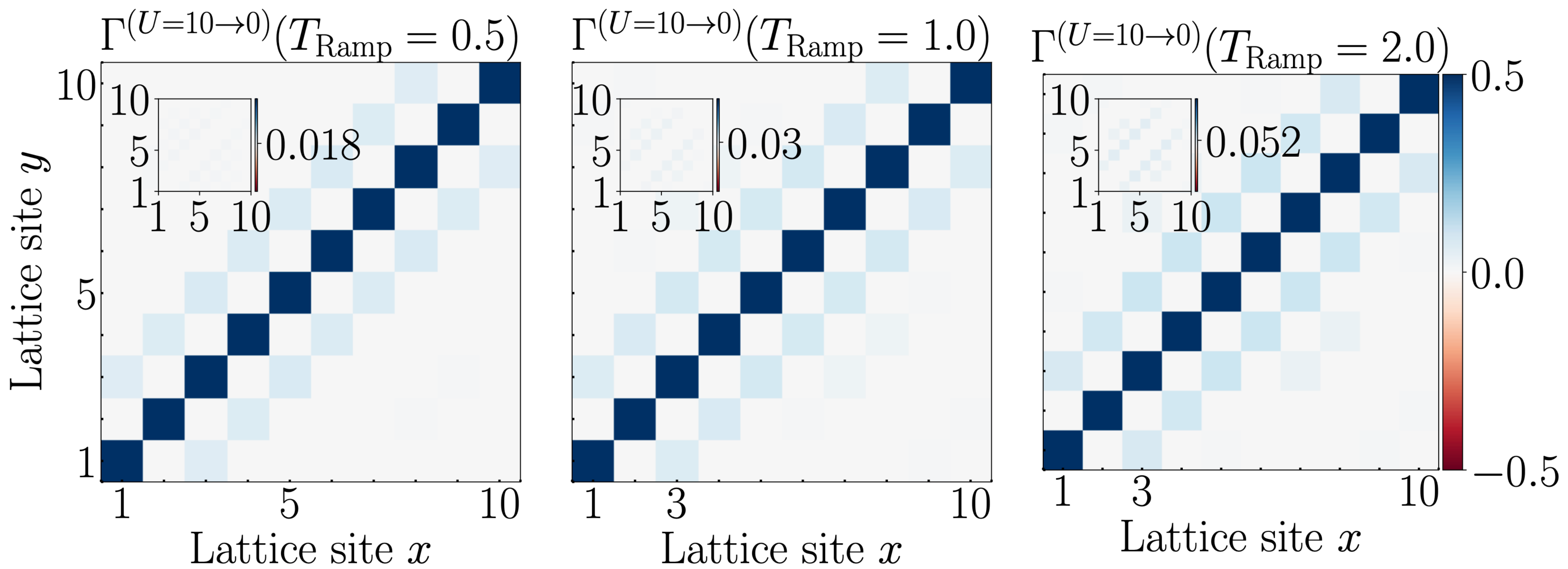}
\caption{\new{\textbf{Influence of ramping interactions down in a finite time.}  
We find that the deviation of correlations  after the ramp $\Gamma^{(U\rightarrow 0)}$  from the 
actual second moments $\Gamma^{(U)}$ increase monotonically together with the ramp duration 
$T_\text{Ramp}$ but are small enough such that the correlation patterns are not qualitatively 
distorted. Hence, we conclude that the tomographic read out would clearly show the effect of the presence of interaction on the second moments and could explore the single-body reduced density matrix for the entire phase diagram of the Hubbard model.
}}
\label{apprampint}
\end{figure}

}

\section{Fermionic mode entanglement}
\label{sec:entanglement}
In this section, we provide further insights into the operational meaning of the entanglement quantified based on our 
reconstructions, and what \emph{fermionic mode entanglement} operationally means.
We denote the fermionic Fock state vectors by
\begin{align}
  \ket{ \mu}_\text{F} := (\fd_1)^{\mu_1}\ldots (\fd_L)^{\mu_L}\vacket
\end{align}
with $\mu \in \{0,1\}^{\times L}$ and
where $\vacket$ denotes the vacuum state vector defined as satisfying $\f_x \vacket =0$ for all $x=1,\ldots,L$. For
a given order (so basically a symmetric group element $S\in S_L$ that captures the order
of the fermionic modes when mapping them to spins),
we 
can define a Jordan-Wigner transformation by the relations
\begin{align}
  \f_x = Z^{\otimes x-1}\otimes \begin{pmatrix} 0&1\\0&0\end{pmatrix} \otimes \id_2^{\otimes L-x}.
  \label{eq:JW}
\end{align}
This choice of the Pauli-$z$ matrix fixes its eigenbasis so that we have the convention $Z\ket{\nu}_\text{S} = (-1)^\nu \ket{\nu}_\text{S}$ with $\nu =0,1$ and then Eq.~\eqref{eq:JW} imply directly that
\begin{align}
  \ket{ \mu}_\text{F} = \otimes_{x=1}^L\ket{\mu_x}_\text{S}.
  \label{eq:basis} 
\end{align}
Note that any ordering of modes can give rise to such a qubit representation where fermionic operators and state vectors can be expressed in an explicit matrix form, but once chosen it should remain fixed throughout calculations.
Importantly, occupation states of modes after a non-trivial Bogoliubov transformation will not anymore have such a tensor-product form and one should consider the decomposition into antisymmetric subspaces.

Having fixed the reference ordering of modes, we have a natural notion of fermionic subsystems: A subsystem $A$ consists of 
a collection of modes, i.e., positions in the lattice, while $B$ is constituted by the complementing modes.
In all what follows, we are perfectly free to take the ordering so that modes labeled $1,\dots, |A|$ 
give rise to subsystem $A$, while $|A|+1,\dots, L$ give rise to $B$.

We now turn to quantum states in such a fermionic setting.
Given a quantum state $\hat \varrho$ supported on the Hilbert space of the entire system,
the reduced state $\hat \varrho_A$ is defined via functionals of observable algebras: Specifically,
it will reflect correlation functions involving operators acting in $A$ equal to those of the global state. 
That is to say, in what 
follows, we can treat the system as a system of $|A|$ qubits held by $A$ and
$L-|A|$ qubits held by $B$.

Any allowed quantum operation performed within an isolated fermionic system must respect the 
\emph{parity of fermion number super-selection rule} \cite{Earman,WignerWightmanWick}, which implies that at any time, 
any correlation function involving an odd number of fermionic creation and annihilation operators must vanish.
This means that if one performs operations and measurements locally in subsystems $A$ and $B$,
then to describe any of such processes it suffices to consider the reduced states $\hat \varrho_A$ and $\hat \varrho_B$ to be a direct sum of
sectors reflecting even and odd particle numbers in $A$ and $B$, respectively.
That is to say, we can first make use of a projection $\pi\otimes \pi$ which projects each of the local subsystems $A$ and $B$ into a direct sum of even and odd particle numbers.

\emph{Local operations with classical communication
reflecting super-selection rules}, referred to as LOCC+SSR, 
can hence be identified with local operations in the qubit systems reflecting $A$ and $B$ under the
Jordan-Wigner transformations, 
respecting 
the local direct sum structure of even and odd particle numbers in
$A$ and $B$ and the total system. 
This is a perfectly operational prescription. We call a quantum
state \emph{mode entangled} throughout this article,
if it is not a convex combination of uncorrelated quantum states in fermionic modes.

\subsection{Single copy fermionic mode entanglement}

There is a subtlety arising in the fermionic context, however: This has to do with the
presence of super-selection rules. This can be seen as follows.
Turning to quantitative prescriptions, one can define \emph{single-copy entanglement} 
\cite{PhysRevLett.83.436,PhysRevA.72.042112}
as the 
maximum probability at which one can -- at least in principle -- extract maximally entangled states of 
distinguishable quantum systems out of the original system of fermionic modes, making use of any
 operation in LOCC+SSR that is allowed by quantum mechanics. In this prescription, one may make use
 of suitable physical interactions or measurements respecting LOCC+SSR. The target distinguishable quantum systems 
 can either be seen as being actually available in the laboratory,
e.g., as spin degrees of freedom, or as a
conceptual tool to precisely think of mode entanglement in the first place.

Interestingly, in this sense,
a state $\rh = \ket \psi\bra \psi$ with 
 $\ket \psi = (|0,1\rangle_\text{F} + |1,0\rangle_\text{F})/\sqrt{2}$ represented as
\begin{equation}
\ket \Phi := \frac 1{\sqrt 2} ( \ket 0_\text S \otimes \ket 1 _\text S+\ket 1_\text S \otimes \ket 0 _\text S)
\label{eq:BellPair}
\end{equation}
 is not \emph{single-copy
entangled}, as no physically allowed protocol can map this state onto an entangled state of
distinguishable quantum systems 
with any non-zero probability. 
 The projection $\pi$ acting as
\begin{equation}
|\Phi\rangle\langle \Phi| \mapsto
(\pi\otimes \pi)  |\Phi\rangle\langle \Phi| (\pi\otimes \pi)
\end{equation}
 will render the state operationally 
indistinguishable from a quantum state that is merely classically correlated and contains no 
quantum entanglement. 
That is to say, for all practical purposes, the state does not contain
any entanglement that can be operationally extracted (i.e., with all operations allowed)
from a single specimen or copy.
It is important to stress that this statement is referring to single-copy entanglement
only: It is perfectly an entangled state if asymptotic state transformations are being allowed for, as explained below.

\subsection{Distillable fermionic mode entanglement and asymptotic state manipulation}

However, the above fermionic two-mode quantum state represented by the state vector
$\ket \psi = (|0,1\rangle_\text{F} + |1,0\rangle_\text{F})/\sqrt{2}$
 is in fact \emph{many-copy entangled}, as the SSR is asymptotically
not detrimental to the entanglement content of the state vector. Such 
asymptotic notions involving \emph{many copies of identically prepared systems}
at the same time are the most commonly applied 
standard
notions of entanglement theory, much inspired by notions of classical information theory.
Here, one still restricts to quantum operations local to $A$ and $B$, coordinated
by classical means by communicating measurement outcomes between $A$ and $B$.
But one assumes
to have available several copies (or specimens)
of the quantum many-body systems available and is allowed
to coherently manipulate the quantum state over the copies. Obviously, in practical
settings, it is implausible to achieve completely general operations of this kind: It is still a highly
convenient abstraction. 

The notion of the 
\emph{distillable entanglement} \cite{PureBipartiteBennett,Horodecki} captures most naturally the resource character of
entanglement in quantum information theory. It is in a way the ``entanglement content'' 
of a state. It 
is defined as the optimal rate at which one could extract
maximally entangled pairs of distinguishable systems in such a hypothetical
optimal state manipulation from many identical copies.
This notion not only allows to detect  the presence of 
fermionic entanglement: It also allows to make \emph{quantitative
estimates}, which is precisely what we are interested in here.
More precisely and specifically put, the distillable entanglement quantifies
the asymptotic rate at which
one can extract (``distill'') distinguishable approximately perfect 
maximally entangled qubit pairs (``Bell pairs'') as in 
Eq.~(\ref{eq:BellPair})
of distinguishable quantum systems from many identical copies of an 
input state composed of fermionic modes, using quantum operations from 
LOCC+SRR \cite{PhysRevA.70.042310}.
 For a fermionic initial quantum state $\rh$
on $L$ modes partitioned into $A$ and $B$, 
the entanglement distillation
 problem involves as initial state the bona fide quantum state of $n$ copies on $nL$ modes. The final state
 is asymptotically in $n$ better and better approximating $m$ copies of a maximally entangled
 pure quantum
 state associated with the state vector
 \begin{equation}
	\ket{\psi_{\rm Final}}= \ket \Phi^{\otimes m},
\end{equation}
in trace-norm $\|.\|_1$,
so will approximate $m$ Bell pairs. The larger $m$ is, the larger is the yield of this procedure.
The distillable entanglement under LOCC+SSR is now
 \begin{equation}
	E_D^{\rm SSR}(\hat \varrho) = \limsup_{n\rightarrow\infty}\frac{n}{m},
\end{equation}
as the supremum over all LOCC+SSR protocols for $n$ input copies and $m$ output copies each. 
Since the super-selection rule is asymptotically not
altering the rate, we have that 
 \begin{equation}
	E_D^{\rm SSR}(\hat \varrho) = E_D(\hat \varrho),
\end{equation}
where the right hand side is the distillable entanglement for the spin equivalent of $\hat \varrho$, possibly not respecting
super-selection rules \cite{PhysRevA.70.042310}. The right hand side can
be bounded from below
by the \emph{hashing bound}, as stated in the main text.
The \emph{entanglement cost} $E_C$ refers to the optimal rate that can be achieved in the converse, starting
from $m$ copies of maximally entangled Bell pairs and achieving approximately perfect $n$ copies of
the anticipated target state. Again, 
 \begin{equation}
	E_C^{\rm SSR}(\hat \varrho) = E_C(\hat \varrho),
\end{equation}
while in general $E_C(\hat \varrho)\geq E_D(\hat \varrho)$, as the process of distillation can be lossy compared to
the process of formation that the entanglement cost captures.

\subsection{Evaluation of the witness}
We present details on how to evaluate the  entanglement witness $E_G(\Gamma)$ in this section.
We shall make use of the fact that the \emph{von Neumann entropy}, 
in general being defined for a quantum state as
\begin{align}
  S(\hat \varrho)=-\tr[\hat \varrho\log(\hat \varrho)],
\end{align}
is unitarily invariant. For this reason, it can be easily and efficiently be evaluated for a 
 fermionic Gaussian state.
In the main text we are using the fact that given the covariance matrix $\Gamma=\Gamma(\hat \varrho)$ of the state $\hat \varrho$, 
we can associate to the state a unique Gaussian state $\hat \varrho^{(\Gamma)}$ with the same second moments. In fact, we have 
\begin{align}
  \hat \varrho^{(\Gamma)} = \text{argmax}_{\hat \sigma} \{ S(\hat \sigma) \text{ s.t. } \Gamma(\hat \sigma)=\Gamma(\hat\varrho) \},
\end{align}
which means that Gaussian states give rise to the maximum possible von Neumann 
entropy, among all quantum states for fixed second moments $\Gamma$~\cite{GKFGE16}.

For thermal states of particle number preserving Hamiltonians and for limits of such finite-temperature states, the entropy of a Gaussian state can be computed from the covariance matrix using standard expressions, 
see, 
e.g., Refs.~\cite{AreViolationWolf,Peschel},
 by obtaining the vector $n$ of the eigenvalues of $\Gamma$. We then have 
\begin{align}
  S^{(\Gamma)}=&S(\hat \varrho^{(\Gamma)})=-\sum_{k=1}^L f(n_k),
  \label{eq:ent_nk}
\end{align}
where the function $f:[0,1]\rightarrow [0,1]$  is defined as
\begin{equation}
  f(x)= \left\{
  \begin{array}{ll}
  - x  \log(x)  -  (1-x) \log(1-x)\ ,& \text{ if } x>0,\\
  0 ,& \text{ if } x=0.\\
  \end{array}
  \right.
  \end{equation}
This formula can be evaluated efficiently in the number of modes.
Finally, let us remark that the second moments of a \emph{reduced} density matrix $\hat \varrho_A$ describing a subsystem $A$ is obtained by restricting the second moments to $A$ which yields the corresponding covariance matrix $\Gamma_A$ and hence for a Gaussian state
\begin{align}
S(\hat \varrho_A) = S^{(\Gamma_A)}\ ,
\end{align}
as an expression in terms of the covariance matrix.\\

\subsection{Optimizing the witness value via local Bogoliubov transformations}

Our goal is to maximize the witness at hand to achieve the tightest possible lower bound bound.
In the main text we gave an example where system $A$ consists of the first 10 modes $A=\{1,2,\ldots,10\}$ and $B=\{11,\ldots,20\}$.
If we would apply the witness directly to systems $A$ and $B$ then it will give non-trivial values only for lowest temperatures where the thermal state is very close to the ground state.
Indeed, in the case if the global quantum state is pure, then we have $E_G(\Gamma) = S^{(\Gamma_A)} - S^{(\Gamma)}\approx S^{(\Gamma_A)}$ and whenever the ground state is effectively described by a conformal field theory the subsystem entropy will scale logarithmically in the size of $A$.
In the other limit of large temperatures, the entropies of whole system and subsystem $A$ will both scale according to the volume and so the witness will have a negative value.
To fix this, one has to make use of the freedom to make local unitary rotations on each subsystem.
The goal here is to find some modes that carry only very little entropy such that the witness has a chance to be non-negative.

In general finding the optimal local unitaries that maximize the witness can be complicated but we found that the following heuristic is helpful.
Denote by $\Gamma_A$ and $\Gamma_B$ the principal sub-matrices of $\Gamma$ which are the covariance matrices of subsystems $A$ and $B$ and let $U_A$ and $U_B$ be the corresponding diagonalization unitaries.
Then the  Gaussian state $\rh'$ with the second moments given by  
\begin{align}
  \Gamma' = U \Gamma U^\dagger = 
  \begin{pmatrix} \Lambda_A Q_{A,B}\\Q_{A,B}^\dagger \Lambda_B\end{pmatrix}
\end{align}
where $U=U_A\oplus U_B$, features an identical amount of entanglement because the entanglement cost is invariant under local unitary transformations $E_C(\rh)=E_C(\rh')$.
We hence can focus on the quantum state $\rh'$.

We observe that this is a viable \emph{distillation} heuristic, as now there can be a mode $a\in A$ and a mode $b\in B$ such that their covariance matrix $\Gamma'_{\{a,b\}}$ features a larger purity than before.
This is the meaning of the sparse off-diagonal structure seen in the figure presented in the main text.
We can restrict to these modes because entanglement cost is monotonous under partial traces 
\begin{equation}
E_C(\rh')\ge E_C(\rh'_{a,b}),
\end{equation}
 i.e., discarding some degrees of freedom can only decrease the available entanglement resources.
This discarding however is crucial because we get a better chance for the witness value to be non-trivial -- for this the discarded modes should take away most of the entropy while the reduced state of the remaining modes should be as pure as possible. 


\begin{thebibliography}{84}%
\makeatletter
\providecommand \@ifxundefined [1]{%
 \@ifx{#1\undefined}
}%
\providecommand \@ifnum [1]{%
 \ifnum #1\expandafter \@firstoftwo
 \else \expandafter \@secondoftwo
 \fi
}%
\providecommand \@ifx [1]{%
 \ifx #1\expandafter \@firstoftwo
 \else \expandafter \@secondoftwo
 \fi
}%
\providecommand \natexlab [1]{#1}%
\providecommand \enquote  [1]{``#1''}%
\providecommand \bibnamefont  [1]{#1}%
\providecommand \bibfnamefont [1]{#1}%
\providecommand \citenamefont [1]{#1}%
\providecommand \href@noop [0]{\@secondoftwo}%
\providecommand \href [0]{\begingroup \@sanitize@url \@href}%
\providecommand \@href[1]{\@@startlink{#1}\@@href}%
\providecommand \@@href[1]{\endgroup#1\@@endlink}%
\providecommand \@sanitize@url [0]{\catcode `\\12\catcode `\$12\catcode
  `\&12\catcode `\#12\catcode `\^12\catcode `\_12\catcode `\%12\relax}%
\providecommand \@@startlink[1]{}%
\providecommand \@@endlink[0]{}%
\providecommand \url  [0]{\begingroup\@sanitize@url \@url }%
\providecommand \@url [1]{\endgroup\@href {#1}{\urlprefix }}%
\providecommand \urlprefix  [0]{URL }%
\providecommand \Eprint [0]{\href }%
\providecommand \doibase [0]{http://dx.doi.org/}%
\providecommand \selectlanguage [0]{\@gobble}%
\providecommand \bibinfo  [0]{\@secondoftwo}%
\providecommand \bibfield  [0]{\@secondoftwo}%
\providecommand \translation [1]{[#1]}%
\providecommand \BibitemOpen [0]{}%
\providecommand \bibitemStop [0]{}%
\providecommand \bibitemNoStop [0]{.\EOS\space}%
\providecommand \EOS [0]{\spacefactor3000\relax}%
\providecommand \BibitemShut  [1]{\csname bibitem#1\endcsname}%
\let\auto@bib@innerbib\@empty
%</preamble>
\bibitem [{\citenamefont {Bloch}\ \emph {et~al.}(2012)\citenamefont {Bloch},
  \citenamefont {Dalibard},\ and\ \citenamefont
  {Nascimbene}}]{BlochSimulation}%
  \BibitemOpen
  \bibfield  {author} {\bibinfo {author} {\bibfnamefont {I.}~\bibnamefont
  {Bloch}}, \bibinfo {author} {\bibfnamefont {J.}~\bibnamefont {Dalibard}}, \
  and\ \bibinfo {author} {\bibfnamefont {S.}~\bibnamefont {Nascimbene}},\
  }\href@noop {} {\bibfield  {journal} {\bibinfo  {journal} {Nature Phys.}\
  }\textbf {\bibinfo {volume} {8}},\ \bibinfo {pages} {267} (\bibinfo {year}
  {2012})}\BibitemShut {NoStop}%
\bibitem [{\citenamefont {Greiner}\ \emph {et~al.}(2001)\citenamefont
  {Greiner}, \citenamefont {Bloch}, \citenamefont {Mandel}, \citenamefont
  {H\"ansch},\ and\ \citenamefont {Esslinger}}]{Greiner-PRL-2001}%
  \BibitemOpen
  \bibfield  {author} {\bibinfo {author} {\bibfnamefont {M.}~\bibnamefont
  {Greiner}}, \bibinfo {author} {\bibfnamefont {I.}~\bibnamefont {Bloch}},
  \bibinfo {author} {\bibfnamefont {O.}~\bibnamefont {Mandel}}, \bibinfo
  {author} {\bibfnamefont {T.~W.}\ \bibnamefont {H\"ansch}}, \ and\ \bibinfo
  {author} {\bibfnamefont {T.}~\bibnamefont {Esslinger}},\ }\href {\doibase
  10.1103/PhysRevLett.87.160405} {\bibfield  {journal} {\bibinfo  {journal}
  {Phys. Rev. Lett.}\ }\textbf {\bibinfo {volume} {87}},\ \bibinfo {pages}
  {160405} (\bibinfo {year} {2001})}\BibitemShut {NoStop}%
\bibitem [{\citenamefont {Endres}\ \emph {et~al.}(2011)\citenamefont {Endres},
  \citenamefont {Cheneau}, \citenamefont {Fukuhara}, \citenamefont
  {Weitenberg}, \citenamefont {Schauss}, \citenamefont {Gross}, \citenamefont
  {Mazza}, \citenamefont {Banuls}, \citenamefont {Pollet}, \citenamefont
  {Bloch},\ and\ \citenamefont {Kuhr}}]{StringOrder}%
  \BibitemOpen
  \bibfield  {author} {\bibinfo {author} {\bibfnamefont {M.}~\bibnamefont
  {Endres}}, \bibinfo {author} {\bibfnamefont {M.}~\bibnamefont {Cheneau}},
  \bibinfo {author} {\bibfnamefont {T.}~\bibnamefont {Fukuhara}}, \bibinfo
  {author} {\bibfnamefont {C.}~\bibnamefont {Weitenberg}}, \bibinfo {author}
  {\bibfnamefont {P.}~\bibnamefont {Schauss}}, \bibinfo {author} {\bibfnamefont
  {C.}~\bibnamefont {Gross}}, \bibinfo {author} {\bibfnamefont
  {L.}~\bibnamefont {Mazza}}, \bibinfo {author} {\bibfnamefont {M.~C.}\
  \bibnamefont {Banuls}}, \bibinfo {author} {\bibfnamefont {L.}~\bibnamefont
  {Pollet}}, \bibinfo {author} {\bibfnamefont {I.}~\bibnamefont {Bloch}}, \
  and\ \bibinfo {author} {\bibfnamefont {S.}~\bibnamefont {Kuhr}},\ }\href@noop
  {} {\bibfield  {journal} {\bibinfo  {journal} {Science}\ }\textbf {\bibinfo
  {volume} {334}},\ \bibinfo {pages} {200} (\bibinfo {year}
  {2011})}\BibitemShut {NoStop}%
\bibitem [{\citenamefont {Tai}\ \emph {et~al.}(2017)\citenamefont {Tai},
  \citenamefont {Lukin}, \citenamefont {Rispoli}, \citenamefont {Schittko},
  \citenamefont {Menke}, \citenamefont {Borgnia}, \citenamefont {Preiss},
  \citenamefont {Grusdt}, \citenamefont {Kaufman},\ and\ \citenamefont
  {Greiner}}]{HarperHofstadter}%
  \BibitemOpen
  \bibfield  {author} {\bibinfo {author} {\bibfnamefont {M.~E.}\ \bibnamefont
  {Tai}}, \bibinfo {author} {\bibfnamefont {A.}~\bibnamefont {Lukin}}, \bibinfo
  {author} {\bibfnamefont {M.}~\bibnamefont {Rispoli}}, \bibinfo {author}
  {\bibfnamefont {R.}~\bibnamefont {Schittko}}, \bibinfo {author}
  {\bibfnamefont {T.}~\bibnamefont {Menke}}, \bibinfo {author} {\bibfnamefont
  {D.}~\bibnamefont {Borgnia}}, \bibinfo {author} {\bibfnamefont {P.~M.}\
  \bibnamefont {Preiss}}, \bibinfo {author} {\bibfnamefont {F.}~\bibnamefont
  {Grusdt}}, \bibinfo {author} {\bibfnamefont {A.~M.}\ \bibnamefont {Kaufman}},
  \ and\ \bibinfo {author} {\bibfnamefont {M.}~\bibnamefont {Greiner}},\
  }\href@noop {} {\bibfield  {journal} {\bibinfo  {journal} {Nature}\ }\textbf
  {\bibinfo {volume} {546}},\ \bibinfo {pages} {519} (\bibinfo {year}
  {2017})}\BibitemShut {NoStop}%
\bibitem [{\citenamefont {Trotzky}\ \emph {et~al.}(2010)\citenamefont
  {Trotzky}, \citenamefont {Pollet}, \citenamefont {Gerbier}, \citenamefont
  {Schnorrberger}, \citenamefont {Bloch}, \citenamefont {Prokof'ev},
  \citenamefont {Svistunov},\ and\ \citenamefont
  {Troyer}}]{MonteCarloValidator}%
  \BibitemOpen
  \bibfield  {author} {\bibinfo {author} {\bibfnamefont {S.}~\bibnamefont
  {Trotzky}}, \bibinfo {author} {\bibfnamefont {L.}~\bibnamefont {Pollet}},
  \bibinfo {author} {\bibfnamefont {F.}~\bibnamefont {Gerbier}}, \bibinfo
  {author} {\bibfnamefont {U.}~\bibnamefont {Schnorrberger}}, \bibinfo {author}
  {\bibfnamefont {I.}~\bibnamefont {Bloch}}, \bibinfo {author} {\bibfnamefont
  {N.}~\bibnamefont {Prokof'ev}}, \bibinfo {author} {\bibfnamefont
  {B.}~\bibnamefont {Svistunov}}, \ and\ \bibinfo {author} {\bibfnamefont
  {M.}~\bibnamefont {Troyer}},\ }\href@noop {} {\bibfield  {journal} {\bibinfo
  {journal} {Nature Phys.}\ }\textbf {\bibinfo {volume} {6}},\ \bibinfo {pages}
  {998} (\bibinfo {year} {2010})}\BibitemShut {NoStop}%
\bibitem [{\citenamefont {Trotzky}\ \emph {et~al.}(2012)\citenamefont
  {Trotzky}, \citenamefont {Chen}, \citenamefont {Flesch}, \citenamefont
  {McCulloch}, \citenamefont {Schollw{\"o}ck}, \citenamefont {Eisert},\ and\
  \citenamefont {Bloch}}]{Trotzky_etal12}%
  \BibitemOpen
  \bibfield  {author} {\bibinfo {author} {\bibfnamefont {S.}~\bibnamefont
  {Trotzky}}, \bibinfo {author} {\bibfnamefont {Y.-A.}\ \bibnamefont {Chen}},
  \bibinfo {author} {\bibfnamefont {A.}~\bibnamefont {Flesch}}, \bibinfo
  {author} {\bibfnamefont {I.~P.}\ \bibnamefont {McCulloch}}, \bibinfo {author}
  {\bibfnamefont {U.}~\bibnamefont {Schollw{\"o}ck}}, \bibinfo {author}
  {\bibfnamefont {J.}~\bibnamefont {Eisert}}, \ and\ \bibinfo {author}
  {\bibfnamefont {I.}~\bibnamefont {Bloch}},\ }\href {\doibase
  10.1038/nphys2232} {\bibfield  {journal} {\bibinfo  {journal} {Nature Phys.}\
  }\textbf {\bibinfo {volume} {8}},\ \bibinfo {pages} {325} (\bibinfo {year}
  {2012})}\BibitemShut {NoStop}%
\bibitem [{\citenamefont {Kaufman}\ \emph {et~al.}(2016)\citenamefont
  {Kaufman}, \citenamefont {Tai}, \citenamefont {Lukin}, \citenamefont
  {Rispoli}, \citenamefont {Schittko}, \citenamefont {Preiss},\ and\
  \citenamefont {Greiner}}]{Kaufman}%
  \BibitemOpen
  \bibfield  {author} {\bibinfo {author} {\bibfnamefont {A.~M.}\ \bibnamefont
  {Kaufman}}, \bibinfo {author} {\bibfnamefont {M.~E.}\ \bibnamefont {Tai}},
  \bibinfo {author} {\bibfnamefont {A.}~\bibnamefont {Lukin}}, \bibinfo
  {author} {\bibfnamefont {M.}~\bibnamefont {Rispoli}}, \bibinfo {author}
  {\bibfnamefont {R.}~\bibnamefont {Schittko}}, \bibinfo {author}
  {\bibfnamefont {P.~M.}\ \bibnamefont {Preiss}}, \ and\ \bibinfo {author}
  {\bibfnamefont {M.}~\bibnamefont {Greiner}},\ }\href@noop {} {\bibfield
  {journal} {\bibinfo  {journal} {Science}\ }\textbf {\bibinfo {volume}
  {353}},\ \bibinfo {pages} {794} (\bibinfo {year} {2016})}\BibitemShut
  {NoStop}%
\bibitem [{\citenamefont {Braun}\ \emph {et~al.}(2015)\citenamefont {Braun},
  \citenamefont {Friesdorf}, \citenamefont {Hodgman}, \citenamefont
  {Schreiber}, \citenamefont {Ronzheimer}, \citenamefont {Riera}, \citenamefont
  {del Rey}, \citenamefont {Bloch}, \citenamefont {Eisert},\ and\ \citenamefont
  {Schneider}}]{Emergence}%
  \BibitemOpen
  \bibfield  {author} {\bibinfo {author} {\bibfnamefont {S.}~\bibnamefont
  {Braun}}, \bibinfo {author} {\bibfnamefont {M.}~\bibnamefont {Friesdorf}},
  \bibinfo {author} {\bibfnamefont {S.~S.}\ \bibnamefont {Hodgman}}, \bibinfo
  {author} {\bibfnamefont {M.}~\bibnamefont {Schreiber}}, \bibinfo {author}
  {\bibfnamefont {J.~P.~P.}\ \bibnamefont {Ronzheimer}}, \bibinfo {author}
  {\bibfnamefont {A.}~\bibnamefont {Riera}}, \bibinfo {author} {\bibfnamefont
  {M.}~\bibnamefont {del Rey}}, \bibinfo {author} {\bibfnamefont
  {I.}~\bibnamefont {Bloch}}, \bibinfo {author} {\bibfnamefont
  {J.}~\bibnamefont {Eisert}}, \ and\ \bibinfo {author} {\bibfnamefont
  {U.}~\bibnamefont {Schneider}},\ }\href@noop {} {\bibfield  {journal}
  {\bibinfo  {journal} {Proc. Natl. Ac. Sc.}\ }\textbf {\bibinfo {volume}
  {112}},\ \bibinfo {pages} {3641} (\bibinfo {year} {2015})}\BibitemShut
  {NoStop}%
\bibitem [{\citenamefont {Ronzheimer}\ \emph {et~al.}(2013)\citenamefont
  {Ronzheimer}, \citenamefont {Schreiber}, \citenamefont {Braun}, \citenamefont
  {Hodgman}, \citenamefont {Langer}, \citenamefont {McCulloch}, \citenamefont
  {Heidrich-Meisner}, \citenamefont {Bloch},\ and\ \citenamefont
  {Schneider}}]{Expansion}%
  \BibitemOpen
  \bibfield  {author} {\bibinfo {author} {\bibfnamefont {J.~P.}\ \bibnamefont
  {Ronzheimer}}, \bibinfo {author} {\bibfnamefont {M.}~\bibnamefont
  {Schreiber}}, \bibinfo {author} {\bibfnamefont {S.}~\bibnamefont {Braun}},
  \bibinfo {author} {\bibfnamefont {S.~S.}\ \bibnamefont {Hodgman}}, \bibinfo
  {author} {\bibfnamefont {S.}~\bibnamefont {Langer}}, \bibinfo {author}
  {\bibfnamefont {I.~P.}\ \bibnamefont {McCulloch}}, \bibinfo {author}
  {\bibfnamefont {F.}~\bibnamefont {Heidrich-Meisner}}, \bibinfo {author}
  {\bibfnamefont {I.}~\bibnamefont {Bloch}}, \ and\ \bibinfo {author}
  {\bibfnamefont {U.}~\bibnamefont {Schneider}},\ }\href@noop {} {\bibfield
  {journal} {\bibinfo  {journal} {Phys. Rev. Lett.}\ }\textbf {\bibinfo
  {volume} {110}},\ \bibinfo {pages} {205301} (\bibinfo {year}
  {2013})}\BibitemShut {NoStop}%
\bibitem [{\citenamefont {Hofferberth}\ \emph {et~al.}(2007)\citenamefont
  {Hofferberth}, \citenamefont {Lesanovsky}, \citenamefont {Fischer},
  \citenamefont {Schumm},\ and\ \citenamefont
  {Schmiedmayer}}]{Hofferberth_etal07}%
  \BibitemOpen
  \bibfield  {author} {\bibinfo {author} {\bibfnamefont {S.}~\bibnamefont
  {Hofferberth}}, \bibinfo {author} {\bibfnamefont {I.}~\bibnamefont
  {Lesanovsky}}, \bibinfo {author} {\bibfnamefont {B.}~\bibnamefont {Fischer}},
  \bibinfo {author} {\bibfnamefont {T.}~\bibnamefont {Schumm}}, \ and\ \bibinfo
  {author} {\bibfnamefont {J.}~\bibnamefont {Schmiedmayer}},\ }\href@noop {}
  {\bibfield  {journal} {\bibinfo  {journal} {Nature}\ }\textbf {\bibinfo
  {volume} {449}},\ \bibinfo {pages} {324} (\bibinfo {year}
  {2007})}\BibitemShut {NoStop}%
\bibitem [{\citenamefont {Gring}\ \emph {et~al.}(2012)\citenamefont {Gring},
  \citenamefont {Kuhnert}, \citenamefont {Langen}, \citenamefont {Kitagawa},
  \citenamefont {Rauer}, \citenamefont {Schreitl}, \citenamefont {Mazets},
  \citenamefont {Smith}, \citenamefont {Demler},\ and\ \citenamefont
  {Schmiedmayer}}]{Gring_etal12}%
  \BibitemOpen
  \bibfield  {author} {\bibinfo {author} {\bibfnamefont {M.}~\bibnamefont
  {Gring}}, \bibinfo {author} {\bibfnamefont {M.}~\bibnamefont {Kuhnert}},
  \bibinfo {author} {\bibfnamefont {T.}~\bibnamefont {Langen}}, \bibinfo
  {author} {\bibfnamefont {T.}~\bibnamefont {Kitagawa}}, \bibinfo {author}
  {\bibfnamefont {B.}~\bibnamefont {Rauer}}, \bibinfo {author} {\bibfnamefont
  {M.}~\bibnamefont {Schreitl}}, \bibinfo {author} {\bibfnamefont
  {I.}~\bibnamefont {Mazets}}, \bibinfo {author} {\bibfnamefont {D.~A.}\
  \bibnamefont {Smith}}, \bibinfo {author} {\bibfnamefont {E.}~\bibnamefont
  {Demler}}, \ and\ \bibinfo {author} {\bibfnamefont {J.}~\bibnamefont
  {Schmiedmayer}},\ }\href@noop {} {\bibfield  {journal} {\bibinfo  {journal}
  {Science}\ }\textbf {\bibinfo {volume} {337}},\ \bibinfo {pages} {1318}
  (\bibinfo {year} {2012})}\BibitemShut {NoStop}%
\bibitem [{\citenamefont {Langen}\ \emph {et~al.}(2015)\citenamefont {Langen},
  \citenamefont {Erne}, \citenamefont {Geiger}, \citenamefont {Rauer},
  \citenamefont {Schweigler}, \citenamefont {Kuhnert}, \citenamefont
  {Rohringer}, \citenamefont {Mazets}, \citenamefont {Gasenzer},\ and\
  \citenamefont {Schmiedmayer}}]{SchmiedmayerGGE}%
  \BibitemOpen
  \bibfield  {author} {\bibinfo {author} {\bibfnamefont {T.}~\bibnamefont
  {Langen}}, \bibinfo {author} {\bibfnamefont {S.}~\bibnamefont {Erne}},
  \bibinfo {author} {\bibfnamefont {R.}~\bibnamefont {Geiger}}, \bibinfo
  {author} {\bibfnamefont {B.}~\bibnamefont {Rauer}}, \bibinfo {author}
  {\bibfnamefont {T.}~\bibnamefont {Schweigler}}, \bibinfo {author}
  {\bibfnamefont {M.}~\bibnamefont {Kuhnert}}, \bibinfo {author} {\bibfnamefont
  {W.}~\bibnamefont {Rohringer}}, \bibinfo {author} {\bibfnamefont {I.~E.}\
  \bibnamefont {Mazets}}, \bibinfo {author} {\bibfnamefont {T.}~\bibnamefont
  {Gasenzer}}, \ and\ \bibinfo {author} {\bibfnamefont {J.}~\bibnamefont
  {Schmiedmayer}},\ }\href@noop {} {\bibfield  {journal} {\bibinfo  {journal}
  {Science}\ }\textbf {\bibinfo {volume} {348}},\ \bibinfo {pages} {207}
  (\bibinfo {year} {2015})}\BibitemShut {NoStop}%
\bibitem [{\citenamefont {Cheneau}\ \emph {et~al.}(2012)\citenamefont
  {Cheneau}, \citenamefont {Barmettler}, \citenamefont {Poletti}, \citenamefont
  {Endres}, \citenamefont {Schauss}, \citenamefont {Fukuhara}, \citenamefont
  {Gross}, \citenamefont {Bloch}, \citenamefont {Kollath},\ and\ \citenamefont
  {Kuhr}}]{1111.0776}%
  \BibitemOpen
  \bibfield  {author} {\bibinfo {author} {\bibfnamefont {M.}~\bibnamefont
  {Cheneau}}, \bibinfo {author} {\bibfnamefont {P.}~\bibnamefont {Barmettler}},
  \bibinfo {author} {\bibfnamefont {D.}~\bibnamefont {Poletti}}, \bibinfo
  {author} {\bibfnamefont {M.}~\bibnamefont {Endres}}, \bibinfo {author}
  {\bibfnamefont {P.}~\bibnamefont {Schauss}}, \bibinfo {author} {\bibfnamefont
  {T.}~\bibnamefont {Fukuhara}}, \bibinfo {author} {\bibfnamefont
  {C.}~\bibnamefont {Gross}}, \bibinfo {author} {\bibfnamefont
  {I.}~\bibnamefont {Bloch}}, \bibinfo {author} {\bibfnamefont
  {C.}~\bibnamefont {Kollath}}, \ and\ \bibinfo {author} {\bibfnamefont
  {S.}~\bibnamefont {Kuhr}},\ }\href@noop {} {\bibfield  {journal} {\bibinfo
  {journal} {Nature}\ }\textbf {\bibinfo {volume} {481}},\ \bibinfo {pages}
  {484} (\bibinfo {year} {2012})}\BibitemShut {NoStop}%
\bibitem [{\citenamefont {Schweigler}\ \emph {et~al.}(2017)\citenamefont
  {Schweigler}, \citenamefont {Kasper}, \citenamefont {Erne}, \citenamefont
  {Mazets}, \citenamefont {Rauer}, \citenamefont {Cataldini}, \citenamefont
  {Langen}, \citenamefont {Gasenzer}, \citenamefont {Berges},\ and\
  \citenamefont {Schmiedmayer}}]{Schweigler2017}%
  \BibitemOpen
  \bibfield  {author} {\bibinfo {author} {\bibfnamefont {T.}~\bibnamefont
  {Schweigler}}, \bibinfo {author} {\bibfnamefont {V.}~\bibnamefont {Kasper}},
  \bibinfo {author} {\bibfnamefont {S.}~\bibnamefont {Erne}}, \bibinfo {author}
  {\bibfnamefont {I.~E.}\ \bibnamefont {Mazets}}, \bibinfo {author}
  {\bibfnamefont {B.}~\bibnamefont {Rauer}}, \bibinfo {author} {\bibfnamefont
  {F.}~\bibnamefont {Cataldini}}, \bibinfo {author} {\bibfnamefont
  {T.}~\bibnamefont {Langen}}, \bibinfo {author} {\bibfnamefont
  {T.}~\bibnamefont {Gasenzer}}, \bibinfo {author} {\bibfnamefont
  {J.}~\bibnamefont {Berges}}, \ and\ \bibinfo {author} {\bibfnamefont
  {J.}~\bibnamefont {Schmiedmayer}},\ }\href {\doibase 10.1038/nature22310}
  {\bibfield  {journal} {\bibinfo  {journal} {Nature}\ }\textbf {\bibinfo
  {volume} {545}},\ \bibinfo {pages} {323} (\bibinfo {year}
  {2017})}\BibitemShut {NoStop}%
\bibitem [{\citenamefont {Acin}\ \emph {et~al.}(2018)\citenamefont {Acin},
  \citenamefont {Bloch}, \citenamefont {Buhrman}, \citenamefont {Calarco},
  \citenamefont {Eichler}, \citenamefont {Eisert}, \citenamefont {Esteve},
  \citenamefont {Gisin}, \citenamefont {Glaser}, \citenamefont {Jelezko},
  \citenamefont {Kuhr}, \citenamefont {Lewenstein}, \citenamefont {Riedel},
  \citenamefont {Schmidt}, \citenamefont {Thew}, \citenamefont {Wallraff},
  \citenamefont {Walmsley},\ and\ \citenamefont {Wilhelm}}]{Roadmap}%
  \BibitemOpen
  \bibfield  {author} {\bibinfo {author} {\bibfnamefont {A.}~\bibnamefont
  {Acin}}, \bibinfo {author} {\bibfnamefont {I.}~\bibnamefont {Bloch}},
  \bibinfo {author} {\bibfnamefont {H.}~\bibnamefont {Buhrman}}, \bibinfo
  {author} {\bibfnamefont {T.}~\bibnamefont {Calarco}}, \bibinfo {author}
  {\bibfnamefont {C.}~\bibnamefont {Eichler}}, \bibinfo {author} {\bibfnamefont
  {J.}~\bibnamefont {Eisert}}, \bibinfo {author} {\bibfnamefont
  {D.}~\bibnamefont {Esteve}}, \bibinfo {author} {\bibfnamefont
  {N.}~\bibnamefont {Gisin}}, \bibinfo {author} {\bibfnamefont {S.~J.}\
  \bibnamefont {Glaser}}, \bibinfo {author} {\bibfnamefont {F.}~\bibnamefont
  {Jelezko}}, \bibinfo {author} {\bibfnamefont {S.}~\bibnamefont {Kuhr}},
  \bibinfo {author} {\bibfnamefont {M.}~\bibnamefont {Lewenstein}}, \bibinfo
  {author} {\bibfnamefont {M.~F.}\ \bibnamefont {Riedel}}, \bibinfo {author}
  {\bibfnamefont {P.~O.}\ \bibnamefont {Schmidt}}, \bibinfo {author}
  {\bibfnamefont {R.}~\bibnamefont {Thew}}, \bibinfo {author} {\bibfnamefont
  {A.}~\bibnamefont {Wallraff}}, \bibinfo {author} {\bibfnamefont
  {I.}~\bibnamefont {Walmsley}}, \ and\ \bibinfo {author} {\bibfnamefont
  {F.~K.}\ \bibnamefont {Wilhelm}},\ }\href@noop {} {\bibfield  {journal}
  {\bibinfo  {journal} {New J. Phys.}\ }\textbf {\bibinfo {volume} {20}},\
  \bibinfo {pages} {080201} (\bibinfo {year} {2018})}\BibitemShut {NoStop}%
\bibitem [{\citenamefont {Mazurenko}\ \emph {et~al.}(2017)\citenamefont
  {Mazurenko}, \citenamefont {Chiu}, \citenamefont {Ji}, \citenamefont
  {Parsons}, \citenamefont {Kanasz-Nagy}, \citenamefont {Schmidt},
  \citenamefont {E.}, \citenamefont {Demler}, , \citenamefont {Greif},\ and\
  \citenamefont {Greiner}}]{Mazurenko}%
  \BibitemOpen
  \bibfield  {author} {\bibinfo {author} {\bibfnamefont {A.}~\bibnamefont
  {Mazurenko}}, \bibinfo {author} {\bibfnamefont {C.~S.}\ \bibnamefont {Chiu}},
  \bibinfo {author} {\bibfnamefont {G.}~\bibnamefont {Ji}}, \bibinfo {author}
  {\bibfnamefont {M.~F.}\ \bibnamefont {Parsons}}, \bibinfo {author}
  {\bibfnamefont {M.}~\bibnamefont {Kanasz-Nagy}}, \bibinfo {author}
  {\bibfnamefont {R.}~\bibnamefont {Schmidt}}, \bibinfo {author} {\bibfnamefont
  {F.~G.}\ \bibnamefont {E.}}, \bibinfo {author} {\bibnamefont {Demler}}, ,
  \bibinfo {author} {\bibnamefont {Greif}}, \ and\ \bibinfo {author}
  {\bibfnamefont {M.}~\bibnamefont {Greiner}},\ }\href@noop {} {\bibfield
  {journal} {\bibinfo  {journal} {Nature}\ }\textbf {\bibinfo {volume} {545}},\
  \bibinfo {pages} {462} (\bibinfo {year} {2017})}\BibitemShut {NoStop}%
\bibitem [{\citenamefont {K{\"o}hl}\ \emph {et~al.}(2005)\citenamefont
  {K{\"o}hl}, \citenamefont {Moritz}, \citenamefont {St{\"o}ferle},
  \citenamefont {G{\"u}nter},\ and\ \citenamefont {Esslinger}}]{ColdFermions}%
  \BibitemOpen
  \bibfield  {author} {\bibinfo {author} {\bibfnamefont {M.}~\bibnamefont
  {K{\"o}hl}}, \bibinfo {author} {\bibfnamefont {H.}~\bibnamefont {Moritz}},
  \bibinfo {author} {\bibfnamefont {T.}~\bibnamefont {St{\"o}ferle}}, \bibinfo
  {author} {\bibfnamefont {K.}~\bibnamefont {G{\"u}nter}}, \ and\ \bibinfo
  {author} {\bibfnamefont {T.}~\bibnamefont {Esslinger}},\ }\href@noop {}
  {\bibfield  {journal} {\bibinfo  {journal} {Phys. Rev. Lett.}\ }\textbf
  {\bibinfo {volume} {94}},\ \bibinfo {pages} {080403} (\bibinfo {year}
  {2005})}\BibitemShut {NoStop}%
\bibitem [{\citenamefont {Esslinger}(2010)}]{EsslingerReview}%
  \BibitemOpen
  \bibfield  {author} {\bibinfo {author} {\bibfnamefont {T.}~\bibnamefont
  {Esslinger}},\ }\href@noop {} {\bibfield  {journal} {\bibinfo  {journal}
  {Ann. Rev. Con and Mat. Phys.}\ }\textbf {\bibinfo {volume} {1}},\ \bibinfo
  {pages} {129} (\bibinfo {year} {2010})}\BibitemShut {NoStop}%
\bibitem [{\citenamefont {Schneider}\ \emph {et~al.}(2012)\citenamefont
  {Schneider}, \citenamefont {Hackerm{\"u}ller}, \citenamefont {Ronzheimer},
  \citenamefont {Will}, \citenamefont {S.~Braun}, \citenamefont {Bloch},
  \citenamefont {Demler}, \citenamefont {Mandt}, \citenamefont {Rasch},\ and\
  \citenamefont {Rosch}}]{RoschTransport}%
  \BibitemOpen
  \bibfield  {author} {\bibinfo {author} {\bibfnamefont {U.}~\bibnamefont
  {Schneider}}, \bibinfo {author} {\bibfnamefont {L.}~\bibnamefont
  {Hackerm{\"u}ller}}, \bibinfo {author} {\bibfnamefont {J.~P.}\ \bibnamefont
  {Ronzheimer}}, \bibinfo {author} {\bibfnamefont {S.}~\bibnamefont {Will}},
  \bibinfo {author} {\bibfnamefont {T.~B.}\ \bibnamefont {S.~Braun}}, \bibinfo
  {author} {\bibfnamefont {I.}~\bibnamefont {Bloch}}, \bibinfo {author}
  {\bibfnamefont {E.}~\bibnamefont {Demler}}, \bibinfo {author} {\bibfnamefont
  {S.}~\bibnamefont {Mandt}}, \bibinfo {author} {\bibfnamefont
  {D.}~\bibnamefont {Rasch}}, \ and\ \bibinfo {author} {\bibfnamefont
  {A.}~\bibnamefont {Rosch}},\ }\href@noop {} {\bibfield  {journal} {\bibinfo
  {journal} {Nature Phys.}\ }\textbf {\bibinfo {volume} {8}},\ \bibinfo {pages}
  {213} (\bibinfo {year} {2012})}\BibitemShut {NoStop}%
\bibitem [{\citenamefont {Rom}\ \emph {et~al.}(2006)\citenamefont {Rom},
  \citenamefont {Best}, \citenamefont {van Oosten}, \citenamefont {Schneider},
  \citenamefont {Foelling}, \citenamefont {Paredes},\ and\ \citenamefont
  {Bloch}}]{Rom}%
  \BibitemOpen
  \bibfield  {author} {\bibinfo {author} {\bibfnamefont {T.}~\bibnamefont
  {Rom}}, \bibinfo {author} {\bibfnamefont {T.}~\bibnamefont {Best}}, \bibinfo
  {author} {\bibfnamefont {D.}~\bibnamefont {van Oosten}}, \bibinfo {author}
  {\bibfnamefont {U.}~\bibnamefont {Schneider}}, \bibinfo {author}
  {\bibfnamefont {S.}~\bibnamefont {Foelling}}, \bibinfo {author}
  {\bibfnamefont {B.}~\bibnamefont {Paredes}}, \ and\ \bibinfo {author}
  {\bibfnamefont {I.}~\bibnamefont {Bloch}},\ }\href@noop {} {\bibfield
  {journal} {\bibinfo  {journal} {Nature}\ }\textbf {\bibinfo {volume} {444}},\
  \bibinfo {pages} {733} (\bibinfo {year} {2006})}\BibitemShut {NoStop}%
\bibitem [{\citenamefont {Chiu}\ \emph {et~al.}(2018)\citenamefont {Chiu},
  \citenamefont {Ji}, \citenamefont {Mazurenko}, \citenamefont {Greif},\ and\
  \citenamefont {Greiner}}]{PhysRevLett.120.243201}%
  \BibitemOpen
  \bibfield  {author} {\bibinfo {author} {\bibfnamefont {C.~S.}\ \bibnamefont
  {Chiu}}, \bibinfo {author} {\bibfnamefont {G.}~\bibnamefont {Ji}}, \bibinfo
  {author} {\bibfnamefont {A.}~\bibnamefont {Mazurenko}}, \bibinfo {author}
  {\bibfnamefont {D.}~\bibnamefont {Greif}}, \ and\ \bibinfo {author}
  {\bibfnamefont {M.}~\bibnamefont {Greiner}},\ }\href {\doibase
  10.1103/PhysRevLett.120.243201} {\bibfield  {journal} {\bibinfo  {journal}
  {Phys. Rev. Lett.}\ }\textbf {\bibinfo {volume} {120}},\ \bibinfo {pages}
  {243201} (\bibinfo {year} {2018})}\BibitemShut {NoStop}%
\bibitem [{\citenamefont {Behrle}\ \emph {et~al.}(2018)\citenamefont {Behrle},
  \citenamefont {Harrison}, \citenamefont {Kombe}, \citenamefont {Gao},
  \citenamefont {Link}, \citenamefont {Bernier}, \citenamefont {Kollath},\ and\
  \citenamefont {K{\"o}hl}}]{Koehl}%
  \BibitemOpen
  \bibfield  {author} {\bibinfo {author} {\bibfnamefont {A.}~\bibnamefont
  {Behrle}}, \bibinfo {author} {\bibfnamefont {T.}~\bibnamefont {Harrison}},
  \bibinfo {author} {\bibfnamefont {J.}~\bibnamefont {Kombe}}, \bibinfo
  {author} {\bibfnamefont {K.}~\bibnamefont {Gao}}, \bibinfo {author}
  {\bibfnamefont {M.}~\bibnamefont {Link}}, \bibinfo {author} {\bibfnamefont
  {J.-S.}\ \bibnamefont {Bernier}}, \bibinfo {author} {\bibfnamefont
  {C.}~\bibnamefont {Kollath}}, \ and\ \bibinfo {author} {\bibfnamefont
  {M.}~\bibnamefont {K{\"o}hl}},\ }\href@noop {} {\bibfield  {journal}
  {\bibinfo  {journal} {Nature Phys.}\ }\textbf {\bibinfo {volume} {14}},\
  \bibinfo {pages} {781} (\bibinfo {year} {2018})}\BibitemShut {NoStop}%
\bibitem [{\citenamefont {Viebahn}\ \emph {et~al.}(2019)\citenamefont
  {Viebahn}, \citenamefont {Sbroscia}, \citenamefont {Carter}, \citenamefont
  {Yu},\ and\ \citenamefont {Schneider}}]{PhysRevLett.122.110404}%
  \BibitemOpen
  \bibfield  {author} {\bibinfo {author} {\bibfnamefont {K.}~\bibnamefont
  {Viebahn}}, \bibinfo {author} {\bibfnamefont {M.}~\bibnamefont {Sbroscia}},
  \bibinfo {author} {\bibfnamefont {E.}~\bibnamefont {Carter}}, \bibinfo
  {author} {\bibfnamefont {J.-C.}\ \bibnamefont {Yu}}, \ and\ \bibinfo {author}
  {\bibfnamefont {U.}~\bibnamefont {Schneider}},\ }\href {\doibase
  10.1103/PhysRevLett.122.110404} {\bibfield  {journal} {\bibinfo  {journal}
  {Phys. Rev. Lett.}\ }\textbf {\bibinfo {volume} {122}},\ \bibinfo {pages}
  {110404} (\bibinfo {year} {2019})}\BibitemShut {NoStop}%
\bibitem [{\citenamefont {Chiu}\ \emph {et~al.}(2019)\citenamefont {Chiu},
  \citenamefont {Ji}, \citenamefont {Bohrdt}, \citenamefont {Xu}, \citenamefont
  {Knap}, \citenamefont {Demler}, \citenamefont {Grusdt}, \citenamefont
  {Greiner},\ and\ \citenamefont {Greif}}]{Chiu251}%
  \BibitemOpen
  \bibfield  {author} {\bibinfo {author} {\bibfnamefont {C.~S.}\ \bibnamefont
  {Chiu}}, \bibinfo {author} {\bibfnamefont {G.}~\bibnamefont {Ji}}, \bibinfo
  {author} {\bibfnamefont {A.}~\bibnamefont {Bohrdt}}, \bibinfo {author}
  {\bibfnamefont {M.}~\bibnamefont {Xu}}, \bibinfo {author} {\bibfnamefont
  {M.}~\bibnamefont {Knap}}, \bibinfo {author} {\bibfnamefont {E.}~\bibnamefont
  {Demler}}, \bibinfo {author} {\bibfnamefont {F.}~\bibnamefont {Grusdt}},
  \bibinfo {author} {\bibfnamefont {M.}~\bibnamefont {Greiner}}, \ and\
  \bibinfo {author} {\bibfnamefont {D.}~\bibnamefont {Greif}},\ }\href
  {\doibase 10.1126/science.aav3587} {\bibfield  {journal} {\bibinfo  {journal}
  {Science}\ }\textbf {\bibinfo {volume} {365}},\ \bibinfo {pages} {251}
  (\bibinfo {year} {2019})}\BibitemShut {NoStop}%
\bibitem [{\citenamefont {Yang}\ \emph {et~al.}(2020)\citenamefont {Yang},
  \citenamefont {Sun}, \citenamefont {Huang}, \citenamefont {Wang},
  \citenamefont {Deng}, \citenamefont {Dai}, \citenamefont {Yuan},\ and\
  \citenamefont {Pan}}]{yang2020cooling}%
  \BibitemOpen
  \bibfield  {author} {\bibinfo {author} {\bibfnamefont {B.}~\bibnamefont
  {Yang}}, \bibinfo {author} {\bibfnamefont {H.}~\bibnamefont {Sun}}, \bibinfo
  {author} {\bibfnamefont {C.-J.}\ \bibnamefont {Huang}}, \bibinfo {author}
  {\bibfnamefont {H.-Y.}\ \bibnamefont {Wang}}, \bibinfo {author}
  {\bibfnamefont {Y.}~\bibnamefont {Deng}}, \bibinfo {author} {\bibfnamefont
  {H.-N.}\ \bibnamefont {Dai}}, \bibinfo {author} {\bibfnamefont {Z.-S.}\
  \bibnamefont {Yuan}}, \ and\ \bibinfo {author} {\bibfnamefont {J.-W.}\
  \bibnamefont {Pan}},\ }\href {\doibase 10.1126/science.aaz6801} {\bibfield
  {journal} {\bibinfo  {journal} {Science}\ }\textbf {\bibinfo {volume}
  {369}},\ \bibinfo {pages} {550} (\bibinfo {year} {2020})}\BibitemShut
  {NoStop}%
\bibitem [{\citenamefont {Eisert}\ \emph {et~al.}(2020)\citenamefont {Eisert},
  \citenamefont {Hangleiter}, \citenamefont {Walk}, \citenamefont {Roth},
  \citenamefont {Markham}, \citenamefont {Parekh}, \citenamefont {Chabaud},\
  and\ \citenamefont {Kashefi}}]{eisert2019quantum}%
  \BibitemOpen
  \bibfield  {author} {\bibinfo {author} {\bibfnamefont {J.}~\bibnamefont
  {Eisert}}, \bibinfo {author} {\bibfnamefont {D.}~\bibnamefont {Hangleiter}},
  \bibinfo {author} {\bibfnamefont {N.}~\bibnamefont {Walk}}, \bibinfo {author}
  {\bibfnamefont {I.}~\bibnamefont {Roth}}, \bibinfo {author} {\bibfnamefont
  {D.}~\bibnamefont {Markham}}, \bibinfo {author} {\bibfnamefont
  {R.}~\bibnamefont {Parekh}}, \bibinfo {author} {\bibfnamefont
  {U.}~\bibnamefont {Chabaud}}, \ and\ \bibinfo {author} {\bibfnamefont
  {E.}~\bibnamefont {Kashefi}},\ }\href@noop {} {\bibfield  {journal} {\bibinfo
   {journal} {Nature Rev. Phys.}\ }\textbf {\bibinfo {volume} {2}},\ \bibinfo
  {pages} {382} (\bibinfo {year} {2020})}\BibitemShut {NoStop}%
\bibitem [{\citenamefont {Sherson}\ \emph {et~al.}(2010)\citenamefont
  {Sherson}, \citenamefont {Weitenberg}, \citenamefont {Endres}, \citenamefont
  {Cheneau}, \citenamefont {Bloch},\ and\ \citenamefont
  {Kuhr}}]{Sherson-Nature-2010}%
  \BibitemOpen
  \bibfield  {author} {\bibinfo {author} {\bibfnamefont {J.~F.}\ \bibnamefont
  {Sherson}}, \bibinfo {author} {\bibfnamefont {C.}~\bibnamefont {Weitenberg}},
  \bibinfo {author} {\bibfnamefont {M.}~\bibnamefont {Endres}}, \bibinfo
  {author} {\bibfnamefont {M.}~\bibnamefont {Cheneau}}, \bibinfo {author}
  {\bibfnamefont {I.}~\bibnamefont {Bloch}}, \ and\ \bibinfo {author}
  {\bibfnamefont {S.}~\bibnamefont {Kuhr}},\ }\href {\doibase
  10.1038/nature09378} {\bibfield  {journal} {\bibinfo  {journal} {Nature}\
  }\textbf {\bibinfo {volume} {467}},\ \bibinfo {pages} {68 } (\bibinfo {year}
  {2010})}\BibitemShut {NoStop}%
\bibitem [{\citenamefont {Bakr}\ \emph {et~al.}(2010)\citenamefont {Bakr},
  \citenamefont {Peng}, \citenamefont {Tai}, \citenamefont {Ma}, \citenamefont
  {Simon}, \citenamefont {Gillen}, \citenamefont {F{\"o}lling}, \citenamefont
  {Pollet},\ and\ \citenamefont {Greiner}}]{Bakr-Science-2010}%
  \BibitemOpen
  \bibfield  {author} {\bibinfo {author} {\bibfnamefont {W.~S.}\ \bibnamefont
  {Bakr}}, \bibinfo {author} {\bibfnamefont {A.}~\bibnamefont {Peng}}, \bibinfo
  {author} {\bibfnamefont {M.~E.}\ \bibnamefont {Tai}}, \bibinfo {author}
  {\bibfnamefont {R.}~\bibnamefont {Ma}}, \bibinfo {author} {\bibfnamefont
  {J.}~\bibnamefont {Simon}}, \bibinfo {author} {\bibfnamefont {J.~I.}\
  \bibnamefont {Gillen}}, \bibinfo {author} {\bibfnamefont {S.}~\bibnamefont
  {F{\"o}lling}}, \bibinfo {author} {\bibfnamefont {L.}~\bibnamefont {Pollet}},
  \ and\ \bibinfo {author} {\bibfnamefont {M.}~\bibnamefont {Greiner}},\ }\href
  {\doibase 10.1126/science.1192368} {\bibfield  {journal} {\bibinfo  {journal}
  {Science}\ }\textbf {\bibinfo {volume} {329}},\ \bibinfo {pages} {547}
  (\bibinfo {year} {2010})}\BibitemShut {NoStop}%
\bibitem [{\citenamefont {Weitenberg}\ \emph {et~al.}(2011)\citenamefont
  {Weitenberg}, \citenamefont {Endres}, \citenamefont {Sherson}, \citenamefont
  {Cheneau}, \citenamefont {Schau{\ss}}, \citenamefont {Fukuhara},
  \citenamefont {Bloch},\ and\ \citenamefont {Kuhr}}]{Weitenberg-nature-2011}%
  \BibitemOpen
  \bibfield  {author} {\bibinfo {author} {\bibfnamefont {C.}~\bibnamefont
  {Weitenberg}}, \bibinfo {author} {\bibfnamefont {M.}~\bibnamefont {Endres}},
  \bibinfo {author} {\bibfnamefont {J.~F.}\ \bibnamefont {Sherson}}, \bibinfo
  {author} {\bibfnamefont {M.}~\bibnamefont {Cheneau}}, \bibinfo {author}
  {\bibfnamefont {P.}~\bibnamefont {Schau{\ss}}}, \bibinfo {author}
  {\bibfnamefont {T.}~\bibnamefont {Fukuhara}}, \bibinfo {author}
  {\bibfnamefont {I.}~\bibnamefont {Bloch}}, \ and\ \bibinfo {author}
  {\bibfnamefont {S.}~\bibnamefont {Kuhr}},\ }\href {\doibase
  10.1038/nature09827} {\bibfield  {journal} {\bibinfo  {journal} {Nature}\
  }\textbf {\bibinfo {volume} {471}},\ \bibinfo {pages} {319} (\bibinfo {year}
  {2011})}\BibitemShut {NoStop}%
\bibitem [{\citenamefont {Eisert}\ \emph {et~al.}(2015)\citenamefont {Eisert},
  \citenamefont {Friesdorf},\ and\ \citenamefont {Gogolin}}]{1408.5148}%
  \BibitemOpen
  \bibfield  {author} {\bibinfo {author} {\bibfnamefont {J.}~\bibnamefont
  {Eisert}}, \bibinfo {author} {\bibfnamefont {M.}~\bibnamefont {Friesdorf}}, \
  and\ \bibinfo {author} {\bibfnamefont {C.}~\bibnamefont {Gogolin}},\
  }\href@noop {} {\bibfield  {journal} {\bibinfo  {journal} {Nature Phys.}\
  }\textbf {\bibinfo {volume} {11}},\ \bibinfo {pages} {124} (\bibinfo {year}
  {2015})}\BibitemShut {NoStop}%
\bibitem [{\citenamefont {Schreiber}\ \emph {et~al.}(2015)\citenamefont
  {Schreiber}, \citenamefont {Hodgman}, \citenamefont {Bordia}, \citenamefont
  {L{\"u}schen}, \citenamefont {Fischer}, \citenamefont {Vosk}, \citenamefont
  {Altman}, \citenamefont {Schneider},\ and\ \citenamefont {Bloch}}]{BlochMBL}%
  \BibitemOpen
  \bibfield  {author} {\bibinfo {author} {\bibfnamefont {M.}~\bibnamefont
  {Schreiber}}, \bibinfo {author} {\bibfnamefont {S.~S.}\ \bibnamefont
  {Hodgman}}, \bibinfo {author} {\bibfnamefont {P.}~\bibnamefont {Bordia}},
  \bibinfo {author} {\bibfnamefont {H.~P.}\ \bibnamefont {L{\"u}schen}},
  \bibinfo {author} {\bibfnamefont {M.~H.}\ \bibnamefont {Fischer}}, \bibinfo
  {author} {\bibfnamefont {R.}~\bibnamefont {Vosk}}, \bibinfo {author}
  {\bibfnamefont {E.}~\bibnamefont {Altman}}, \bibinfo {author} {\bibfnamefont
  {U.}~\bibnamefont {Schneider}}, \ and\ \bibinfo {author} {\bibfnamefont
  {I.}~\bibnamefont {Bloch}},\ }\href@noop {} {\bibfield  {journal} {\bibinfo
  {journal} {Science}\ }\textbf {\bibinfo {volume} {349}},\ \bibinfo {pages}
  {842} (\bibinfo {year} {2015})}\BibitemShut {NoStop}%
\bibitem [{\citenamefont {Bergschneider}\ \emph {et~al.}(2019)\citenamefont
  {Bergschneider}, \citenamefont {Klinkhamer}, \citenamefont {Becher},
  \citenamefont {Klemt}, \citenamefont {Palm}, \citenamefont {Z{\"u}rn},
  \citenamefont {Jochim},\ and\ \citenamefont
  {Preiss}}]{bergschneider2019experimental}%
  \BibitemOpen
  \bibfield  {author} {\bibinfo {author} {\bibfnamefont {A.}~\bibnamefont
  {Bergschneider}}, \bibinfo {author} {\bibfnamefont {V.~M.}\ \bibnamefont
  {Klinkhamer}}, \bibinfo {author} {\bibfnamefont {J.~H.}\ \bibnamefont
  {Becher}}, \bibinfo {author} {\bibfnamefont {R.}~\bibnamefont {Klemt}},
  \bibinfo {author} {\bibfnamefont {L.}~\bibnamefont {Palm}}, \bibinfo {author}
  {\bibfnamefont {G.}~\bibnamefont {Z{\"u}rn}}, \bibinfo {author}
  {\bibfnamefont {S.}~\bibnamefont {Jochim}}, \ and\ \bibinfo {author}
  {\bibfnamefont {P.~M.}\ \bibnamefont {Preiss}},\ }\href@noop {} {\bibfield
  {journal} {\bibinfo  {journal} {Nature Physics}\ }\textbf {\bibinfo {volume}
  {15}},\ \bibinfo {pages} {640} (\bibinfo {year} {2019})}\BibitemShut
  {NoStop}%
\bibitem [{\citenamefont {Gluza}\ \emph {et~al.}(2020)\citenamefont {Gluza},
  \citenamefont {Schweigler}, \citenamefont {Rauer}, \citenamefont {Krumnow},
  \citenamefont {Schmiedmayer},\ and\ \citenamefont {Eisert}}]{QuantumReadout}%
  \BibitemOpen
  \bibfield  {author} {\bibinfo {author} {\bibfnamefont {M.}~\bibnamefont
  {Gluza}}, \bibinfo {author} {\bibfnamefont {T.}~\bibnamefont {Schweigler}},
  \bibinfo {author} {\bibfnamefont {B.}~\bibnamefont {Rauer}}, \bibinfo
  {author} {\bibfnamefont {C.}~\bibnamefont {Krumnow}}, \bibinfo {author}
  {\bibfnamefont {J.}~\bibnamefont {Schmiedmayer}}, \ and\ \bibinfo {author}
  {\bibfnamefont {J.}~\bibnamefont {Eisert}},\ }\href@noop {} {\bibfield
  {journal} {\bibinfo  {journal} {Comm. Phys.}\ }\textbf {\bibinfo {volume}
  {3}},\ \bibinfo {pages} {12} (\bibinfo {year} {2020})}\BibitemShut {NoStop}%
\bibitem [{\citenamefont {Ohliger}\ \emph {et~al.}(2013)\citenamefont
  {Ohliger}, \citenamefont {Nesme},\ and\ \citenamefont {Eisert}}]{Efficient}%
  \BibitemOpen
  \bibfield  {author} {\bibinfo {author} {\bibfnamefont {M.}~\bibnamefont
  {Ohliger}}, \bibinfo {author} {\bibfnamefont {V.}~\bibnamefont {Nesme}}, \
  and\ \bibinfo {author} {\bibfnamefont {J.}~\bibnamefont {Eisert}},\
  }\href@noop {} {\bibfield  {journal} {\bibinfo  {journal} {New J. Phys.}\
  }\textbf {\bibinfo {volume} {15}},\ \bibinfo {pages} {015024} (\bibinfo
  {year} {2013})}\BibitemShut {NoStop}%
\bibitem [{\citenamefont {Merkel}\ \emph {et~al.}(2010)\citenamefont {Merkel},
  \citenamefont {Riofr\'{\i}o}, \citenamefont {Flammia},\ and\ \citenamefont
  {Deutsch}}]{PhysRevA.81.032126}%
  \BibitemOpen
  \bibfield  {author} {\bibinfo {author} {\bibfnamefont {S.~T.}\ \bibnamefont
  {Merkel}}, \bibinfo {author} {\bibfnamefont {C.~A.}\ \bibnamefont
  {Riofr\'{\i}o}}, \bibinfo {author} {\bibfnamefont {S.~T.}\ \bibnamefont
  {Flammia}}, \ and\ \bibinfo {author} {\bibfnamefont {I.~H.}\ \bibnamefont
  {Deutsch}},\ }\href {\doibase 10.1103/PhysRevA.81.032126} {\bibfield
  {journal} {\bibinfo  {journal} {Phys. Rev. A}\ }\textbf {\bibinfo {volume}
  {81}},\ \bibinfo {pages} {032126} (\bibinfo {year} {2010})}\BibitemShut
  {NoStop}%
\bibitem [{\citenamefont {Elben}\ \emph {et~al.}(2018)\citenamefont {Elben},
  \citenamefont {Vermersch}, \citenamefont {Dalmonte}, \citenamefont {Cirac},\
  and\ \citenamefont {Zoller}}]{PhysRevLett.120.050406}%
  \BibitemOpen
  \bibfield  {author} {\bibinfo {author} {\bibfnamefont {A.}~\bibnamefont
  {Elben}}, \bibinfo {author} {\bibfnamefont {B.}~\bibnamefont {Vermersch}},
  \bibinfo {author} {\bibfnamefont {M.}~\bibnamefont {Dalmonte}}, \bibinfo
  {author} {\bibfnamefont {J.~I.}\ \bibnamefont {Cirac}}, \ and\ \bibinfo
  {author} {\bibfnamefont {P.}~\bibnamefont {Zoller}},\ }\href {\doibase
  10.1103/PhysRevLett.120.050406} {\bibfield  {journal} {\bibinfo  {journal}
  {Phys. Rev. Lett.}\ }\textbf {\bibinfo {volume} {120}},\ \bibinfo {pages}
  {050406} (\bibinfo {year} {2018})}\BibitemShut {NoStop}%
\bibitem [{\citenamefont {Hauke}\ \emph {et~al.}(2014)\citenamefont {Hauke},
  \citenamefont {Lewenstein},\ and\ \citenamefont
  {Eckardt}}]{PhysRevLett.113.045303}%
  \BibitemOpen
  \bibfield  {author} {\bibinfo {author} {\bibfnamefont {P.}~\bibnamefont
  {Hauke}}, \bibinfo {author} {\bibfnamefont {M.}~\bibnamefont {Lewenstein}}, \
  and\ \bibinfo {author} {\bibfnamefont {A.}~\bibnamefont {Eckardt}},\ }\href
  {\doibase 10.1103/PhysRevLett.113.045303} {\bibfield  {journal} {\bibinfo
  {journal} {Phys. Rev. Lett.}\ }\textbf {\bibinfo {volume} {113}},\ \bibinfo
  {pages} {045303} (\bibinfo {year} {2014})}\BibitemShut {NoStop}%
\bibitem [{\citenamefont {Pena~Ardila}\ \emph {et~al.}(2018)\citenamefont
  {Pena~Ardila}, \citenamefont {Heyl},\ and\ \citenamefont
  {Eckardt}}]{ardila2018}%
  \BibitemOpen
  \bibfield  {author} {\bibinfo {author} {\bibfnamefont {L.~A.}\ \bibnamefont
  {Pena~Ardila}}, \bibinfo {author} {\bibfnamefont {M.}~\bibnamefont {Heyl}}, \
  and\ \bibinfo {author} {\bibfnamefont {A.}~\bibnamefont {Eckardt}},\ }\href
  {\doibase 10.1103/PhysRevLett.121.260401} {\bibfield  {journal} {\bibinfo
  {journal} {Phys. Rev. Lett.}\ }\textbf {\bibinfo {volume} {121}},\ \bibinfo
  {pages} {260401} (\bibinfo {year} {2018})}\BibitemShut {NoStop}%
\bibitem [{\citenamefont {Qin}\ \emph {et~al.}(2018)\citenamefont {Qin},
  \citenamefont {Schnell}, \citenamefont {Sengstock}, \citenamefont
  {Weitenberg}, \citenamefont {Eckardt},\ and\ \citenamefont
  {Hofstetter}}]{qin2018charge}%
  \BibitemOpen
  \bibfield  {author} {\bibinfo {author} {\bibfnamefont {T.}~\bibnamefont
  {Qin}}, \bibinfo {author} {\bibfnamefont {A.}~\bibnamefont {Schnell}},
  \bibinfo {author} {\bibfnamefont {K.}~\bibnamefont {Sengstock}}, \bibinfo
  {author} {\bibfnamefont {C.}~\bibnamefont {Weitenberg}}, \bibinfo {author}
  {\bibfnamefont {A.}~\bibnamefont {Eckardt}}, \ and\ \bibinfo {author}
  {\bibfnamefont {W.}~\bibnamefont {Hofstetter}},\ }\href@noop {} {\bibfield
  {journal} {\bibinfo  {journal} {arXiv preprint arXiv:1804.03200}\ } (\bibinfo
  {year} {2018})}\BibitemShut {NoStop}%
\bibitem [{\citenamefont {Tarnowski}\ \emph {et~al.}(2017)\citenamefont
  {Tarnowski}, \citenamefont {{\"U}nal}, \citenamefont {Fl{\"a}schner},
  \citenamefont {Rem}, \citenamefont {Eckardt}, \citenamefont {Sengstock},\
  and\ \citenamefont {Weitenberg}}]{tarnowski2017characterizing}%
  \BibitemOpen
  \bibfield  {author} {\bibinfo {author} {\bibfnamefont {M.}~\bibnamefont
  {Tarnowski}}, \bibinfo {author} {\bibfnamefont {F.~N.}\ \bibnamefont
  {{\"U}nal}}, \bibinfo {author} {\bibfnamefont {N.}~\bibnamefont
  {Fl{\"a}schner}}, \bibinfo {author} {\bibfnamefont {B.~S.}\ \bibnamefont
  {Rem}}, \bibinfo {author} {\bibfnamefont {A.}~\bibnamefont {Eckardt}},
  \bibinfo {author} {\bibfnamefont {K.}~\bibnamefont {Sengstock}}, \ and\
  \bibinfo {author} {\bibfnamefont {C.}~\bibnamefont {Weitenberg}},\
  }\href@noop {} {\bibfield  {journal} {\bibinfo  {journal} {arXiv:1709.01046}\
  } (\bibinfo {year} {2017})}\BibitemShut {NoStop}%
\bibitem [{\citenamefont {Ke\ss{}ler}\ and\ \citenamefont
  {Marquardt}(2014)}]{PhysRevA.89.061601}%
  \BibitemOpen
  \bibfield  {author} {\bibinfo {author} {\bibfnamefont {S.}~\bibnamefont
  {Ke\ss{}ler}}\ and\ \bibinfo {author} {\bibfnamefont {F.}~\bibnamefont
  {Marquardt}},\ }\href {\doibase 10.1103/PhysRevA.89.061601} {\bibfield
  {journal} {\bibinfo  {journal} {Phys. Rev. A}\ }\textbf {\bibinfo {volume}
  {89}},\ \bibinfo {pages} {061601} (\bibinfo {year} {2014})}\BibitemShut
  {NoStop}%
\bibitem [{\citenamefont {Loida}\ \emph {et~al.}(2018)\citenamefont {Loida},
  \citenamefont {Bernier}, \citenamefont {Citro}, \citenamefont {Orignac},\
  and\ \citenamefont {Kollath}}]{PhysRevA.98.033605Kollath}%
  \BibitemOpen
  \bibfield  {author} {\bibinfo {author} {\bibfnamefont {K.}~\bibnamefont
  {Loida}}, \bibinfo {author} {\bibfnamefont {J.-S.}\ \bibnamefont {Bernier}},
  \bibinfo {author} {\bibfnamefont {R.}~\bibnamefont {Citro}}, \bibinfo
  {author} {\bibfnamefont {E.}~\bibnamefont {Orignac}}, \ and\ \bibinfo
  {author} {\bibfnamefont {C.}~\bibnamefont {Kollath}},\ }\href {\doibase
  10.1103/PhysRevA.98.033605} {\bibfield  {journal} {\bibinfo  {journal} {Phys.
  Rev. A}\ }\textbf {\bibinfo {volume} {98}},\ \bibinfo {pages} {033605}
  (\bibinfo {year} {2018})}\BibitemShut {NoStop}%
\bibitem [{\citenamefont {Atala}\ \emph {et~al.}(2014)\citenamefont {Atala},
  \citenamefont {Aidelsburger}, \citenamefont {Lohse}, \citenamefont
  {Barreiro}, \citenamefont {Paredes},\ and\ \citenamefont
  {Bloch}}]{atala2014observation}%
  \BibitemOpen
  \bibfield  {author} {\bibinfo {author} {\bibfnamefont {M.}~\bibnamefont
  {Atala}}, \bibinfo {author} {\bibfnamefont {M.}~\bibnamefont {Aidelsburger}},
  \bibinfo {author} {\bibfnamefont {M.}~\bibnamefont {Lohse}}, \bibinfo
  {author} {\bibfnamefont {J.~T.}\ \bibnamefont {Barreiro}}, \bibinfo {author}
  {\bibfnamefont {B.}~\bibnamefont {Paredes}}, \ and\ \bibinfo {author}
  {\bibfnamefont {I.}~\bibnamefont {Bloch}},\ }\href@noop {} {\bibfield
  {journal} {\bibinfo  {journal} {Nature Physics}\ }\textbf {\bibinfo {volume}
  {10}},\ \bibinfo {pages} {588} (\bibinfo {year} {2014})}\BibitemShut
  {NoStop}%
\bibitem [{\citenamefont {Schweizer}\ \emph {et~al.}(2016)\citenamefont
  {Schweizer}, \citenamefont {Lohse}, \citenamefont {Citro},\ and\
  \citenamefont {Bloch}}]{PhysRevLett.117.170405}%
  \BibitemOpen
  \bibfield  {author} {\bibinfo {author} {\bibfnamefont {C.}~\bibnamefont
  {Schweizer}}, \bibinfo {author} {\bibfnamefont {M.}~\bibnamefont {Lohse}},
  \bibinfo {author} {\bibfnamefont {R.}~\bibnamefont {Citro}}, \ and\ \bibinfo
  {author} {\bibfnamefont {I.}~\bibnamefont {Bloch}},\ }\href {\doibase
  10.1103/PhysRevLett.117.170405} {\bibfield  {journal} {\bibinfo  {journal}
  {Phys. Rev. Lett.}\ }\textbf {\bibinfo {volume} {117}},\ \bibinfo {pages}
  {170405} (\bibinfo {year} {2016})}\BibitemShut {NoStop}%
\bibitem [{\citenamefont {Huang}\ \emph {et~al.}(2020)\citenamefont {Huang},
  \citenamefont {Kueng},\ and\ \citenamefont {Preskill}}]{Shadows}%
  \BibitemOpen
  \bibfield  {author} {\bibinfo {author} {\bibfnamefont {H.-Y.}\ \bibnamefont
  {Huang}}, \bibinfo {author} {\bibfnamefont {R.}~\bibnamefont {Kueng}}, \ and\
  \bibinfo {author} {\bibfnamefont {J.}~\bibnamefont {Preskill}},\ }\href@noop
  {} {\bibfield  {journal} {\bibinfo  {journal} {Nature Phys.}\ } (\bibinfo
  {year} {2020})}\BibitemShut {NoStop}%
\bibitem [{\citenamefont {Diamond}\ and\ \citenamefont {Boyd}(2016)}]{cvxpy}%
  \BibitemOpen
  \bibfield  {author} {\bibinfo {author} {\bibfnamefont {S.}~\bibnamefont
  {Diamond}}\ and\ \bibinfo {author} {\bibfnamefont {S.}~\bibnamefont {Boyd}},\
  }\href {http://stanford.edu/~boyd/papers/pdf/cvxpy_paper.pdf} {\bibfield
  {journal} {\bibinfo  {journal} {J. Mach. Learn. Res.}\ }\textbf {\bibinfo
  {volume} {17}},\ \bibinfo {pages} {1} (\bibinfo {year} {2016})}\BibitemShut
  {NoStop}%
\bibitem [{\citenamefont {Gluza}()}]{github_tomography}%
  \BibitemOpen
  \bibfield  {author} {\bibinfo {author} {\bibfnamefont {M.}~\bibnamefont
  {Gluza}},\ }\href@noop {} {}\bibinfo {note} {The numerical code is freely
  available at {\url{https://github.com/marekgluza/hopping_tomography}} and
  includes interactive Python notebooks allowing to reproduce the
  figures.}\BibitemShut {Stop}%
\bibitem [{\citenamefont {Preiss}\ \emph {et~al.}(2015)\citenamefont {Preiss},
  \citenamefont {Ma}, \citenamefont {Tai}, \citenamefont {Lukin}, \citenamefont
  {Rispoli}, \citenamefont {Zupancic}, \citenamefont {Lahini}, \citenamefont
  {Islam},\ and\ \citenamefont {Greiner}}]{Preiss1229}%
  \BibitemOpen
  \bibfield  {author} {\bibinfo {author} {\bibfnamefont {P.~M.}\ \bibnamefont
  {Preiss}}, \bibinfo {author} {\bibfnamefont {R.}~\bibnamefont {Ma}}, \bibinfo
  {author} {\bibfnamefont {M.~E.}\ \bibnamefont {Tai}}, \bibinfo {author}
  {\bibfnamefont {A.}~\bibnamefont {Lukin}}, \bibinfo {author} {\bibfnamefont
  {M.}~\bibnamefont {Rispoli}}, \bibinfo {author} {\bibfnamefont
  {P.}~\bibnamefont {Zupancic}}, \bibinfo {author} {\bibfnamefont
  {Y.}~\bibnamefont {Lahini}}, \bibinfo {author} {\bibfnamefont
  {R.}~\bibnamefont {Islam}}, \ and\ \bibinfo {author} {\bibfnamefont
  {M.}~\bibnamefont {Greiner}},\ }\href {\doibase 10.1126/science.1260364}
  {\bibfield  {journal} {\bibinfo  {journal} {Science}\ }\textbf {\bibinfo
  {volume} {347}},\ \bibinfo {pages} {1229} (\bibinfo {year}
  {2015})}\BibitemShut {NoStop}%
\bibitem [{\citenamefont {Rauer}\ \emph {et~al.}(2018)\citenamefont {Rauer},
  \citenamefont {Erne}, \citenamefont {Schweigler}, \citenamefont {Cataldini},
  \citenamefont {Tajik},\ and\ \citenamefont {Schmiedmayer}}]{Recurrence}%
  \BibitemOpen
  \bibfield  {author} {\bibinfo {author} {\bibfnamefont {B.}~\bibnamefont
  {Rauer}}, \bibinfo {author} {\bibfnamefont {S.}~\bibnamefont {Erne}},
  \bibinfo {author} {\bibfnamefont {T.}~\bibnamefont {Schweigler}}, \bibinfo
  {author} {\bibfnamefont {F.}~\bibnamefont {Cataldini}}, \bibinfo {author}
  {\bibfnamefont {M.}~\bibnamefont {Tajik}}, \ and\ \bibinfo {author}
  {\bibfnamefont {J.}~\bibnamefont {Schmiedmayer}},\ }\href@noop {} {\bibfield
  {journal} {\bibinfo  {journal} {Science}\ }\textbf {\bibinfo {volume}
  {360}},\ \bibinfo {pages} {307} (\bibinfo {year} {2018})}\BibitemShut
  {NoStop}%
\bibitem [{\citenamefont {Lieb}\ and\ \citenamefont {Robinson}(1972)}]{LR}%
  \BibitemOpen
  \bibfield  {author} {\bibinfo {author} {\bibfnamefont {E.~H.}\ \bibnamefont
  {Lieb}}\ and\ \bibinfo {author} {\bibfnamefont {D.~W.}\ \bibnamefont
  {Robinson}},\ }\href@noop {} {\bibfield  {journal} {\bibinfo  {journal}
  {Commun. Math. Phys.}\ }\textbf {\bibinfo {volume} {28}},\ \bibinfo {pages}
  {251} (\bibinfo {year} {1972})}\BibitemShut {NoStop}%
\bibitem [{\citenamefont {Hartke}\ \emph {et~al.}(2020)\citenamefont {Hartke},
  \citenamefont {Oreg}, \citenamefont {Jia},\ and\ \citenamefont
  {Zwierlein}}]{PhysRevLett.125.113601}%
  \BibitemOpen
  \bibfield  {author} {\bibinfo {author} {\bibfnamefont {T.}~\bibnamefont
  {Hartke}}, \bibinfo {author} {\bibfnamefont {B.}~\bibnamefont {Oreg}},
  \bibinfo {author} {\bibfnamefont {N.}~\bibnamefont {Jia}}, \ and\ \bibinfo
  {author} {\bibfnamefont {M.}~\bibnamefont {Zwierlein}},\ }\href {\doibase
  10.1103/PhysRevLett.125.113601} {\bibfield  {journal} {\bibinfo  {journal}
  {Phys. Rev. Lett.}\ }\textbf {\bibinfo {volume} {125}},\ \bibinfo {pages}
  {113601} (\bibinfo {year} {2020})}\BibitemShut {NoStop}%
\bibitem [{\citenamefont {Nichols}\ \emph {et~al.}(2019)\citenamefont
  {Nichols}, \citenamefont {Cheuk}, \citenamefont {Okan}, \citenamefont
  {Hartke}, \citenamefont {Mendez}, \citenamefont {Senthil}, \citenamefont
  {Khatami}, \citenamefont {Zhang},\ and\ \citenamefont
  {Zwierlein}}]{Nichols383}%
  \BibitemOpen
  \bibfield  {author} {\bibinfo {author} {\bibfnamefont {M.~A.}\ \bibnamefont
  {Nichols}}, \bibinfo {author} {\bibfnamefont {L.~W.}\ \bibnamefont {Cheuk}},
  \bibinfo {author} {\bibfnamefont {M.}~\bibnamefont {Okan}}, \bibinfo {author}
  {\bibfnamefont {T.~R.}\ \bibnamefont {Hartke}}, \bibinfo {author}
  {\bibfnamefont {E.}~\bibnamefont {Mendez}}, \bibinfo {author} {\bibfnamefont
  {T.}~\bibnamefont {Senthil}}, \bibinfo {author} {\bibfnamefont
  {E.}~\bibnamefont {Khatami}}, \bibinfo {author} {\bibfnamefont
  {H.}~\bibnamefont {Zhang}}, \ and\ \bibinfo {author} {\bibfnamefont {M.~W.}\
  \bibnamefont {Zwierlein}},\ }\href {\doibase 10.1126/science.aat4387}
  {\bibfield  {journal} {\bibinfo  {journal} {Science}\ }\textbf {\bibinfo
  {volume} {363}},\ \bibinfo {pages} {383} (\bibinfo {year}
  {2019})}\BibitemShut {NoStop}%
\bibitem [{\citenamefont {Brown}\ \emph {et~al.}(2019)\citenamefont {Brown},
  \citenamefont {Mitra}, \citenamefont {Guardado-Sanchez}, \citenamefont
  {Nourafkan}, \citenamefont {Reymbaut}, \citenamefont {H{\'e}bert},
  \citenamefont {Bergeron}, \citenamefont {Tremblay}, \citenamefont {Kokalj},
  \citenamefont {Huse}, \citenamefont {Schau{\ss}},\ and\ \citenamefont
  {Bakr}}]{Brown379}%
  \BibitemOpen
  \bibfield  {author} {\bibinfo {author} {\bibfnamefont {P.~T.}\ \bibnamefont
  {Brown}}, \bibinfo {author} {\bibfnamefont {D.}~\bibnamefont {Mitra}},
  \bibinfo {author} {\bibfnamefont {E.}~\bibnamefont {Guardado-Sanchez}},
  \bibinfo {author} {\bibfnamefont {R.}~\bibnamefont {Nourafkan}}, \bibinfo
  {author} {\bibfnamefont {A.}~\bibnamefont {Reymbaut}}, \bibinfo {author}
  {\bibfnamefont {C.-D.}\ \bibnamefont {H{\'e}bert}}, \bibinfo {author}
  {\bibfnamefont {S.}~\bibnamefont {Bergeron}}, \bibinfo {author}
  {\bibfnamefont {A.-M.~S.}\ \bibnamefont {Tremblay}}, \bibinfo {author}
  {\bibfnamefont {J.}~\bibnamefont {Kokalj}}, \bibinfo {author} {\bibfnamefont
  {D.~A.}\ \bibnamefont {Huse}}, \bibinfo {author} {\bibfnamefont
  {P.}~\bibnamefont {Schau{\ss}}}, \ and\ \bibinfo {author} {\bibfnamefont
  {W.~S.}\ \bibnamefont {Bakr}},\ }\href {\doibase 10.1126/science.aat4134}
  {\bibfield  {journal} {\bibinfo  {journal} {Science}\ }\textbf {\bibinfo
  {volume} {363}},\ \bibinfo {pages} {379} (\bibinfo {year}
  {2019})}\BibitemShut {NoStop}%
\bibitem [{\citenamefont {Takasu}\ \emph {et~al.}(2020)\citenamefont {Takasu},
  \citenamefont {Yagami}, \citenamefont {Asaka}, \citenamefont {Fukushima},
  \citenamefont {Nagao}, \citenamefont {Goto}, \citenamefont {Danshita},\ and\
  \citenamefont {Takahashi}}]{Takasueaba9255}%
  \BibitemOpen
  \bibfield  {author} {\bibinfo {author} {\bibfnamefont {Y.}~\bibnamefont
  {Takasu}}, \bibinfo {author} {\bibfnamefont {T.}~\bibnamefont {Yagami}},
  \bibinfo {author} {\bibfnamefont {H.}~\bibnamefont {Asaka}}, \bibinfo
  {author} {\bibfnamefont {Y.}~\bibnamefont {Fukushima}}, \bibinfo {author}
  {\bibfnamefont {K.}~\bibnamefont {Nagao}}, \bibinfo {author} {\bibfnamefont
  {S.}~\bibnamefont {Goto}}, \bibinfo {author} {\bibfnamefont {I.}~\bibnamefont
  {Danshita}}, \ and\ \bibinfo {author} {\bibfnamefont {Y.}~\bibnamefont
  {Takahashi}},\ }\href {\doibase 10.1126/sciadv.aba9255} {\bibfield  {journal}
  {\bibinfo  {journal} {Science Advances}\ }\textbf {\bibinfo {volume} {6}}
  (\bibinfo {year} {2020}),\ 10.1126/sciadv.aba9255}\BibitemShut {NoStop}%
\bibitem [{\citenamefont {Nakamura}\ \emph {et~al.}(2019)\citenamefont
  {Nakamura}, \citenamefont {Takasu}, \citenamefont {Kobayashi}, \citenamefont
  {Asaka}, \citenamefont {Fukushima}, \citenamefont {Inaba}, \citenamefont
  {Yamashita},\ and\ \citenamefont {Takahashi}}]{PhysRevA.99.033609}%
  \BibitemOpen
  \bibfield  {author} {\bibinfo {author} {\bibfnamefont {Y.}~\bibnamefont
  {Nakamura}}, \bibinfo {author} {\bibfnamefont {Y.}~\bibnamefont {Takasu}},
  \bibinfo {author} {\bibfnamefont {J.}~\bibnamefont {Kobayashi}}, \bibinfo
  {author} {\bibfnamefont {H.}~\bibnamefont {Asaka}}, \bibinfo {author}
  {\bibfnamefont {Y.}~\bibnamefont {Fukushima}}, \bibinfo {author}
  {\bibfnamefont {K.}~\bibnamefont {Inaba}}, \bibinfo {author} {\bibfnamefont
  {M.}~\bibnamefont {Yamashita}}, \ and\ \bibinfo {author} {\bibfnamefont
  {Y.}~\bibnamefont {Takahashi}},\ }\href {\doibase 10.1103/PhysRevA.99.033609}
  {\bibfield  {journal} {\bibinfo  {journal} {Phys. Rev. A}\ }\textbf {\bibinfo
  {volume} {99}},\ \bibinfo {pages} {033609} (\bibinfo {year}
  {2019})}\BibitemShut {NoStop}%
\bibitem [{\citenamefont {Cocchi}\ \emph {et~al.}(2017)\citenamefont {Cocchi},
  \citenamefont {Miller}, \citenamefont {Drewes}, \citenamefont {Chan},
  \citenamefont {Pertot}, \citenamefont {Brennecke},\ and\ \citenamefont
  {K\"ohl}}]{PhysRevX.7.031025}%
  \BibitemOpen
  \bibfield  {author} {\bibinfo {author} {\bibfnamefont {E.}~\bibnamefont
  {Cocchi}}, \bibinfo {author} {\bibfnamefont {L.~A.}\ \bibnamefont {Miller}},
  \bibinfo {author} {\bibfnamefont {J.~H.}\ \bibnamefont {Drewes}}, \bibinfo
  {author} {\bibfnamefont {C.~F.}\ \bibnamefont {Chan}}, \bibinfo {author}
  {\bibfnamefont {D.}~\bibnamefont {Pertot}}, \bibinfo {author} {\bibfnamefont
  {F.}~\bibnamefont {Brennecke}}, \ and\ \bibinfo {author} {\bibfnamefont
  {M.}~\bibnamefont {K\"ohl}},\ }\href {\doibase 10.1103/PhysRevX.7.031025}
  {\bibfield  {journal} {\bibinfo  {journal} {Phys. Rev. X}\ }\textbf {\bibinfo
  {volume} {7}},\ \bibinfo {pages} {031025} (\bibinfo {year}
  {2017})}\BibitemShut {NoStop}%
\bibitem [{\citenamefont {Chan}\ \emph {et~al.}(2020)\citenamefont {Chan},
  \citenamefont {Gall}, \citenamefont {Wurz},\ and\ \citenamefont
  {K\"ohl}}]{PhysRevResearch.2.023210}%
  \BibitemOpen
  \bibfield  {author} {\bibinfo {author} {\bibfnamefont {C.~F.}\ \bibnamefont
  {Chan}}, \bibinfo {author} {\bibfnamefont {M.}~\bibnamefont {Gall}}, \bibinfo
  {author} {\bibfnamefont {N.}~\bibnamefont {Wurz}}, \ and\ \bibinfo {author}
  {\bibfnamefont {M.}~\bibnamefont {K\"ohl}},\ }\href {\doibase
  10.1103/PhysRevResearch.2.023210} {\bibfield  {journal} {\bibinfo  {journal}
  {Phys. Rev. Research}\ }\textbf {\bibinfo {volume} {2}},\ \bibinfo {pages}
  {023210} (\bibinfo {year} {2020})}\BibitemShut {NoStop}%
\bibitem [{\citenamefont {Guardado-Sanchez}\ \emph {et~al.}(2020)\citenamefont
  {Guardado-Sanchez}, \citenamefont {Morningstar}, \citenamefont {Spar},
  \citenamefont {Brown}, \citenamefont {Huse},\ and\ \citenamefont
  {Bakr}}]{PhysRevX.10.011042}%
  \BibitemOpen
  \bibfield  {author} {\bibinfo {author} {\bibfnamefont {E.}~\bibnamefont
  {Guardado-Sanchez}}, \bibinfo {author} {\bibfnamefont {A.}~\bibnamefont
  {Morningstar}}, \bibinfo {author} {\bibfnamefont {B.~M.}\ \bibnamefont
  {Spar}}, \bibinfo {author} {\bibfnamefont {P.~T.}\ \bibnamefont {Brown}},
  \bibinfo {author} {\bibfnamefont {D.~A.}\ \bibnamefont {Huse}}, \ and\
  \bibinfo {author} {\bibfnamefont {W.~S.}\ \bibnamefont {Bakr}},\ }\href
  {\doibase 10.1103/PhysRevX.10.011042} {\bibfield  {journal} {\bibinfo
  {journal} {Phys. Rev. X}\ }\textbf {\bibinfo {volume} {10}},\ \bibinfo
  {pages} {011042} (\bibinfo {year} {2020})}\BibitemShut {NoStop}%
\bibitem [{\citenamefont {Kokail}\ \emph {et~al.}(2019)\citenamefont {Kokail},
  \citenamefont {Maier}, \citenamefont {van Bijnen}, \citenamefont {Brydges},
  \citenamefont {Joshi}, \citenamefont {Jurcevic}, \citenamefont {Muschik},
  \citenamefont {Silvi}, \citenamefont {Blatt}, \citenamefont {Roos} \emph
  {et~al.}}]{kokail2019self}%
  \BibitemOpen
  \bibfield  {author} {\bibinfo {author} {\bibfnamefont {C.}~\bibnamefont
  {Kokail}}, \bibinfo {author} {\bibfnamefont {C.}~\bibnamefont {Maier}},
  \bibinfo {author} {\bibfnamefont {R.}~\bibnamefont {van Bijnen}}, \bibinfo
  {author} {\bibfnamefont {T.}~\bibnamefont {Brydges}}, \bibinfo {author}
  {\bibfnamefont {M.~K.}\ \bibnamefont {Joshi}}, \bibinfo {author}
  {\bibfnamefont {P.}~\bibnamefont {Jurcevic}}, \bibinfo {author}
  {\bibfnamefont {C.~A.}\ \bibnamefont {Muschik}}, \bibinfo {author}
  {\bibfnamefont {P.}~\bibnamefont {Silvi}}, \bibinfo {author} {\bibfnamefont
  {R.}~\bibnamefont {Blatt}}, \bibinfo {author} {\bibfnamefont {C.~F.}\
  \bibnamefont {Roos}},  \emph {et~al.},\ }\href@noop {} {\bibfield  {journal}
  {\bibinfo  {journal} {Nature}\ }\textbf {\bibinfo {volume} {569}},\ \bibinfo
  {pages} {355} (\bibinfo {year} {2019})}\BibitemShut {NoStop}%
\bibitem [{\citenamefont {Bennett}\ \emph {et~al.}(1996)\citenamefont
  {Bennett}, \citenamefont {Bernstein}, \citenamefont {Popescu},\ and\
  \citenamefont {Schumacher}}]{PureBipartiteBennett}%
  \BibitemOpen
  \bibfield  {author} {\bibinfo {author} {\bibfnamefont {C.~H.}\ \bibnamefont
  {Bennett}}, \bibinfo {author} {\bibfnamefont {H.~J.}\ \bibnamefont
  {Bernstein}}, \bibinfo {author} {\bibfnamefont {S.}~\bibnamefont {Popescu}},
  \ and\ \bibinfo {author} {\bibfnamefont {B.}~\bibnamefont {Schumacher}},\
  }\href@noop {} {\bibfield  {journal} {\bibinfo  {journal} {Phys. Rev. A}\
  }\textbf {\bibinfo {volume} {53}},\ \bibinfo {pages} {2046} (\bibinfo {year}
  {1996})}\BibitemShut {NoStop}%
\bibitem [{\citenamefont {Horodecki}\ \emph {et~al.}(2009)\citenamefont
  {Horodecki}, \citenamefont {Horodecki}, \citenamefont {Horodecki},\ and\
  \citenamefont {Horodecki}}]{Horodecki}%
  \BibitemOpen
  \bibfield  {author} {\bibinfo {author} {\bibfnamefont {R.}~\bibnamefont
  {Horodecki}}, \bibinfo {author} {\bibfnamefont {P.}~\bibnamefont
  {Horodecki}}, \bibinfo {author} {\bibfnamefont {M.}~\bibnamefont
  {Horodecki}}, \ and\ \bibinfo {author} {\bibfnamefont {K.}~\bibnamefont
  {Horodecki}},\ }\href@noop {} {\bibfield  {journal} {\bibinfo  {journal}
  {Rev. Mod. Phys.}\ }\textbf {\bibinfo {volume} {81}},\ \bibinfo {pages} {865}
  (\bibinfo {year} {2009})}\BibitemShut {NoStop}%
\bibitem [{\citenamefont {Eisert}\ \emph {et~al.}(2007)\citenamefont {Eisert},
  \citenamefont {Brandao},\ and\ \citenamefont {Audenaert}}]{quant-ph/0607167}%
  \BibitemOpen
  \bibfield  {author} {\bibinfo {author} {\bibfnamefont {J.}~\bibnamefont
  {Eisert}}, \bibinfo {author} {\bibfnamefont {F.~G.}\ \bibnamefont {Brandao}},
  \ and\ \bibinfo {author} {\bibfnamefont {K.~M.}\ \bibnamefont {Audenaert}},\
  }\href@noop {} {\bibfield  {journal} {\bibinfo  {journal} {New J. Phys.}\
  }\textbf {\bibinfo {volume} {9}} (\bibinfo {year} {2007})}\BibitemShut
  {NoStop}%
\bibitem [{\citenamefont {Audenaert}\ and\ \citenamefont
  {Plenio}(2006)}]{Audenaert06}%
  \BibitemOpen
  \bibfield  {author} {\bibinfo {author} {\bibfnamefont {K.~M.~R.}\
  \bibnamefont {Audenaert}}\ and\ \bibinfo {author} {\bibfnamefont {M.~B.}\
  \bibnamefont {Plenio}},\ }\href@noop {} {\bibfield  {journal} {\bibinfo
  {journal} {New J. Phys.}\ }\textbf {\bibinfo {volume} {8}},\ \bibinfo {pages}
  {266} (\bibinfo {year} {2006})}\BibitemShut {NoStop}%
\bibitem [{\citenamefont {Guehne}\ \emph {et~al.}(2007)\citenamefont {Guehne},
  \citenamefont {Reimpell},\ and\ \citenamefont {Werner}}]{Guehne}%
  \BibitemOpen
  \bibfield  {author} {\bibinfo {author} {\bibfnamefont {O.}~\bibnamefont
  {Guehne}}, \bibinfo {author} {\bibfnamefont {M.}~\bibnamefont {Reimpell}}, \
  and\ \bibinfo {author} {\bibfnamefont {R.~F.}\ \bibnamefont {Werner}},\
  }\href@noop {} {\bibfield  {journal} {\bibinfo  {journal} {Phys. Rev. Lett.}\
  }\textbf {\bibinfo {volume} {98}},\ \bibinfo {pages} {110502} (\bibinfo
  {year} {2007})}\BibitemShut {NoStop}%
\bibitem [{\citenamefont {Cramer}\ \emph {et~al.}(2013)\citenamefont {Cramer},
  \citenamefont {Bernard}, \citenamefont {Fabbri}, \citenamefont {Fallani},
  \citenamefont {Fort}, \citenamefont {Rosi}, \citenamefont {Caruso},
  \citenamefont {Inguscio},\ and\ \citenamefont {Plenio}}]{1302.4897}%
  \BibitemOpen
  \bibfield  {author} {\bibinfo {author} {\bibfnamefont {M.}~\bibnamefont
  {Cramer}}, \bibinfo {author} {\bibfnamefont {A.}~\bibnamefont {Bernard}},
  \bibinfo {author} {\bibfnamefont {N.}~\bibnamefont {Fabbri}}, \bibinfo
  {author} {\bibfnamefont {L.}~\bibnamefont {Fallani}}, \bibinfo {author}
  {\bibfnamefont {C.}~\bibnamefont {Fort}}, \bibinfo {author} {\bibfnamefont
  {S.}~\bibnamefont {Rosi}}, \bibinfo {author} {\bibfnamefont {F.}~\bibnamefont
  {Caruso}}, \bibinfo {author} {\bibfnamefont {M.}~\bibnamefont {Inguscio}}, \
  and\ \bibinfo {author} {\bibfnamefont {M.}~\bibnamefont {Plenio}},\
  }\href@noop {} {\bibfield  {journal} {\bibinfo  {journal} {Nature Comm.}\
  }\textbf {\bibinfo {volume} {4}},\ \bibinfo {pages} {3161} (\bibinfo {year}
  {2013})}\BibitemShut {NoStop}%
\bibitem [{\citenamefont {Earman}(2008)}]{Earman}%
  \BibitemOpen
  \bibfield  {author} {\bibinfo {author} {\bibfnamefont {J.}~\bibnamefont
  {Earman}},\ }\href {\doibase 10.1007/s10670-008-9124-z} {\bibfield  {journal}
  {\bibinfo  {journal} {Erkenntnis}\ }\textbf {\bibinfo {volume} {69}},\
  \bibinfo {pages} {377} (\bibinfo {year} {2008})}\BibitemShut {NoStop}%
\bibitem [{\citenamefont {Wick}\ \emph {et~al.}(1952)\citenamefont {Wick},
  \citenamefont {Wightman},\ and\ \citenamefont {Wigner}}]{WignerWightmanWick}%
  \BibitemOpen
  \bibfield  {author} {\bibinfo {author} {\bibfnamefont {G.~C.}\ \bibnamefont
  {Wick}}, \bibinfo {author} {\bibfnamefont {A.~S.}\ \bibnamefont {Wightman}},
  \ and\ \bibinfo {author} {\bibfnamefont {E.~P.}\ \bibnamefont {Wigner}},\
  }\href {\doibase 10.1103/PhysRev.88.101} {\bibfield  {journal} {\bibinfo
  {journal} {Phys. Rev.}\ }\textbf {\bibinfo {volume} {88}},\ \bibinfo {pages}
  {101} (\bibinfo {year} {1952})}\BibitemShut {NoStop}%
\bibitem [{\citenamefont {Schuch}\ \emph
  {et~al.}(2004{\natexlab{a}})\citenamefont {Schuch}, \citenamefont
  {Verstraete},\ and\ \citenamefont {Cirac}}]{quant-ph/0404079}%
  \BibitemOpen
  \bibfield  {author} {\bibinfo {author} {\bibfnamefont {N.}~\bibnamefont
  {Schuch}}, \bibinfo {author} {\bibfnamefont {F.}~\bibnamefont {Verstraete}},
  \ and\ \bibinfo {author} {\bibfnamefont {J.~I.}\ \bibnamefont {Cirac}},\
  }\href@noop {} {\bibfield  {journal} {\bibinfo  {journal} {Phys. Rev. A}\
  }\textbf {\bibinfo {volume} {70}},\ \bibinfo {pages} {042310} (\bibinfo
  {year} {2004}{\natexlab{a}})}\BibitemShut {NoStop}%
\bibitem [{\citenamefont {Devetak}\ and\ \citenamefont
  {Winter}(2005)}]{HashingBound}%
  \BibitemOpen
  \bibfield  {author} {\bibinfo {author} {\bibfnamefont {I.}~\bibnamefont
  {Devetak}}\ and\ \bibinfo {author} {\bibfnamefont {A.}~\bibnamefont
  {Winter}},\ }\href@noop {} {\bibfield  {journal} {\bibinfo  {journal} {Proc.
  R. Soc. Lon and A}\ }\textbf {\bibinfo {volume} {461}},\ \bibinfo {pages}
  {207} (\bibinfo {year} {2005})}\BibitemShut {NoStop}%
\bibitem [{\citenamefont {Pastawski}\ \emph {et~al.}(2017)\citenamefont
  {Pastawski}, \citenamefont {Eisert},\ and\ \citenamefont
  {Wilming}}]{Pastawski2016}%
  \BibitemOpen
  \bibfield  {author} {\bibinfo {author} {\bibfnamefont {F.}~\bibnamefont
  {Pastawski}}, \bibinfo {author} {\bibfnamefont {J.}~\bibnamefont {Eisert}}, \
  and\ \bibinfo {author} {\bibfnamefont {H.}~\bibnamefont {Wilming}},\ }\href
  {\doibase 10.1103/PhysRevLett.119.020501} {\bibfield  {journal} {\bibinfo
  {journal} {Phys. Rev. Lett.}\ }\textbf {\bibinfo {volume} {119}},\ \bibinfo
  {pages} {020501} (\bibinfo {year} {2017})}\BibitemShut {NoStop}%
\bibitem [{\citenamefont {Eisert}\ and\ \citenamefont
  {Wolf}(2007)}]{GaussianChannel}%
  \BibitemOpen
  \bibfield  {author} {\bibinfo {author} {\bibfnamefont {J.}~\bibnamefont
  {Eisert}}\ and\ \bibinfo {author} {\bibfnamefont {M.~W.}\ \bibnamefont
  {Wolf}},\ }\enquote {\bibinfo {title} {{Gaussian quantum channels}},}\ in\
  \href@noop {} {\emph {\bibinfo {booktitle} {Quantum Information with
  Continous Variables of Atoms and Light}}}\ (\bibinfo  {publisher} {Imperial
  College Press},\ \bibinfo {address} {London},\ \bibinfo {year} {2007})\ pp.\
  \bibinfo {pages} {23--42},\ \bibinfo {note}
  {arXiv:quant-ph/0505151}\BibitemShut {NoStop}%
\bibitem [{\citenamefont {Kraus}\ and\ \citenamefont
  {Cirac}(2010)}]{kraus2010generalized}%
  \BibitemOpen
  \bibfield  {author} {\bibinfo {author} {\bibfnamefont {C.~V.}\ \bibnamefont
  {Kraus}}\ and\ \bibinfo {author} {\bibfnamefont {J.~I.}\ \bibnamefont
  {Cirac}},\ }\href@noop {} {\bibfield  {journal} {\bibinfo  {journal} {New J.
  Phys.}\ }\textbf {\bibinfo {volume} {12}},\ \bibinfo {pages} {113004}
  (\bibinfo {year} {2010})}\BibitemShut {NoStop}%
\bibitem [{\citenamefont {Brydges}\ \emph {et~al.}(2019)\citenamefont
  {Brydges}, \citenamefont {Elben}, \citenamefont {Jurcevic}, \citenamefont
  {Vermersch}, \citenamefont {Maier}, \citenamefont {Lanyon}, \citenamefont
  {Zoller}, \citenamefont {Blatt},\ and\ \citenamefont {Roos}}]{Elben}%
  \BibitemOpen
  \bibfield  {author} {\bibinfo {author} {\bibfnamefont {T.}~\bibnamefont
  {Brydges}}, \bibinfo {author} {\bibfnamefont {A.}~\bibnamefont {Elben}},
  \bibinfo {author} {\bibfnamefont {P.}~\bibnamefont {Jurcevic}}, \bibinfo
  {author} {\bibfnamefont {B.}~\bibnamefont {Vermersch}}, \bibinfo {author}
  {\bibfnamefont {C.}~\bibnamefont {Maier}}, \bibinfo {author} {\bibfnamefont
  {B.~P.}\ \bibnamefont {Lanyon}}, \bibinfo {author} {\bibfnamefont
  {P.}~\bibnamefont {Zoller}}, \bibinfo {author} {\bibfnamefont
  {R.}~\bibnamefont {Blatt}}, \ and\ \bibinfo {author} {\bibfnamefont {C.~F.}\
  \bibnamefont {Roos}},\ }\href@noop {} {\bibfield  {journal} {\bibinfo
  {journal} {Science}\ }\textbf {\bibinfo {volume} {364}},\ \bibinfo {pages}
  {260} (\bibinfo {year} {2019})}\BibitemShut {NoStop}%
\bibitem [{\citenamefont {Hastings}\ and\ \citenamefont
  {Koma}(2006)}]{HastingsKoma06}%
  \BibitemOpen
  \bibfield  {author} {\bibinfo {author} {\bibfnamefont {M.~B.}\ \bibnamefont
  {Hastings}}\ and\ \bibinfo {author} {\bibfnamefont {T.}~\bibnamefont
  {Koma}},\ }\href {\doibase 10.1007/s00220-006-0030-4} {\bibfield  {journal}
  {\bibinfo  {journal} {Commun. Math. Phys.}\ }\textbf {\bibinfo {volume}
  {265}},\ \bibinfo {pages} {781} (\bibinfo {year} {2006})}\BibitemShut
  {NoStop}%
\bibitem [{\citenamefont {Bera}\ \emph {et~al.}(2015)\citenamefont {Bera},
  \citenamefont {Schomerus}, \citenamefont {Heidrich-Meisner},\ and\
  \citenamefont {Bardarson}}]{FHM}%
  \BibitemOpen
  \bibfield  {author} {\bibinfo {author} {\bibfnamefont {S.}~\bibnamefont
  {Bera}}, \bibinfo {author} {\bibfnamefont {H.}~\bibnamefont {Schomerus}},
  \bibinfo {author} {\bibfnamefont {F.}~\bibnamefont {Heidrich-Meisner}}, \
  and\ \bibinfo {author} {\bibfnamefont {J.~H.}\ \bibnamefont {Bardarson}},\
  }\href {\doibase 10.1103/PhysRevLett.115.046603} {\bibfield  {journal}
  {\bibinfo  {journal} {Phys. Rev. Lett.}\ }\textbf {\bibinfo {volume} {115}},\
  \bibinfo {pages} {046603} (\bibinfo {year} {2015})}\BibitemShut {NoStop}%
\bibitem [{\citenamefont {Gluza}\ \emph {et~al.}(2018)\citenamefont {Gluza},
  \citenamefont {Kliesch}, \citenamefont {Eisert},\ and\ \citenamefont
  {Aolita}}]{FW}%
  \BibitemOpen
  \bibfield  {author} {\bibinfo {author} {\bibfnamefont {M.}~\bibnamefont
  {Gluza}}, \bibinfo {author} {\bibfnamefont {M.}~\bibnamefont {Kliesch}},
  \bibinfo {author} {\bibfnamefont {J.}~\bibnamefont {Eisert}}, \ and\ \bibinfo
  {author} {\bibfnamefont {L.}~\bibnamefont {Aolita}},\ }\href {\doibase
  10.1103/PhysRevLett.120.190501} {\bibfield  {journal} {\bibinfo  {journal}
  {Phys. Rev. Lett.}\ }\textbf {\bibinfo {volume} {120}},\ \bibinfo {pages}
  {190501} (\bibinfo {year} {2018})}\BibitemShut {NoStop}%
\bibitem [{\citenamefont {Arute}\ \emph {et~al.}(2020)\citenamefont {Arute},
  \citenamefont {Arya}, \citenamefont {Babbush}, \citenamefont {Bacon},
  \citenamefont {Bardin}, \citenamefont {Barends}, \citenamefont {Boixo},
  \citenamefont {Broughton}, \citenamefont {Buckley}, \citenamefont {Buell}
  \emph {et~al.}}]{arute2020hartree}%
  \BibitemOpen
  \bibfield  {author} {\bibinfo {author} {\bibfnamefont {F.}~\bibnamefont
  {Arute}}, \bibinfo {author} {\bibfnamefont {K.}~\bibnamefont {Arya}},
  \bibinfo {author} {\bibfnamefont {R.}~\bibnamefont {Babbush}}, \bibinfo
  {author} {\bibfnamefont {D.}~\bibnamefont {Bacon}}, \bibinfo {author}
  {\bibfnamefont {J.~C.}\ \bibnamefont {Bardin}}, \bibinfo {author}
  {\bibfnamefont {R.}~\bibnamefont {Barends}}, \bibinfo {author} {\bibfnamefont
  {S.}~\bibnamefont {Boixo}}, \bibinfo {author} {\bibfnamefont
  {M.}~\bibnamefont {Broughton}}, \bibinfo {author} {\bibfnamefont {B.~B.}\
  \bibnamefont {Buckley}}, \bibinfo {author} {\bibfnamefont {D.~A.}\
  \bibnamefont {Buell}},  \emph {et~al.},\ }\href@noop {} {\bibfield  {journal}
  {\bibinfo  {journal} {arXiv:2004.04174}\ } (\bibinfo {year}
  {2020})}\BibitemShut {NoStop}%
\bibitem [{\citenamefont {Flammia}\ and\ \citenamefont
  {Liu}(2011)}]{FidelityEstimation}%
  \BibitemOpen
  \bibfield  {author} {\bibinfo {author} {\bibfnamefont {S.~T.}\ \bibnamefont
  {Flammia}}\ and\ \bibinfo {author} {\bibfnamefont {Y.-K.}\ \bibnamefont
  {Liu}},\ }\href@noop {} {\bibfield  {journal} {\bibinfo  {journal} {Phys.
  Rev. Lett.}\ }\textbf {\bibinfo {volume} {106}},\ \bibinfo {pages} {230501}
  (\bibinfo {year} {2011})}\BibitemShut {NoStop}%
\bibitem [{\citenamefont {Nielsen}(1999)}]{PhysRevLett.83.436}%
  \BibitemOpen
  \bibfield  {author} {\bibinfo {author} {\bibfnamefont {M.~A.}\ \bibnamefont
  {Nielsen}},\ }\href {\doibase 10.1103/PhysRevLett.83.436} {\bibfield
  {journal} {\bibinfo  {journal} {Phys. Rev. Lett.}\ }\textbf {\bibinfo
  {volume} {83}},\ \bibinfo {pages} {436} (\bibinfo {year} {1999})}\BibitemShut
  {NoStop}%
\bibitem [{\citenamefont {Eisert}\ and\ \citenamefont
  {Cramer}(2005)}]{PhysRevA.72.042112}%
  \BibitemOpen
  \bibfield  {author} {\bibinfo {author} {\bibfnamefont {J.}~\bibnamefont
  {Eisert}}\ and\ \bibinfo {author} {\bibfnamefont {M.}~\bibnamefont
  {Cramer}},\ }\href {\doibase 10.1103/PhysRevA.72.042112} {\bibfield
  {journal} {\bibinfo  {journal} {Phys. Rev. A}\ }\textbf {\bibinfo {volume}
  {72}},\ \bibinfo {pages} {042112} (\bibinfo {year} {2005})}\BibitemShut
  {NoStop}%
\bibitem [{\citenamefont {Schuch}\ \emph
  {et~al.}(2004{\natexlab{b}})\citenamefont {Schuch}, \citenamefont
  {Verstraete},\ and\ \citenamefont {Cirac}}]{PhysRevA.70.042310}%
  \BibitemOpen
  \bibfield  {author} {\bibinfo {author} {\bibfnamefont {N.}~\bibnamefont
  {Schuch}}, \bibinfo {author} {\bibfnamefont {F.}~\bibnamefont {Verstraete}},
  \ and\ \bibinfo {author} {\bibfnamefont {J.~I.}\ \bibnamefont {Cirac}},\
  }\href {\doibase 10.1103/PhysRevA.70.042310} {\bibfield  {journal} {\bibinfo
  {journal} {Phys. Rev. A}\ }\textbf {\bibinfo {volume} {70}},\ \bibinfo
  {pages} {042310} (\bibinfo {year} {2004}{\natexlab{b}})}\BibitemShut
  {NoStop}%
\bibitem [{\citenamefont {Gluza}\ \emph {et~al.}(2016)\citenamefont {Gluza},
  \citenamefont {Krumnow}, \citenamefont {Friesdorf}, \citenamefont {Gogolin},\
  and\ \citenamefont {Eisert}}]{GKFGE16}%
  \BibitemOpen
  \bibfield  {author} {\bibinfo {author} {\bibfnamefont {M.}~\bibnamefont
  {Gluza}}, \bibinfo {author} {\bibfnamefont {C.}~\bibnamefont {Krumnow}},
  \bibinfo {author} {\bibfnamefont {M.}~\bibnamefont {Friesdorf}}, \bibinfo
  {author} {\bibfnamefont {C.}~\bibnamefont {Gogolin}}, \ and\ \bibinfo
  {author} {\bibfnamefont {J.}~\bibnamefont {Eisert}},\ }\href {\doibase
  10.1103/PhysRevLett.117.190602} {\bibfield  {journal} {\bibinfo  {journal}
  {Phys. Rev. Lett.}\ }\textbf {\bibinfo {volume} {117}},\ \bibinfo {pages}
  {190602} (\bibinfo {year} {2016})}\BibitemShut {NoStop}%
\bibitem [{\citenamefont {Wolf}(2006)}]{AreViolationWolf}%
  \BibitemOpen
  \bibfield  {author} {\bibinfo {author} {\bibfnamefont {M.~M.}\ \bibnamefont
  {Wolf}},\ }\href {\doibase 10.1103/PhysRevLett.96.010404} {\bibfield
  {journal} {\bibinfo  {journal} {Phys. Rev. Lett.}\ }\textbf {\bibinfo
  {volume} {96}},\ \bibinfo {pages} {010404} (\bibinfo {year}
  {2006})}\BibitemShut {NoStop}%
\bibitem [{\citenamefont {Peschel}(2004)}]{Peschel}%
  \BibitemOpen
  \bibfield  {author} {\bibinfo {author} {\bibfnamefont {I.}~\bibnamefont
  {Peschel}},\ }\href@noop {} {\bibfield  {journal} {\bibinfo  {journal} {J.
  Stat. Mech.}\ ,\ \bibinfo {pages} {P12005}} (\bibinfo {year}
  {2004})}\BibitemShut {NoStop}%
\end{thebibliography}
\end{document}